\documentclass[11pt]{article}
\usepackage{setspace}

\pdfoutput=1
\usepackage[utf8]{inputenc}
\usepackage[icelandic, english]{babel}
\usepackage{t1enc}
\usepackage{graphicx}
\usepackage[intoc]{nomencl}
\usepackage{enumerate,color}
\usepackage{url}
\usepackage{amsmath}
\usepackage{amsfonts}
\usepackage{thmtools}
\usepackage{amssymb}
\usepackage[nottoc]{tocbibind}
\usepackage[sf,normalsize]{subfigure}
\usepackage[format=plain,labelformat=simple,labelsep=colon]{caption}
\usepackage{placeins}
\usepackage{tabularx}
\usepackage[T1]{fontenc}
\usepackage{algorithm}
\usepackage[noend]{algpseudocode}
\usepackage{natbib}
\usepackage{booktabs}
\usepackage[table,dvipsnames]{xcolor}
\usepackage{aps-bibstyle}   

\newcommand{\fat}[1]{\boldsymbol{#1}}
\newcommand{\trp}{^\mathsf{T}}

\newcounter{lastnote}

\title{Max-and-Smooth: a two-step approach for approximate Bayesian inference in latent Gaussian models} 

\author
{Birgir Hrafnkelsson,$^{1\ast}$ Stefan Siegert,$^{2}$ Rapha\"el Huser,$^{3}$     \\
Haakon Bakka,$^{4}$ \'Arni V. Johannesson$^{1}$  \\
\normalsize{$^{1}$University of Iceland,}\\
\normalsize{$^{2}$University of Exeter}\\
\normalsize{$^{3}$King Abdullah University of Science and Technology}\\
\normalsize{$^{4}$University of Oslo}\\
}

\date{}

\begin{document} 




\baselineskip12pt


\maketitle 


\begin{abstract}
With modern high-dimensional data, complex statistical models are necessary, requiring computationally feasible inference schemes. We introduce Max-and-Smooth, an approximate Bayesian inference scheme for a flexible class of latent Gaussian models (LGMs) where one or more of the likelihood parameters are modeled by latent additive Gaussian processes. Our proposed inference scheme is a two-step approach. In the first step (Max), the likelihood function is approximated by a Gaussian density with mean and covariance equal to either (a) the maximum likelihood estimate and the inverse observed information, respectively, or (b) the mean and covariance of the normalized likelihood function. In the second step (Smooth), the latent parameters and hyperparameters are inferred and smoothed with the approximated likelihood function. The proposed method ensures that the uncertainty from the first step is correctly propagated to the second step. Because the prior density for the latent parameters is assumed to be Gaussian and the approximated likelihood function is Gaussian, the approximate posterior density of the latent parameters (conditional on the hyperparameters) is also Gaussian, thus facilitating efficient posterior inference in high dimensions. Furthermore, the approximate marginal posterior distribution of the hyperparameters is tractable, and as a result, the hyperparameters can be sampled independently of the latent parameters. We show that the computational cost of Max-and-Smooth
is close to being insensitive to the number of independent data replicates, and that it scales well with increased dimension of the latent parameter vector provided that its Gaussian prior density is specified with a sparse precision matrix. In the case of a large number of independent data replicates, sparse precision matrices, and high-dimensional latent vectors, the speedup is substantial in comparison to an MCMC scheme that infers the posterior density from the exact likelihood function. The accuracy of the Gaussian approximation to the likelihood function increases with the number of data replicates per latent model parameter. The proposed inference scheme is demonstrated on one spatially referenced real dataset and on simulated data mimicking spatial, temporal, and spatio-temporal inference problems. Our results show that Max-and-Smooth is accurate and fast.
\end{abstract}


\newpage

\section{Introduction}
\pagenumbering{arabic}

Data are being generated today at an unprecedented rate. 
Many datasets are large and exhibit complex marginal behaviors and dependence structures.
In particular, data that are indexed in space and time may indicate non-linear time trends and spatial patterns, and may be driven by complex space-time interactions.
Statistical modeling and inference for high-dimensional spatio-temporal datasets becomes increasingly computationally demanding as the spatial and/or temporal dimension increases \citep[see, e.g.,][]{Heaton.etal:2019}. The same is true for, e.g., multinomial data \citep[see, e.g.,][]{gustafson2008bayesian} and survival data \citep[see, e.g.,][]{lee2011bayesian}, i.e., as the number of units or individuals increases the computational cost increases.

In this paper we focus on latent Gaussian models (LGMs), which form a general and very flexible class of models that has proven to be useful in a wide range of concrete applications \citep[see, e.g.,][]{gelfand2007multilevel,cooley2007bayesian,rue2009approximate,margeirsson2010decision,sigurdarson2016precip,zinszer2017ebola,opitz2018inla,lombardo2018landslides,lombardo2019landslides}.
We here introduce Max-and-Smooth, 
a novel approximate Bayesian inference procedure for LGMs with independent data replicates that is both accurate and fast, providing significant speedups in high dimensions. Our approach has superficial similarities with the recent contribution by \citet{risser2019extremeprecip}, who propose a frequentist two-step inference approach and focus on the GEV distribution. In contrast, our proposed inference scheme is more general and designed for fully Bayesian inference. LGMs are Bayesian hierarchical models \citep{cressie2011spatiotemporal,banerjee2014hierarchical} that consist of three levels: the data level, the latent level and the hyperparameter level. Each level is specified with a probability distribution, the one at the latent level being Gaussian.
For computational reasons, it is common to assume that the data are conditionally independent, given the latent process, and we also assume this here. The role of the latent process is to capture the underlying dynamics of the data (such as space-time dependence interactions).
Our focus in this paper is mainly on three types of spatio-temporal LGMs that are useful in different settings, although our proposed method applies more generally, e.g., with (replicated) multinomial or survival data.
The first type of LGMs that we consider assumes that the spatio-temporal dynamics of the data is  
described by latent parameters that vary spatially but are constant in time, and that (a potentially different number of) data time replicates are available at each spatial location. 
This type of LGMs focuses on capturing the data's spatial behavior, although datasets with temporal covariates or slowly-varying temporal trends may also be modeled in the same framework. 
The second type of LGMs assumes that the latent parameters vary in time, and that several spatial replicates of the data are available at each time point. In this setting the latent parameters are usually constant in space, although they may also refer to the main effects of distinct regions that vary over time. The focus is therefore on capturing the data's temporal behavior.
Finally, the third type of LGMs assumes that the latent process varies in both space and time, and that several replicates for each spatio-temporal observation are available.


Several strategies have been proposed to fit LGMs. Simulation-based Markov chain Monte Carlo (MCMC) methods can be used \citep[see][]{cressie2011spatiotemporal,banerjee2014hierarchical}, although their application in high-dimensional settings (i.e., situations with either space-rich and/or time-rich data, or with many parameters involved at the latent level) may be limited by the computational complexity.
In order to make MCMC sampling more efficient, \cite{knorr2002block} proposed a single block updating strategy for LGMs characterized by a univariate link function. 
Their strategy reduces the cross-correlation between the hyperparameters and the latent parameters within the posterior samples. 
A detailed comparison of several sampling strategies for LGMs in \cite{filippone2013comparative} showed that the single block updating strategy of \cite{knorr2002block} has larger effective sample size compared to sufficient augmentation, ancillary augmentation and ancillarity-sufficiency interweaving strategy \citep{yu2011center}, and the surrogate method \citep{murray2010slice}. 
Another approach to infer LGMs was proposed by \cite{filippone2014pseudo}, who suggested using a pseudo-marginal sampling procedure for the marginal posterior density of the hyperparameters, which relies on the Metropolis-Hastings algorithm and importance sampling. 
Essentially, samples from the marginal posterior density of the hyperparameters are obtained first, and the latent parameters are sampled from the conditional posterior density of the latent parameters. 
\cite{filippone2014pseudo} compared their pseudo-marginal approach to ancillary augmentation \citep{yu2011center} and the surrogate method \citep{murray2010slice}, and found that the effective sample size of the hyperparameters was much lower for ancillary augmentation and the surrogate method than for the pseudo-marginal approach. 
\cite{filippone2014pseudo} concluded that this was due to ancillary augmentation and the surrogate method not being fully capable of breaking down the correlation between hyperparameters and latent parameters. 
The findings of \cite{filippone2013comparative} and \cite{filippone2014pseudo} underline that sampling from the marginal posterior density of the hyperparameters in an LGM leads to more effective sampling schemes.

Alternatively, the integrated nested Laplace approximation (INLA) has proven to be very fast and accurate for approximate Bayesian inference in LGMs \citep{rue2009approximate}. The INLA methodology essentially bypasses MCMC sampling by performing a numerical approximation of the posterior density. Due to its computational efficiency and its convenient implementation in the package {\tt INLA} for the {\tt R} statistical computing environment,
the INLA method has found widespread interest, and has been applied in numerous settings; see the review papers by \citet{rue2017bayesian} and \citet{bakka2018spatial}, and references therein. However, the current implementation in the {\tt INLA} package only supports LGMs characterized by a univariate link function (i.e., with one single Gaussian linear predictor at the latent level), and with a small number (typically less than $20$) of hyperparameters. 
In Section~\ref{sec:application}, we discuss a linear regression model for spatio-temporal meteorological data, where the intercept, the covariate effect, and the residual variance vary spatially, thus requiring a trivariate link function in the likelihood. 
Generally speaking, it is common to assume that spatio-temporal data are described by LGMs of type (i), (ii) or (iii) above with multiple parameters (e.g., intercept, multiple covariate effects, scale and shape parameters, etc.) that vary spatially and/or temporally. 
These types of LGMs usually require multivariate link functions, and we will hereafter refer to models of this type as \emph{extended LGMs}
in line with \cite{geirsson2020mcmc}. 
Although it might be possible in principle to extend the {\tt INLA} software to LGMs with multivariate link functions, this has not been implemented yet.

Posterior inference for extended LGMs in moderate or high dimensions is known to be challenging. \cite{geirsson2020mcmc} developed an efficient block Gibbs sampling scheme, referred to as the LGM sampler, which was shown to significantly reduce the autocorrelation in posterior samples. 
In this paper, we propose using Max-and-Smooth, a novel two-step approximate Bayesian inference approach for extended LGMs, which borrows ideas from the INLA method and the LGM sampler, in order to be both fast and accurate in high dimensions. 
Essentially, our approach approximates the likelihood function by a Gaussian likelihood, similar to the Laplace approximation used in the INLA method. This allows us to perform fast inference with a correct propagation of the uncertainty. The two steps of the inference scheme are as follows: (i) In the first step (Max), we compute the maximum likelihood (ML) estimates of the latent parameters at each spatial, temporal, or spatio-temporal point (depending on the type of LGM considered), and we approximate the variance of the Gaussian approximation using the inverse observed information evaluated at the ML estimate. We also consider an alternative Gaussian approximation that uses the mean and covariance of the normalized likelihood function.
(ii) In the second step (Smooth), we treat the ML estimate (or the mean of the normalized likelihood) as the observed data of the latent parameters, with a Gaussian likelihood (thus taking their estimated variance into account). We then fit the latent Gaussian model by taking advantage of the conjugacy properties of the approximate Gaussian--Gaussian model, which is hereafter referred to as the pseudo model. In other words, we essentially consider that the parameter estimates from the first step are noisy measurements of the latent field (with known noise variance) and we smooth them jointly in the second step. Notice that although our proposed approach has two consecutive steps, it properly propagates the uncertainty, and thus provides a valid approximate procedure to sample from the full posterior density. 
Our proposed procedure is very fast for a variety of reasons. First, the (approximate) sampling scheme is such that the unnormalized marginal posterior density of the hyperparameters can be expressed analytically, making it straightforward to sample from it. 
Second, the conditional density of the latent parameters given the hyperparameters is Gaussian, which is straightforward to sample from. 
Third, because the hyperparameters can be sampled independently from the latent parameters, similarly to \citet{geirsson2020mcmc}, their cross-correlation is reduced, which yields better MCMC mixing properties with a higher effective sample size. 
Fourth, as the computational cost of the second step (i.e., fitting the pseudo model) does not depend on the number of data replicates, our proposed procedure is especially well-suited for datasets with a large number of independent replicates.
Finally, further speed-ups can be obtained by specifying the Gaussian prior density for the parameter vector at the latent level to be a Gaussian Markov random field \citep[GMRF,][]{rue2005gaussian} with a sparse precision matrix. When such GMRFs are used, Max-and-Smooth  
scales well with increasing dimension of the latent parameter vector. 

Our proposed methodology involves approximations at two levels. 
First, the likelihood function is approximated by a Gaussian likelihood and may therefore be misspecified. 
Second, the variance of the Gaussian approximation to the likelihood has to be estimated from data in the first step, but is then treated as exact in the second step. 
Intuitively, if the shape of the ``true'' likelihood function is close to a Gaussian likelihood, our inference approach will be accurate. With a perfectly (or nearly) Gaussian likelihood, our inference approach will be very close to being exact provided the variance of the Gaussian approximation is properly estimated. In contrast, when the number of data replicates per latent model parameter is low, the Gaussian approximation may become a poor approximation to the likelihood, which might negatively impact the posterior inference. 
However, owing to the asymptotic behavior of the likelihood function in posterior inference, see, e.g., \citet{schervish1995theory}, the errors of these two levels of approximation will typically become negligible as the number of data replicates per latent model parameter grows. 
In other words, Max-and-Smooth 
is expected to perform increasingly well as the number of data replicates gets larger, with a negligible effect on the overall computational time. 
The Gaussian approximation based on the normalized likelihood function is more suitable for highly non-Gaussian likelihood functions, because the mean and variance are propagated more precisely compared to the Gaussian approximation based on the ML estimate.

The paper is organized as follows. 
In Section~\ref{sec:ExtendedLGMs}, we introduce the extended latent Gaussian modeling framework, and in Section~\ref{sec:LGMInf} we detail our proposed approximate Bayesian inference methodology and introduce Max-and-Smooth. 
In Section~\ref{sec:simulated_data} we illustrate the strengths and weaknesses of our approach by simulation studies using different types of extended LGMs. 
We apply the proposed methodology to a real dataset in Section~\ref{sec:application}. 
Finally, Section~\ref{sec:discussion} concludes with a discussion and directions for future research.

\section{Extended latent Gaussian models}\label{sec:ExtendedLGMs}

\subsection{LGMs with a univariate link function}
\label{sec:lgms1}

Latent Gaussian models are a subset of Bayesian hierarchical models in which parameters at the latent level have a joint Gaussian prior distribution,
conditional on hyperparameters. LGMs are a 
subclass of structured additive regression models where the observations, $y_i$ ($i = 1, \dots, n$),
are assumed to have a density from the exponential family and the mean or a particular quantile, 
$\mu_i$, is then linked to a structured additive predictor, $\eta_i$, through 
a univariate link function $g(\cdot)$ such that $g(\mu_i)=\eta_i$; see \cite{rue2009approximate}. 
The structured additive predictor can then accommodate covariates 
and random effects in an additive way, namely,
\begin{equation}
\label{eq:RegrAdd}
\eta_i=\beta_0 + \sum_{k=1}^K\beta_k z_{i,k} + \sum_{j=1}^J u_j a_{i,j}  +\epsilon_i,
\end{equation}
where $\beta_0$ is an intercept, $\{\beta_k\}$ are linear fixed effects of covariates $z_{i,1},\ldots,z_{i,K}$, 
$\{u_j\}$ are unknown random effects with some specified dependency structure, and with known weights $a_{i,1},\ldots,a_{i,J}$,
and $\epsilon_{1},\ldots,\epsilon_n$ are unstructured model errors.  
In LGMs, the terms $\beta_0$, $\{\beta_k\}$, 
$\{u_j\}$ and $\{\epsilon_i\}$ 
all have Gaussian prior distributions. Let $\fat{x}$ contain the latent 
parameters, namely, $\beta_0$, $\{\beta_k\}$, $\{u_j\}$  
and $\{\epsilon_i\}$. Sometimes, the vector $\fat{x}$ consists of 
$\beta_0$, $\{\beta_k\}$, $\{u_j\}$ and $\{\eta_i\}$, i.e., $\{\eta_i\}$ is 
included instead of $\{\epsilon_i\}$. Either way, the parameters in 
$\fat{x}$ have a joint Gaussian prior distribution, conditional on hyperparameters, $\fat{\theta}=(\theta_1,\ldots,\theta_p)\trp$. 
The hyperparameters usually do not have a Gaussian prior density. 
Typically, hyperparameters specify the marginal
variance, correlation range, smoothness, and/or other correlation parameters of
the random effects.
Schematically, LGMs with a univariate link function may be represented hierarchically as follows in terms of the data level, the latent level and the hyperparameter level:

\paragraph{Data level: }The observations $\fat{y}=(y_1,\ldots,y_n)\trp$ are assumed to be dependent 
on the latent parameters $\fat{x}$ and have a density $\pi(\fat{y}|\fat{x},\fat{\theta})$.
Often, conditional independence is assumed for simplicity, that is, $\pi(\fat{y}|\fat{x},\fat{\theta})$ factorizes as
$\prod_{i}\pi(y_i|\fat{x},\fat{\theta})$ 
where $\pi(y_i|\fat x,\fat \theta) = \pi(y_i|\eta_i,\fat\theta)$
and $\mu_i=\textrm{E}(y_i|\eta_i)=g^{-1}(\eta_i)$ or $\mu_i=\textrm{Q}_p(y_i|\eta_i)=g^{-1}(\eta_i)$ if $\mu_i$ is a quantile defined in terms of an appropriate quantile function $\textrm{Q}_p$.

\paragraph{Latent level: }The latent parameters $\fat{x}$ have a Gaussian prior density and are potentially dependent on some hyperparameters $\fat{\theta}$. The density of $\fat{x}$  may be written as 
$$
\pi(\fat{x}|\fat{\theta})=\textrm{N}(\fat{x}|\fat{\mu}(\fat{\theta}),\Sigma(\fat{\theta})),
$$
where the right-hand side denotes a Gaussian density with mean vector $\fat{\mu}(\fat{\theta})$ and covariance matrix $\Sigma(\fat{\theta})$. This is in line with the notation 
in \cite{gelman2013bayesian}.

\paragraph{Hyperparameter level: } The hyperparameters $\fat{\theta}$ are assigned a prior density $\pi(\fat{\theta})$.

An LGM is fully specified by the definition of these three levels. In the next section, we extend this framework to LGMs characterized by a multivariate link function.

\subsection{LGMs with a multivariate link function}
\label{sec:ELGMModels} 

LGMs with a multivariate link function are referred to as extended LGMs \citep{geirsson2020mcmc}.
We assume here conditional independence at the data level for simplicity,
and that the data can be lined up according to groups, e.g., sites, time points, spatio-temporal elements or categories. 
The models presented here have the same structure as the LGMs in Section \ref{sec:lgms1}, except that the assumption of the data density being in the exponential family is dropped and the vector $\fat{x}$ refers to several 
subsets of parameters found at the data level, each subset with its 
separate set of linear predictors at the latent level. In contrast, in the case of classical LGMs defined in \cite{rue2009approximate}, only one single parameter of the data distribution is modeled at the latent level.  

We assume that each group $i$ has $n_i$ observations. The total number of groups is $G$. Observations from the same group or from different groups are assumed to be conditionally independent given the latent
process. The groups can represent various types of sampling setups. For example, the groups may be geological sites observed over time, i.e., each group corresponds to a site; or the groups may be time points where several observations are made at the same time point. The groups may also be spatio-temporal elements such that multiple observations are collected 
for each spatially and temporally-referred element. Furthermore, the groups may also represent generic categories that do not have any spatial nor temporal reference, yet, several observations are made within each category.    
Each group is described by $M$ parameters. The general setup is such that the $M$ subsets of parameters at the data level are mapped to $M$ subsets of parameters at the latent level through an $M$-variate link function. Each of these subsets at the latent level is modeled with a linear model of the form in (\ref{eq:RegrAdd}).

The probability density function of $y_{i,j}$, the $j$-th observation from group $i$, is denoted by
$\pi(y_{i,j}|\psi_{1,i},\psi_{2,i},\ldots,\psi_{M,i})$ where $\psi_{1,i},\ldots,\psi_{M,i}$ are 
the parameters within group $i$ such that $(\psi_{1,i},\ldots,\psi_{M,i}) \in \mathcal{D}$, and $\mathcal{D}$ is a subspace of $\mathbb{R}^M$.
Let $\fat{y}$ be the vector containing all the observations, 
let $\fat{y}_i$ be the observations from group $i$,
and let
$$
\fat{\psi}_1 = (\psi_{1,1},\ldots,\psi_{1,G})\trp, \ 
\fat{\psi}_2 = (\psi_{2,1},\ldots,\psi_{2,G})\trp, \
\ldots, \ 
\fat{\psi}_M = (\psi_{M,1},\ldots,\psi_{M,G})\trp,
$$
denote the $M$ subsets of parameters at the data level, each vector containing only one type of parameters, e.g., all the location parameters or the regression slope coefficients for all groups.
Conditional on $\fat{\psi}_1$, $\fat{\psi}_2$, \ldots, $\fat{\psi}_M$, the probability density function of $\fat{y}$ is
$$
\pi(\fat{y}|\fat{\psi}_1,\fat{\psi}_2,\ldots,\fat{\psi}_M)
=\prod_{i=1}^{G}\pi(\fat{y}_{i}|\psi_{1,i},\psi_{2,i},\ldots,\psi_{M,i})
= \prod_{i=1}^{G}\prod_{j \in \mathbb{A}_i}
\pi(y_{ij}|\psi_{1,i},\psi_{2,i},\ldots,\psi_{M,i}),
$$
where $\mathbb{A}_i$ is an index set for group $i$. Let  
$g$ be an $M$-variate link function such that $g:\mathcal{D}\rightarrow \mathbb{R}^M$ 
with
$g(\psi_{1,i},\psi_{2,i},\ldots,\psi_{M,i})=(\eta_{1,i},\eta_{2,i},\ldots,\eta_{M,i})\in \mathbb{R}^M$, so the domain of each $\eta_{m,i}$ is the whole real line. Linear models for the $M$ subsets of parameters at the latent level, i.e., the vectors
$$
\fat{\eta}_1 = (\eta_{1,1},\ldots,\eta_{1,G})\trp, \ 
\fat{\eta}_2 = (\eta_{2,1},\ldots,\eta_{2,G})\trp, \ 
\ldots, \ 
\fat{\eta}_M = (\eta_{M,1},\ldots,\eta_{M,G})\trp, \ 
$$
are then specified  
and they may be expressed in vector notation as
\begin{equation}
\label{eq:latentLevelELGM}
\begin{aligned}
\fat{\eta}_1 &= X_{1} \fat{\beta}_{1}
+ A_{1} \fat{u}_{1} + \fat{\epsilon}_{1},\\
\fat{\eta}_2 &= X_{2} \fat{\beta}_{2}
+ A_{2} \fat{u}_{2} + \fat{\epsilon}_{2},\\
\vdots  \\
\fat{\eta}_M &= X_{M} \fat{\beta}_{M}
+ A_{M} \fat{u}_{M} + \fat{\epsilon}_{M},
\end{aligned}
\end{equation}
where $\fat{\beta}_1$, $\fat{\beta}_2$, \ldots, $\fat{\beta}_M$ are fixed effects, $X_{1}$, $X_{2}$, \ldots, $X_{M}$ are the corresponding 
design matrices containing covariate information, $\fat{u}_{1}$, $\fat{u}_{2}$, \ldots, $\fat{u}_{M}$ are random effects,  $A_1$, $A_2$, \ldots, $A_M$ are their corresponding weight matrices, and $\fat{\epsilon}_{1}$, $\fat{\epsilon}_{2}$, \ldots, $\fat{\epsilon}_{M}$ are independent and unstructured error terms, referred to as model errors.  
The terms $\fat{\beta}_m$, $\fat{u}_{m}$, and $\fat{\epsilon}_{m}$, $m=1,2,\ldots, M$, are assigned Gaussian 
prior densities and assumed to be a priori mutually independent.

In the following section, we develop an approximate inference scheme for the LGMs with a multivariate link function.

\section{Approximate Bayesian inference for extended LGMs}
\label{sec:LGMInf}

\subsection{General idea}
In this section, we detail an approximation to the posterior density used to compute inference for extended LGMs 
(recall Section \ref{sec:ELGMModels}). 
We then introduce Max-and-Smooth,  
a two-step approach that is fully Bayesian and utilizes a Gaussian approximation to the likelihood. This approximation together with the conditional Gaussian prior at the latent level results in a Gaussian--Gaussian model (conditional on the hyperparameters) that is referred to here as the pseudo model. We perform inference both for the hyperparameters and for the latent parameters in the original extended LGM by exploiting this Gaussian--Gaussian model.
In Section~\ref{sec:postdensLGM}, we describe the posterior density of extended LGMs, and in Section~\ref{sec:twostepapproach} we detail the two consecutive steps of our inference approach. 
The first step relies on a Gaussian approximation to the likelihood function. We actually propose two different approximations; one is based on the ML estimates and the inverse of the observed information, while the other one is based on the mean and covariance of the normalized likelihood function. The second step consists in inferring the resulting Gaussian--Gaussian pseudo  model. For more details about the approximate inference scheme, see the Supplementary Material.

\subsection{The posterior density of extended LGMs}
\label{sec:postdensLGM}

The vectors for the model in   
(\ref{eq:latentLevelELGM}) are gathered as follows
$$
\fat{\eta} = (\fat{\eta}_1\trp,\fat{\eta}_2\trp,\ldots,\fat{\eta}_M\trp)\trp,
$$
$$
\fat{\nu} = 
(\fat{\beta}_{1}\trp,\fat{u}_{1}\trp,
\fat{\beta}_{2}\trp,\fat{u}_{2}\trp,\ldots,\fat{\beta}_{M}\trp,\fat{u}_{M}\trp)\trp,
$$
$$
\fat{\epsilon}=(\fat{\epsilon}_1\trp,\fat{\epsilon}_{2}\trp,\ldots,\fat{\epsilon}_{M}\trp)\trp.
$$
A priori, the vectors $\fat{\beta}_{1}$, 
$\fat{u}_{1}$,
$\fat{\beta}_{2}$,
$\fat{u}_{2}$,\ldots, 
$\fat{\beta}_{M}$ and
$\fat{u}_{M}$ are assumed to be independent.
Denote the means and precision (i.e., inverse covariance) matrices of these vectors by
$\fat{\mu}_{\beta,1}$, 
$\fat{\mu}_{u,1}$,
$\fat{\mu}_{\beta,2}$,
$\fat{\mu}_{u,2}$,\ldots,
$\fat{\mu}_{\beta,M}$, 
$\fat{\mu}_{u,M}$, 
$Q_{\beta,1}$, 
$Q_{u,1}$,
$Q_{\beta,2}$,
$Q_{u,2}$, \ldots,
$Q_{\beta,M}$ and
$Q_{u,M}$, respectively.
The prior mean of $\fat{\nu}$ is therefore
$$
\fat{\mu}_{\nu} =
(\fat{\mu}_{\beta,1}\trp, 
\fat{\mu}_{u,1}\trp,
\fat{\mu}_{\beta,2}\trp,
\fat{\mu}_{u,2}\trp,\ldots,
\fat{\mu}_{\beta,M}\trp,
\fat{\mu}_{u,M}\trp)\trp,
$$
while the precision matrix 
of $\fat{\nu}$ is a block diagonal matrix, 
$$
Q_{\nu} =
\textrm{bdiag}
(Q_{\beta,1}, 
Q_{u,1},
Q_{\beta,2},
Q_{u,2},\ldots,
Q_{\beta,M},
Q_{u,M}).
$$
The precision matrices of $\fat{\epsilon}_{1}$, $\fat{\epsilon}_{2}$, \ldots, 
$\fat{\epsilon}_{M}$ are diagonal matrices that are denoted by
$Q_{\epsilon,1}=\sigma_{\epsilon,1}^{-2}I$,
$Q_{\epsilon,2}=\sigma_{\epsilon,2}^{-2}I$, \ldots, 
$Q_{\epsilon,M}=\sigma_{\epsilon,M}^{-2}I$, respectively, where $I$ is the identity matrix and $\sigma_{\epsilon,1}$, $\sigma_{\epsilon,2}$, \ldots, $\sigma_{\epsilon,M}$ are the corresponding standard
deviations. The precision matrix of $\fat{\epsilon}$ is thus given by
$$
Q_{\epsilon} =
\textrm{bdiag}
(Q_{\epsilon,1}, 
Q_{\epsilon,2},\ldots,
Q_{\epsilon,M}).
$$
Define the matrix $Z$ based on $X_{1}$, $A_{1}$, $X_{2}$, $A_{2}$,\ldots, $X_{M}$ and $A_{M}$ as  
$$
Z=
 \begin{pmatrix} 
     X_1 &  A_1 & 0 & 0 & 0 & 0 & 0 & 0 & 0 & 0 \\
    0 & 0 & X_2 & A_2 & 0 & 0 & 0 & 0 & 0 & 0 \\
    \vdots & \vdots & \vdots & \vdots & \ddots & \ddots & \vdots & \vdots & \vdots & \vdots \\    
    0 & 0 & 0 & 0 & 0 & 0  & X_{M-1} & A_{M-1} & 0 & 0 \\
    0 & 0 & 0 & 0 & 0 & 0 & 0 & 0 & X_M & A_M \\
  \end{pmatrix},
$$
where the zeros denote zero matrices. Simple matrix multiplication
implies that (\ref{eq:latentLevelELGM}) can be rewritten more compactly as 
$$
\fat{\eta}=Z \fat{\nu} + \fat{\epsilon}.
$$
Since the model is an extended LGM, the prior densities of $\fat{\nu}$, and $\fat{\eta}$ conditional on $\fat{\nu}$, are assumed to be Gaussian. Furthermore, the precision matrices $Q_{\beta,1}$, $Q_{\beta,2}$, \ldots, $Q_{\beta,M}$ are assumed to be fixed, while hyperparameters govern the precision matrices for the random effects $\fat{u}_1$, $\fat{u}_2$, \ldots, $\fat{u}_M$, i.e., $Q_{u,1}$, $Q_{u,2}$, \ldots, $Q_{u,M}$. 
When the dimensions of $\fat{u}_1$, $\fat{u}_2$, \ldots, $\fat{u}_M$ are large, and we require fast computation, then $Q_{u,1}$, $Q_{u,2}$, \ldots, $Q_{u,M}$ need to be sparse and 
thus specifying them with Gaussian Markov random fields \citep[GMRF,][]{rue2005gaussian} becomes crucial.

The joint posterior density of $(\fat{\eta},\fat{\nu},\fat{\theta})$ may be expressed as 
$$
\pi(\fat{\eta},\fat{\nu},\fat{\theta}|\fat{y}) \propto\pi(\fat{\theta})\pi(\fat{\eta},\fat{\nu}|\fat{\theta})\pi(\fat{y}|\fat{\eta}),
$$
where $\fat{y}$ denotes the data vector, $\pi(\fat{y}|\fat{\eta})$ is the data density defined at the data level, $\pi(\fat{\eta},\fat{\nu}|\fat{\theta})$ is the Gaussian prior density defined at the latent level, and $\pi(\fat{\theta})$ is the prior density for the hyperparameters defined at the hyperparameter level. When the data density $\pi(\fat{y}|\fat{\eta})$ is used for inference, it is referred to as 
the likelihood function and viewed as a function of $\fat{\eta}$. We next describe our proposed approximate inference scheme.

\subsection{Max-and-Smooth: a two-step approximate inference approach}
\label{sec:twostepapproach}
Our proposed approximate Bayesian inference scheme (Max-and-Smooth) is based on approximating the likelihood function with a Gaussian density function (Step 1, Max), and on fitting the resulting Gaussian--Gaussian pseudo model (Step 2, Smooth). We now describe each step separately.

\subsubsection{Step 1 (Max): Gaussian approximation of the likelihood function}
\label{sec:step1}
We here propose two different Gaussian likelihood approximations that we then subsequently exploit for fast fully Bayesian inference.

The first Gaussian approximation is based on the mode of the likelihood function, i.e., the ML estimate (hence the term ``Max''), and the observed information evaluated at the ML estimate.
Let $L(\fat{\eta}|\fat{y})$ denote the likelihood function, where 
$L(\fat{\eta}|\fat{y}) = \pi(\fat{y}|\fat{\eta})$,
and let $\hat{L}$ denote the first Gaussian approximation; then, $c\hat{L}(\fat{\eta}|\fat{y})\approx L(\fat{\eta}|\fat{y})$,
where 
$$
\hat{L}(\fat{\eta}|\fat{y})=\textrm{N}(\fat{\eta}|\fat{\hat{\eta}},\Sigma_{\eta y}),
$$  
$c$ is a constant independent of $\fat{\eta}$, 
$\hat{\fat{\eta}}$ is the ML estimate for $\fat{\eta}$, i.e., it is the mode of $L(\fat{\eta}|\fat{y})$, 
and $\Sigma_{\eta y}=(-H_{\eta y})^{-1}$,
where $H_{\eta y}$ denotes the Hessian matrix of $\log(L(\fat{\eta}|\fat{y}))$ evaluated at 
$\fat{\eta}=\hat{\fat{\eta}}$. Furthermore, the observed information evaluated at $\hat{\fat{\eta}}$ is written as $\mathcal{I}_{\eta y}=-H_{\eta y}$. 

Due to the assumed conditional independence, the first Gaussian approximation is straightforward to evaluate because it can be computed for each group separately. More precisely, let $L(\fat{\eta}_i|\fat{y}_i)=\pi(\fat{y}_i|\fat{\eta}_i)$ denote the likelihood contribution of the $i$-th group such that 
$L(\fat{\eta}|\fat{y})=\prod_{i=1}^{G}L(\fat{\eta}_i|\fat{y}_i)$. 
The ML estimate of $\fat{\eta}_i$, $\hat{\fat{\eta}}_i$, is the mode of $L(\fat{\eta}_i|\fat{y}_i)$, and the inverse of the covariance matrix
is equal to the observed information in $L(\fat{\eta}_i|\fat{y}_i)$ evaluated at $\hat{\fat{\eta}}_i$, i.e.,
$$
\mathcal{I}_{\eta y_i}=-\nabla^2 \log (L( \fat{\eta}_i|\fat{y}_i))|_{\eta_i=\hat{\eta}_i},
$$
and the inverse of $\mathcal{I}_{\eta y_i}$ is equal to the covariance matrix $\Sigma_{\eta y_i}$.
Now we can approximate the likelihood contribution of group $i$, $L(\fat{\eta}_i|\fat{y}_i)$, with 
$c_i\hat{L}(\fat{\eta}_i|\fat{y}_i)$, where
$$
\hat{L}(\fat{\eta}_i|\fat{y}_i) = \textrm{N}(\fat{\eta}_i|\hat{\fat{\eta}}_i,
\Sigma_{\eta y_i})
$$
and $c_i$ is a constant independent of $\fat{\eta}_i$.

Therefore, the approximated posterior density $\hat{\pi}(\fat{\eta},\fat{\nu},\fat{\theta}|\fat{y})$ based on the approximated full likelihood $\hat{L}(\fat{\eta}|\fat{y})=\prod_{i=1}^G\hat{L}(\fat{\eta}_i|\fat{y}_i)$,
is such that $\hat{\pi}(\fat{\eta},\fat{\nu},\fat{\theta}|\fat{y}) \approx \pi(\fat{\eta},\fat{\nu},\fat{\theta}|\fat{y})$,
and it is given by
\begin{equation}
\begin{aligned}
\hat{\pi}(\fat{\eta},\fat{\nu},\fat{\theta}|\fat{y}) & \propto \pi(\fat{\theta})\pi(\fat{\eta},\fat{\nu}|\fat{\theta})\hat{L}(\fat{\eta}|\fat{y}) \\
&\propto \pi(\fat{\theta})\pi(\fat{\eta},\fat{\nu}|\fat{\theta})\textrm{N}(\fat{\eta}|\fat{\hat{\eta}},\Sigma_{\eta y}) \\
&\propto \pi(\fat{\theta})\pi(\fat{\eta},\fat{\nu}|\fat{\theta})
\prod_{i=1}^G\textrm{N}(\fat{\eta}_i|\hat{\fat{\eta}}_i,\Sigma_{\eta y_i}).
\end{aligned}
\label{eq:approxpost}
\end{equation}

The second Gaussian approximation relies on normalizing the likelihood function such that the function $d^{-1}L(\fat{\eta}|\fat{y})$ integrates to one (over the domain of $\fat{\eta}$), where $d\in(0,\infty)$ is an appropriate normalization constant that is independent of $\fat{\eta}$. Here we assume that $d$ is finite. If this is not the case, we may either find a more adequate model parametrization or replace the likelihood function by an alternative generalized likelihood that consists of the likelihood times an extra prior density for $\fat\eta$. Bayes' Theorem ensures the finiteness of $d$ under the generalized likelihood since it is proportional to a posterior density. This second Gaussian approximation is also designed to ``maximize'' the match with the true likelihood function, especially in skewed scenarios.

Then, similarly to the first Gaussian approximation, the likelihood function, or the generalized likelihood function, is approximated with a Gaussian density that has mean and covariance matrix equal to those of the normalized likelihood function.
If there is a need for ensuring that the mean and variance are finite then the extra prior density can be given a finite support.      
We exploit again the assumed conditional independence, i.e., $L(\fat{\eta}|\fat{y})=\prod_{i=1}^{G}L(\fat{\eta}_i|\fat{y}_i)$, now approximating the $i$-th likelihood contribution as
$$
L(\fat{\eta}_i|\fat{y}_i) \approx d_i\tilde{L}(\fat{\eta}_i|\fat{y}_i) = d_i\textrm{N}(\fat{\eta}_i|\tilde{\fat{\eta}}_i,
\Omega_{\eta y_i}),
$$
where $d_i$ is a constant independent of $\fat{\eta}_i$, $\tilde{L}(\fat{\eta}_i|\fat{y}_i)$ is the Gaussian approximation, and 
the mean and the covariance matrix are
$$
\tilde{\fat{\eta}}_i = \int \fat{\eta}_i  d_i^{-1} L(\fat{\eta}_i|\fat{y}_i) d \fat{\eta}_i,
$$
$$
\Omega_{\eta y_i} = \int (\fat{\eta}_i - \hat{\fat{\eta}}_i)(\fat{\eta}_i - \hat{\fat{\eta}}_i)\trp  d_i^{-1} L(\fat{\eta}_i|\fat{y}_i) d \fat{\eta}_i.
$$
Similarly to \eqref{eq:approxpost}, the alternative approximation to the posterior density, $\tilde{\pi}(\fat{\eta},\fat{\nu},\fat{\theta}|\fat{y})$, which is based on 
$\tilde{L}(\fat{\eta}|\fat{y})=\prod_{i=1}^G\tilde{L}(\fat{\eta}_i|\fat{y}_i)$, may be expressed as
\begin{equation}
\begin{aligned}
\tilde{\pi}(\fat{\eta},\fat{\nu},\fat{\theta}|\fat{y}) &\propto \pi(\fat{\theta})\pi(\fat{\eta},\fat{\nu}|\fat{\theta})
\prod_{i=1}^G\textrm{N}(\fat{\eta}_i|\tilde{\fat{\eta}}_i,\Omega_{\eta y_i}).
\end{aligned}
\label{eq:approxpost2}
\end{equation}

Because the Gaussian density is the asymptotic form of the likelihood function under mild regularity conditions \citep{schervish1995theory}, these two types of approximations are expected to work increasingly well when the number of data replicates grows. With a low number of replicates (i.e., less than $10$--$20$ per distinct model parameter involved at the data level and further described at the latent level), a small bias might be expected, although we have found it to be relatively negligible in the settings we have considered. More details on the quality of these Gaussian approximations are given in Section~\ref{sec:simulated_data}.

The computational benefit of the approximations in (\ref{eq:approxpost}) and (\ref{eq:approxpost2})
 lies in the fact that $\pi(\fat{\eta},\fat{\nu}|\fat{\theta})$
is Gaussian with respect to $(\fat{\eta},\fat{\nu})$ and the functional form of both $\hat{L}(\fat{\eta}|\fat{y})$ and $\tilde{L}(\fat{\eta}|\fat{y})$ with respect to $\fat{\eta}$ is proportional to a Gaussian density. 
As a result, the conditional posterior density of $(\fat{\eta},\fat{\nu})$ is Gaussian, and posterior samples
can be obtained directly from this density and it is well known how to generate the samples from it. 
The information about $\fat{\eta}$ stemming from the data is quantified with reasonable 
accuracy in $\hat{L}(\fat{\eta}|\fat{y})$ or $\tilde{L}(\fat{\eta}|\fat{y})$ provided that at least one of these two approximations is fairly good. This information is correctly weighted against the prior information about $\fat{\eta}$ which is quantified in $\pi(\fat{\eta},\fat{\nu}|\fat{\theta})$. Since the inference scheme is Bayesian and the parameters are inferred simultaneously, then  
the information about $\fat{\eta}$ is correctly propagated to $\fat{\nu}$ and $\fat{\theta}$ through $\pi(\fat{\eta},\fat{\nu}|\fat{\theta})$. 
Notice that the likelihood approximation $\hat{L}(\fat{\eta}|\fat{y})$ may be faster to compute than $\tilde{L}(\fat{\eta}|\fat{y})$ as it is free from integrals, while $\tilde{L}(\fat{\eta}|\fat{y})$ is likely to provide a more accurate approximation in the case of a small number of replicates within each group. Hereafter, the Gaussian approximation based on the ML estimates and the inverse of the observed information will be referred to as the first Gaussian approximation, while the Gaussian approximation based on the mean and covariance of the normalized likelihood function will be referred to as the second Gaussian approximation.

\subsubsection{Step 2 (Smooth): Inference for the pseudo Gaussian--Gaussian model}

To infer the model presented in Section \ref{sec:ELGMModels} based on the
approximate posterior density in \eqref{eq:approxpost}, we 
consider a pseudo model that is such that $\hat{\fat{\eta}}$ (obtained from the first Gaussian approximation) is treated as noisy measurements of the latent field. Fitting this pseudo model is equivalent to smoothing the parameters $\hat{\fat{\eta}}$ jointly (hence the term ``Smooth''). A similar approach may be used based on \eqref{eq:approxpost2} by treating $\tilde{\fat{\eta}}$ (obtained from the second Gaussian approximation) as the data. The proposed data density 
 of the pseudo model based on \eqref{eq:approxpost} is $\textrm{N}(\hat{\fat{\eta}}|\fat{\eta},Q_{\eta y}^{-1})$ where $Q_{\eta y}=\Sigma_{\eta y}^{-1}$, and $Q_{\eta y}$ is known. Its numerical values are evaluated from the already observed data. This model can be written hierarchically as 
\begin{equation*}
\begin{aligned}
\pi(\fat{\hat{\eta}}|\fat{\eta},Q_{\eta y}, \fat{\theta}) &= \textrm{N}(\fat{\hat{\eta}}|\fat{\eta},Q_{\eta y}^{-1}), \\
\pi(\fat{\eta}|\fat{\nu},\fat{\theta}) &= \textrm{N}(\fat{\eta}|Z\nu,Q_{\epsilon}^{-1}), \\ 
\pi(\fat{\nu}|\fat{\theta}) &= \textrm{N}(\fat{\nu}|\fat{\mu}_\nu,Q_{\nu}^{-1})
\end{aligned}
\end{equation*}
and $\pi(\fat\theta)$ is the prior density for $\fat\theta$ as before.
The posterior density for this model is given by
\begin{equation*}
\begin{aligned}
\pi(\fat{\eta},\fat{\nu},\fat{\theta}|\fat{\hat{\eta}}) & 
\propto \pi(\fat{\theta})\pi(\fat{\eta},\fat{\nu}|\fat{\theta})
\pi(\fat{\hat{\eta}}|\fat{\eta},Q_{\eta y}, \fat{\theta})\\
&\propto \pi(\fat{\theta})\pi(\fat{\eta},\fat{\nu}|\fat{\theta})\textrm{N}(\fat{\hat{\eta}}|\fat{\eta},Q^{-1}_{\eta y}) \\
&\propto \pi(\fat{\theta})\pi(\fat{\eta},\fat{\nu}|\fat{\theta})\hat{L}(\fat{\eta}|\fat{\hat{\eta}},\Sigma_{\eta y}).
\end{aligned}
\end{equation*}
The above posterior density stems from looking at it as a function of $\fat{\eta}$ and taking $\hat{\fat{\eta}}$ as a fixed quantity, which gives 
$\textrm{N}(\fat{\hat{\eta}}|\fat{\eta},Q^{-1}_{\eta y})=\textrm{N}(\fat{\eta}|\fat{\hat{\eta}},Q^{-1}_{\eta y})$.
Thus, the above posterior density is exactly the same as 
the approximated posterior density in (\ref{eq:approxpost}) for the extend LGM in Section \ref{sec:ELGMModels}.
The pseudo model is a Gaussian--Gaussian model and it is convenient to approach the inference for the
unknown parameters through this model.

Samples of $(\fat{x},\fat{\theta})$, where $\fat{x} = (\fat{\eta}\trp,\fat{\nu}\trp)\trp$, are obtained by sampling first from the marginal posterior density of $\fat{\theta}$, and then from the posterior density of $\fat{x}$ conditional on $\fat{\theta}$.
The marginal posterior density of $\fat{\theta}$ given $\hat{\fat{\eta}}$ is
$\pi(\fat{\theta}|\hat{\fat{\eta}})\propto\pi(\fat{\theta})\pi(\hat{\fat{\eta}}|\fat{\theta})$
and it can be represented as
\begin{equation}
\pi(\fat{\theta}|\hat{\fat{\eta}}) 
\propto
\pi(\fat{\theta})
\frac{\pi(\hat{\fat{\eta}}|\fat{x},\fat{\theta})\pi(\fat{x}|\fat{\theta})}
{\pi(\fat{x}|\hat{\fat{\eta}},\fat{\theta})}.
\label{eq:marginaltheta1}
\end{equation}
The densities $\pi(\hat{\fat{\eta}}|\fat{x},\fat{\theta})$ and $\pi(\fat{x}|\fat{\theta})$ have
precision matrices $Q_{\eta y}$ and $Q_{x}$, respectively, and if $Q_{\eta y}$ and $Q_{x}$ are sparse matrices, then the precision matrix of $\pi(\fat{x}|\hat{\fat{\eta}},\fat{\theta})$ is a sparse matrix; see details in the Supplementary Material. Samples from the marginal posterior density of $\fat{\theta}$ can be obtained by using grid sampling if the dimension of $\fat{\theta}$ is small (i.e., four or less), a Metropolis step or a Metropolis--Hastings step, or other samplers that are well suited for densities with non-tractable form. Since the conditional posterior density of $\fat{x}$ is Gaussian, samples of $\fat{x}$ are straightforward to obtain, and if the precision matrix of $\pi(\fat{x}|\hat{\fat{\eta}},\fat{\theta})$ is sparse, the computational cost is relatively low. 

Further methodological and computational details on the inference scheme for the Gaussian--Gaussian pseudo model are presented in the Supplementary Material.

\section{Simulation examples}
\label{sec:simulated_data}

\subsection{Settings}
In this section, we assess the accuracy of our proposed approximate Bayesian inference scheme, Max-and-Smooth, by evaluating by simulation how close the approximate posterior density of an LGM is to its exact posterior density. In particular, in Section \ref{sec:simulated_data_1} we consider an LGM where the data are independent mean zero Gaussian random variables on a lattice, with spatially-varying variance at each lattice point.
The approximate marginal posterior densities of the latent parameters and the hyperparameters are compared to the exact posterior densities inferred by an ``exact'' MCMC sampler. Moreover, as the link function in this specific example is univariate, \texttt{INLA} can also be applied and we include it in our experiments for comparison.

In the Supplementary Material we explore three other models, namely a linear regression model on a lattice with Gaussian error terms, which is also applied to real data in Section 5, a linear regression model with $t$-distributed error terms and temporally varying coefficients, and a spatio-temporal model for Poisson counts. Our results suggest that the Gaussian likelihood approximation is accurate in finite sample sizes (even with just one single replicate in the case of the Poisson distribution with large counts), and therefore, that our Max-and-Smooth approach performs well in a rich variety of realistic settings.

\subsection{Gaussian data with spatially varying log-variance}
\label{sec:simulated_data_1}

In this section, we apply Max-and-Smooth  
to mean-zero Gaussian data with spatially-varying variance.
We first simulate a single realization of a Gaussian Markov random field \citep{rue2005gaussian}, $\{x_{i}\}$, on a regular lattice of size $10\times 10$ where the index $i$ corresponds to the lattice point with horizontal coordinates $i_1$ and vertical coordinates $i_2$. See further details in the Supplementary Material. With the mean fixed to zero, the conditional mean and variance of this GMRF are 
\begin{align*}
{\rm E}(x_{i_1,i_2} | \fat{x}_{-(i_1,i_2)}) & = \frac14 \left(x_{i_1-1,i_2} + x_{i_1+1,i_2} + x_{i_1,i_2-1} + x_{i_1,i_2+1}\right), \\
{\rm var}(x_{i_1,i_2} | \fat{x}_{-(i_1,i_2)}) &  = (4\tau)^{-1}.
\end{align*}
For the simulation, the precision parameter of the GMRF is fixed at $\tau = 1$.
At each of the $N=100$ lattice points, we simulate $T=10,20,50$ independent Gaussian variates with zero mean and log-variance $x_{i}$, i.e., $y_{i,t} \sim \textrm{N}(0, \exp(x_{i}))$ for $t = 1,\dots, T$ and $i = 1,\dots, N$. 
Therefore, in this simulation example, the ``groups'' mentioned in Section \ref{sec:ELGMModels} represent the data replicates available at each lattice point, and thus, the total number of groups is equal to the number of lattice points, i.e., $G=N=100$.
The goal is to exploit Max-and-Smooth  
to infer the latent variables $\fat{x} = \{x_{i}\}$, and the precision hyperparameter $\tau$ from the observed data $\fat{y} = \{y_{i,t}\}$. 
The model is simple enough to also infer the latent variables and hyperparameter using an MCMC sampler that uses the true likelihood function, to compare the approximation with an ``exact'' fully Bayesian procedure. The model parameters are inferred assuming that the mean at the data level is equal to zero. 
As the link function is here univariate (i.e., $M=1$), \texttt{INLA} is also used to infer the model parameters for the sake of comparison (but note that this would not be possible for more complex models with multivariate link functions). The full details of this simulation study are reported in the Supplementary Material.

\begin{figure}[t!]
\centering
\includegraphics[width=0.95\textwidth]{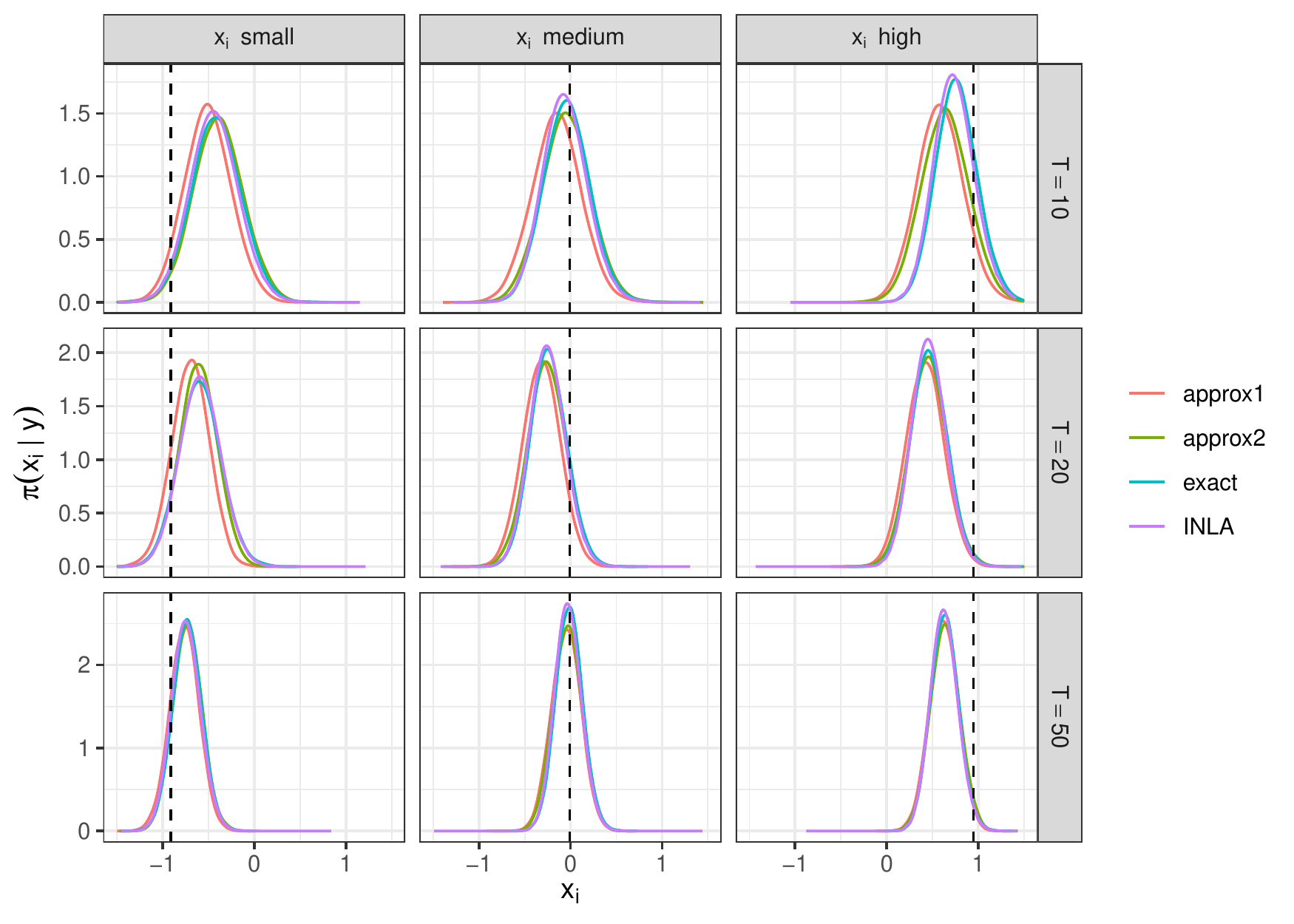}
\caption{Exact and the approximated marginal posterior densities of three latent parameters. The density curves were computed by an ``exact'' MCMC sampler (blue), by Max-and-Smooth based on the first (red) and second (green) Gaussian approximations, and by \texttt{INLA} (purple). We consider $T=10,20,50$ temporal replicates (top to bottom). The true values are indicated by dashed vertical lines.}
\label{fig:compare_post_latent}
\end{figure}

Figure \ref{fig:compare_post_latent} shows the posterior densities of latent parameters $x_{i}$ at three different lattice points, inferred from datasets with $T=10,20,50$ replicates per location.
We chose the three locations with the smallest, closest to zero, and largest element of $\fat{x}$.
Each graph in Figure \ref{fig:compare_post_latent}
shows four posterior densities, based on a fully Bayesian MCMC simulation, based on the two approximate inference schemes from Section \ref{sec:twostepapproach}, and based on \texttt{INLA}. By comparing the density curves, we see that the second Gaussian approximation (with mean and variance derived from the normalized likelihood function) is generally closer to the true posterior density than the first Gaussian approximation (with mean and variance based on the MLE and observed information).
Both approximate posterior inference schemes capture the exact posterior densities very well when $T$ is sufficiently large (greater than $20$). 
In the case of $T=10$, the discrepancy between the approximated and exact posterior density is relatively small for the small and median values of $x_{i}$, while it is more pronounced in the case of large $x_{i}$. 
The reason might be that the Gaussian approximation of the likelihood function does not properly capture the right-skewness of the exact likelihood function.
When the spatial prior for the vector $\fat{x}$ has a strong effect (pulling the estimate toward zero), then the right skewness of the marginal likelihood function will show up most prominently in the case of large $x_{i}$.
The posterior densities estimated by \texttt{INLA} are slightly closer to the exact densities for these selected latent variables than our approximations.

\begin{figure}[t!]
\centering
\includegraphics[width=0.95\textwidth]{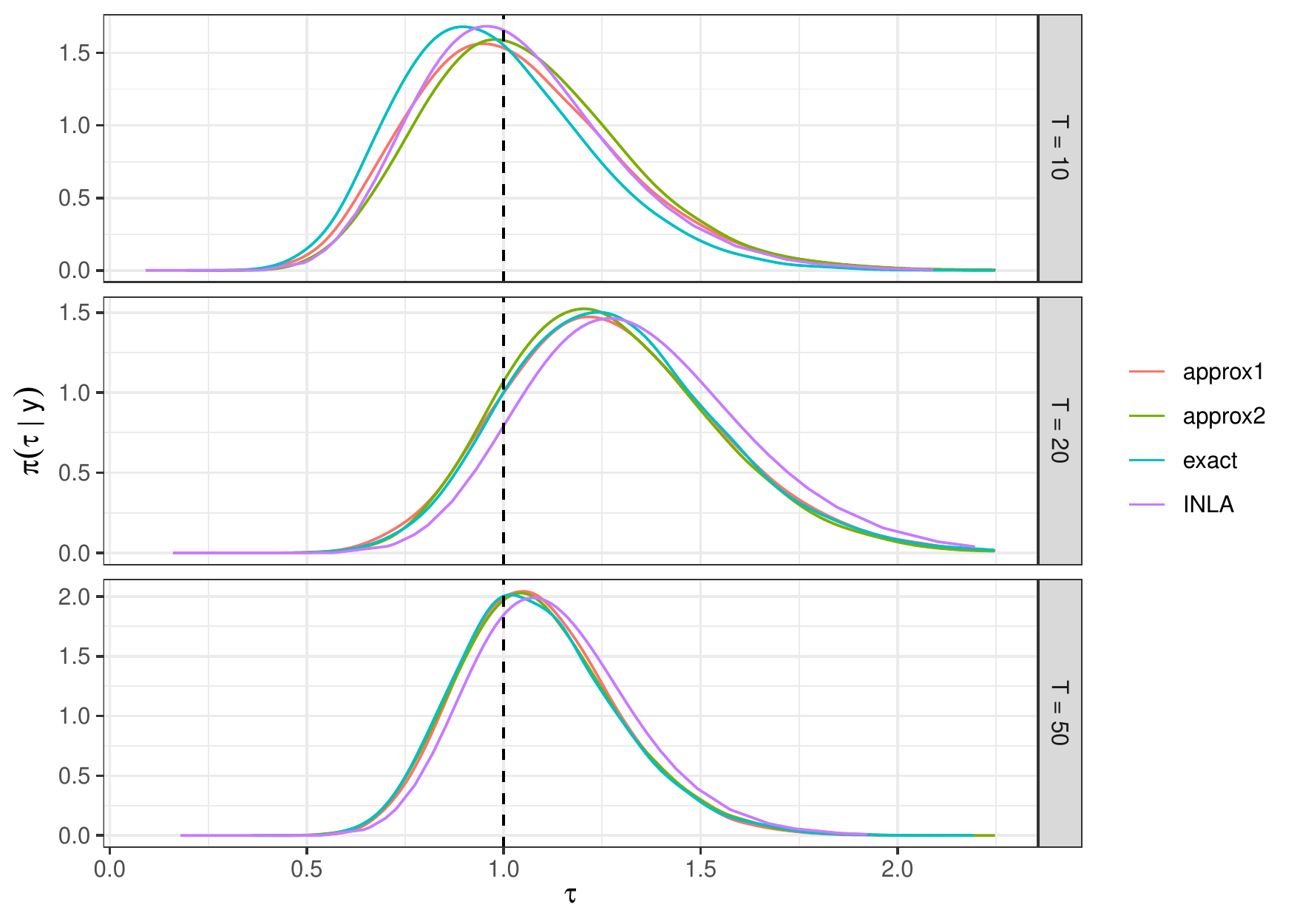}
\caption{The exact and the approximated marginal posterior densities of the hyperparameter $\tau$, inferred from sample sizes $T=10,20,50$ (from top to bottom). Colours are as in Figure \ref{fig:compare_post_latent}. The true value $\tau=1$ is indicated by a dashed vertical line.}
\label{fig:compare_post_hyper}
\end{figure}

The exact and approximate posterior densities of the precision hyperparameter $\tau$ of this model are shown in Figure \ref{fig:compare_post_hyper} for sample sizes $T=10,20,50$.
Figure \ref{fig:compare_post_hyper} shows that when $T\ge 20$ there is a negligible difference between the densities approximated by Max-and-Smooth and the exact posterior densities.
When $T=10$ the difference between the two Max-and-Smooth approximations is small and their difference with respect to the exact density appears reasonably small.
For $T=10$ there is no notable difference between the quality of \texttt{INLA} and Max-and-Smooth, but for $T\ge 20$, the posterior density calculated by \texttt{INLA} is further from the exact posterior than both Max-and-Smooth approximations.

\begin{figure}[t!]
\centering
\includegraphics[width=\textwidth]{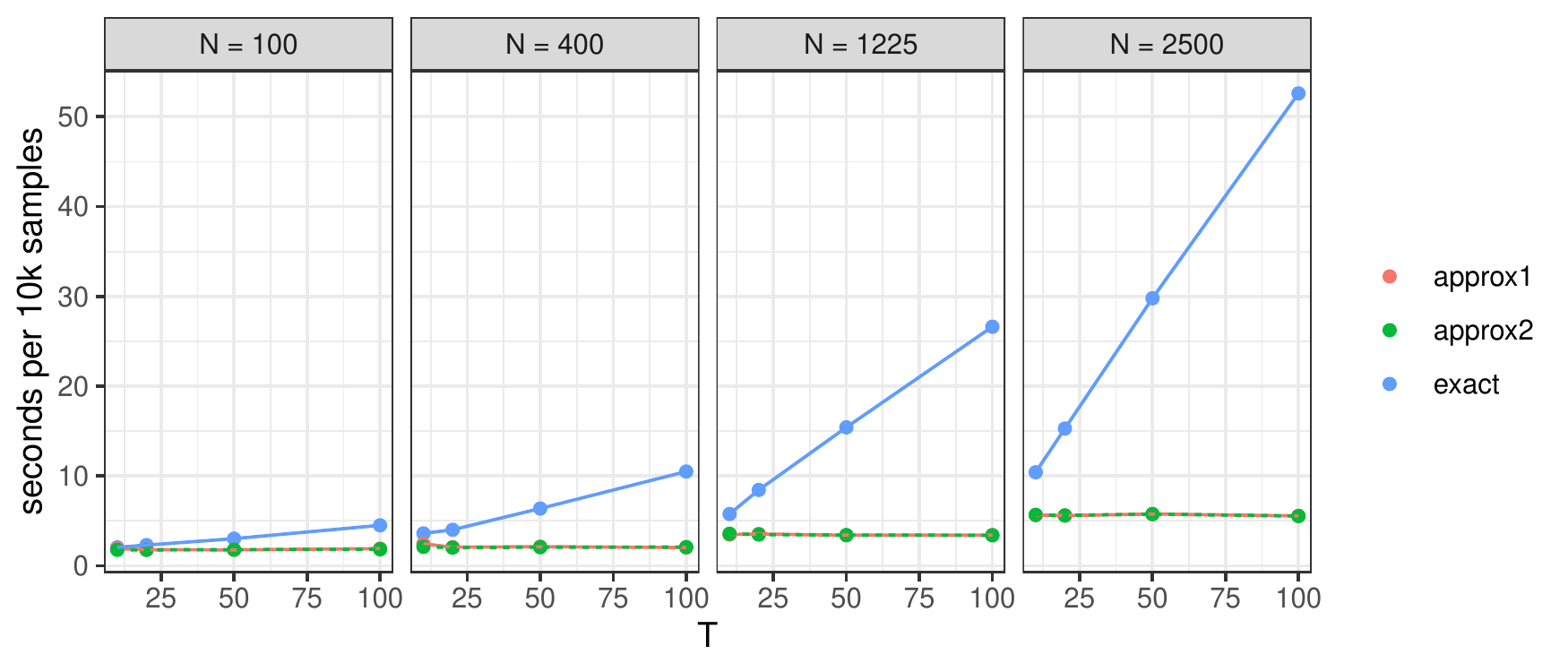}
\caption{The computational cost measured in seconds per ten thousand samples for the spatially varying log-variance model when applying Max-and-Smooth  
based on the first Gaussian approximation (red lines) and the second Gaussian approximation (dashed green line), and when applying an MCMC sampling scheme that uses the exact posterior density (blue line). The four graphs show the computational cost when the dimension of the latent vector is $100$, $400$, $1225$ and $2500$ (left to right).}
\label{fig:timing}
\end{figure}

In addition to the accuracy of the approximation, we also compare the computational speed of each approach.
To calculate the joint density $\pi(\fat{y},\fat{x},\tau) = \pi(\fat{y}|\fat{x},\tau)\pi(\fat{x}|\tau)\pi(\tau)$, the ``exact'' Bayesian inference requires $T$ evaluations of the Gaussian density per lattice point to calculate the likelihood $\pi(\fat{y}|\fat{x},\tau)$.
In contrast, under the approximate inference schemes, calculating the approximation of the likelihood requires only a single evaluation of the Gaussian density per lattice point.
We should thus expect the computational cost of the exact inference to scale linearly with $T$, while the computational cost of Max-and-Smooth  
should be constant in $T$.
Figure \ref{fig:timing} shows the time to draw 10,000 MCMC samples of $\tau$ and $\fat{x}$ for different numbers of temporal replicates $T$ and different numbers of grid points $N$.
Max-and-Smooth has constant computation cost as a function of $T$, and does not suffer from the same slowdown with increasing $T$ as the exact method.
As might be expected, for small sample sizes and grid sizes, the speedup is only moderate.
But for large grid sizes such as $N=50\times50$, and $T=100$ replicates, Max-and-Smooth is faster than the exact method by a factor greater than 10.
Since both approximation methods use a Gaussian approximation for the likelihood function, only with a different mean and variance, there is no difference in computational speedup between them.
As computation of marginal posteriors with \texttt{INLA} is not based on sampling, it was excluded from the comparison of computational cost in Figure \ref{fig:timing}. Inferring the marginal posterior densities of the latent field and the hyperparameters took only a few seconds with \texttt{INLA}, but unlike Max-and-Smooth, \texttt{INLA}'s run time increased with increasing number of temporal replicates.

\section{Predictions of meteorological variables on a lattice}\label{sec:application}

As alluded to in the introduction, a wide variety of Bayesian or frequentist approaches, involving different types of approximations or simplifications, have been proposed to deal with high-dimensional spatio-temporal data, often under the assumption of data being exactly Gaussian. These include low-rank approaches \citep{Cressie.Johannesson:2008}, the predictive process \citep{Banerjee.etal:2008}, covariance tapering \citep{Furrer.etal:2006,Anderes.etal:2013}, multi-resolution models \citep{Nychka.etal:2015,Katzfuss:2017}, hierarchical nearest-neighbor Gaussian processes \citep{Datta.etal:2016}, the Vecchia approximation \citep{Vecchia:1988,Stein.etal:2004,Katzfuss.Guinness:2019}, the integrated nested Laplace approximation \citep{rue2009approximate,bakka2018spatial}, or more recently a frequentist approach for data modeled by a generalized-extreme value (GEV) distribution \citep{risser2019extremeprecip,russell2019harvey}. See \citet{Heaton.etal:2019} for a recent review and comparison of some of these methods. In this section, we apply our proposed Max-and-Smooth approach to analyze a moderately high-dimensional real climate dataset, assuming that the data are well described with an extended LGM.

More precisely, the dataset used in our analysis is from a seasonal climate forecasting
experiment, consisting of retrospective surface temperature forecasts, and
their corresponding ``verifying'' observations.
The forecast data were produced by a global climate model ensemble (28 forecasts started
from perturbed initial conditions) downloaded from the ECMWF C3S Seasonal catalog (\url{https://apps.ecmwf.int/data-catalogues/c3s-seasonal/}) 
via the MARS API on 21 February 2018.
Forecasts were started from perturbed May 1 initial conditions each year from 1993 to
2015, i.e., the sample size is $T=23$ in time. 
Each model run predicts atmospheric conditions several months into the future.
The particular forecast target analyzed for this paper is surface air
temperature on a 1-by-1 degree latitude-longitude grid over a rectangular
region with corners 20W/40N and 40E/60N ($N=1281$ grid points in total, covering most of Europe). Here each grid point forms a group, so the total number of groups is $G=N$.
The forecasts were averaged over the 28 ensemble members, and over the Boreal summer period June/July/August, yielding a single scalar prediction per grid point per year.
Since these forecasts were initialized in May and predict June-August climate, the forecast lead time is 1--3 months.
Verifying observations are from the ERA-Interim reanalysis dataset \citep{dee2011era}, which is available on
the same 1-by-1 degree grid as the forecasts, and can also be downloaded from the ECMWF data base.
Observation data were averaged over the same June--August period as the forecasts.

Due to structural errors and missing physical processes in the climate model, and due to the chaotic nature
of atmospheric dynamics, numerical model forecasts have systematic biases in
the forecast mean. 
Furthermore, for forecasts of the climate system on seasonal time scales, the correlation between forecasts and
verifying observations tends to be low.
The biases in the numerical model forecasts can be partly corrected through a linear
regression of the observations on forecasts.
The adjustment of model forecasts by linear regression, known as model output
statistics (MOS), has a long tradition in the weather forecasting community
\citep{glahn1972use, glahn2009mos}, and is part of an active area of research known as forecast recalibration or forecast post-processing \citep[e.g.,][]{siegert2019forecast}.
In this section, we will infer the post-processing parameters on a spatial grid in an extended LGM framework, using spatial priors for the regression coefficients to reduce their estimation uncertainty, and ultimately improve the predictive skill of the recalibrated model forecasts.

The statistical model linking observed climate $y_{i,t}$ at time $t$ and
location $i$ to the climate model forecast $f_{i,t}$ for the same time and
location is assumed to be
\begin{equation}
y_{i,t} = \alpha_i + \beta_i (f_{i,t} - \bar{f}_{i}) + \epsilon_{i,t},
\label{eq:mos}
\end{equation}
where $\bar{f}_{i}$ denotes the local mean forecast over the data period, and the residuals
$\epsilon_{i,t}$ are independent Gaussian variates with mean zero and variance
$\exp({\tau_{i}})$.
It is common to estimate the regression parameters 
$\fat{\eta}=(\fat{\alpha}\trp,\fat{\beta}\trp,\fat{\tau}\trp)\trp=(\fat{\eta}_{\alpha}\trp,\fat{\eta}_{\beta}\trp,\fat{\eta}_{\tau}\trp)\trp$ individually for each grid point, using maximum likelihood
estimation. However, this may lead to high estimation variability, due to the limited number of samples per grid point.

To exploit the spatial structure in the data, and borrow strength from data
at neighboring grid points when estimating the regression parameters $\fat{\eta}$, we use a
spatial prior distribution for the spatial fields of regression parameters as outlined
in Section 2.2 of the Supplementary Material, and we exploit Max-and-Smooth for Bayesian inference.
The spatial field for each regression parameter is decomposed additively into a spatially correlated component $\fat{u}$ and
an unstructured component $\fat{\epsilon}$, i.e.,
\begin{equation}
\fat{\alpha} =   \fat{\eta}_\alpha = \fat{u}_\alpha + \fat{\epsilon}_\alpha,
\end{equation}
and respectively for $\fat{\beta}=\fat{\eta}_{\beta}$ and $\fat{\tau}=\fat{\eta}_{\tau}$. 
We model the structured term $\fat{u}$
as a first-order intrinsic Gaussian Markov random field on a regular lattice, \citep[Section 3.3.2, pp. 104--108]{rue2005gaussian}
and the unstructured component $\fat{\epsilon}$  
as a mean zero Gaussian process with diagonal covariance
matrix. 
All latent processes $\fat{u}$ and $\fat{\epsilon}$ are mutually independent.

The prior model has a total of six hyperparameters: The precision parameters of
the three spatially correlated fields $\fat{u}_\alpha$, $\fat{u}_\beta$ and $\fat{u}_\tau$, denoted $\{\tau_{u,\alpha}, \tau_{u,\beta}, \tau_{u,\tau}\}$, and the precision
parameters of the three unstructured fields $\fat{\epsilon}_\alpha$, $\fat{\epsilon}_\beta$ and $\fat{\epsilon}_\tau$, denoted $\{\tau_{\epsilon,\alpha}, \tau_{\epsilon,\beta}, \tau_{\epsilon,\tau}\}$.
The hyperparameters $\fat{\theta} = (\tau_{u, \alpha},\tau_{\epsilon, \alpha}, \tau_{u,\beta},\tau_{\epsilon, \beta}, \tau_{u, \tau},  \tau_{\epsilon, \tau})\trp$ are a priori independent, and spatially homogeneous.
We used independent penalized complexity (PC) priors \citep{simpson2017penalising} for all hyperparameters.
Specifically, the PC priors were specified as exponential distributions with rate parameter 1 for the standard deviations, which leads to a prior density for the precision parameter $\tau$ proportional to $\tau^{-1.5} \exp(-1/\sqrt{\tau})$.

We begin by exploring the marginal posterior distributions of the precision hyperparameters
$\fat{\theta}$
obtained using Max-and-Smooth (recall Section \ref{sec:LGMInf}). Preliminary studies suggested that the hyperparameters are
nearly uncorrelated under their joint posterior, which allows us to explore their
posterior distributions individually. We evaluated each unnormalized marginal posterior at 41 points, that are equidistant on the log-scale and centered around
the posterior mode, spanning $\pm 4$ posterior standard deviations (estimated via a Laplace approximation from the numerical second derivative).

\begin{figure}[t!]
\includegraphics[width=0.95\textwidth]{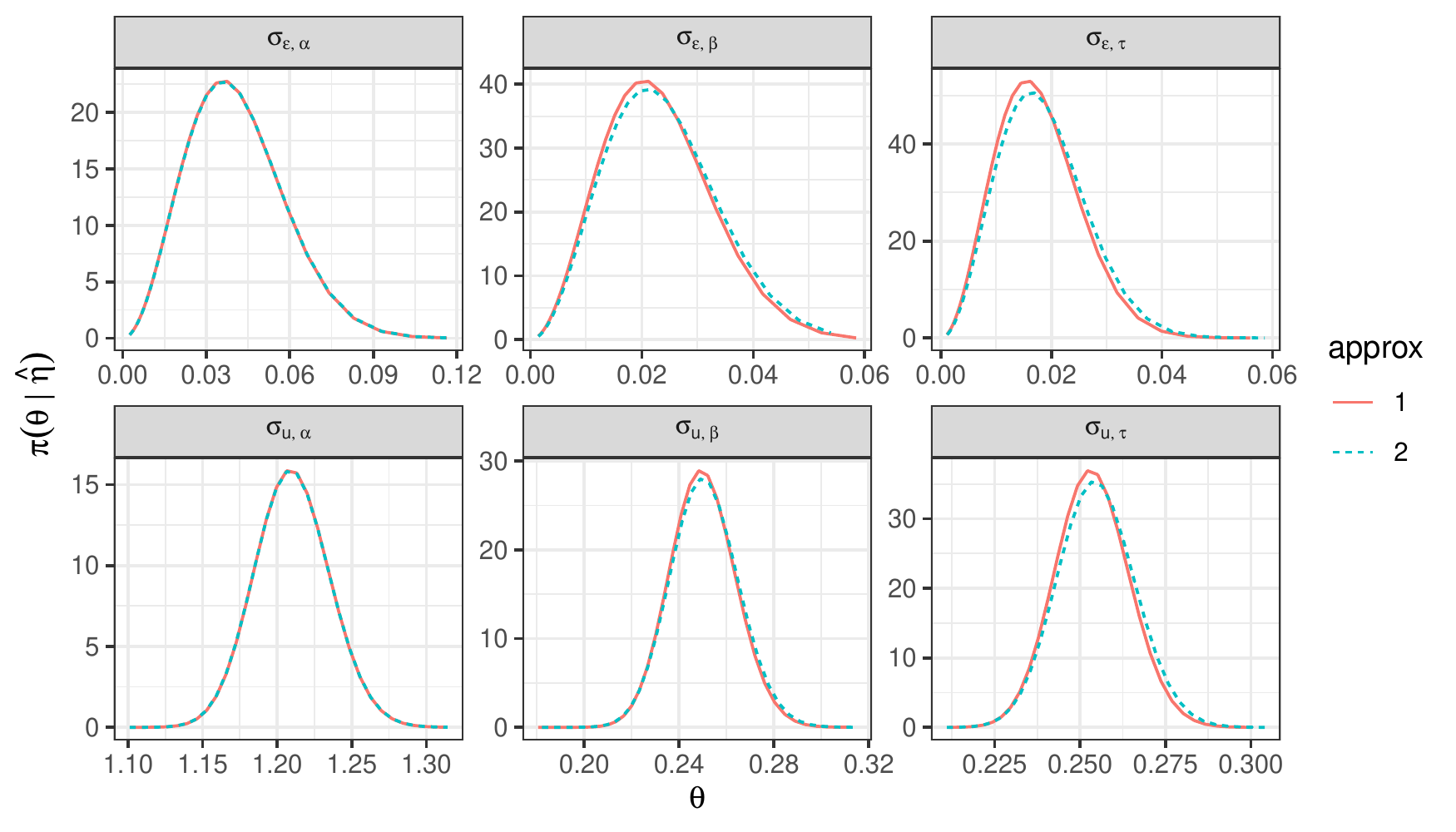}
\caption{Marginal posterior densities of the hyperparameters (transformed to standard deviations).}
\label{fig:hyperpost}
\end{figure}

The (normalized) marginal posteriors of the hyperparameters are shown in Figure
\ref{fig:hyperpost}. 
For better interpretability, the hyperparameters were transformed from precision $\tau_{v,l}$ to standard deviation $\sigma_{v,l} = \tfrac{1}{\sqrt{\tau_{v,l}}}$ for $v \in \{u, \epsilon\}$ and $l \in \{\alpha, \beta, \tau\}$.
The standard deviations of the unstructured components $\sigma_{\epsilon, l}$ are small compared to the spatial variability seen in the ML estimates of the regression coefficients, and compared to the standard deviation parameters of the spatial components $\sigma_{u, l}$. 
This suggests that a simpler model might be fitted to the data that does not include an unstructured component.
The standard deviation corresponding to the spatial effect of the intercept, $\sigma_{u,\alpha}$, is large compared to $\sigma_{u,\beta}$ and $\sigma_{u,\tau}$. Thus, there is a greater spatial variability in the intercept than in the slope and log-variance.
By comparing the posterior densities with their PC prior distributions (not shown), we found no evidence that the posterior is unduly influenced by the prior, indicating that the amount of data available is sufficient to properly constrain the model's hyperparameters.

\begin{figure}[t!]
\centering
\includegraphics[width=0.95\linewidth]{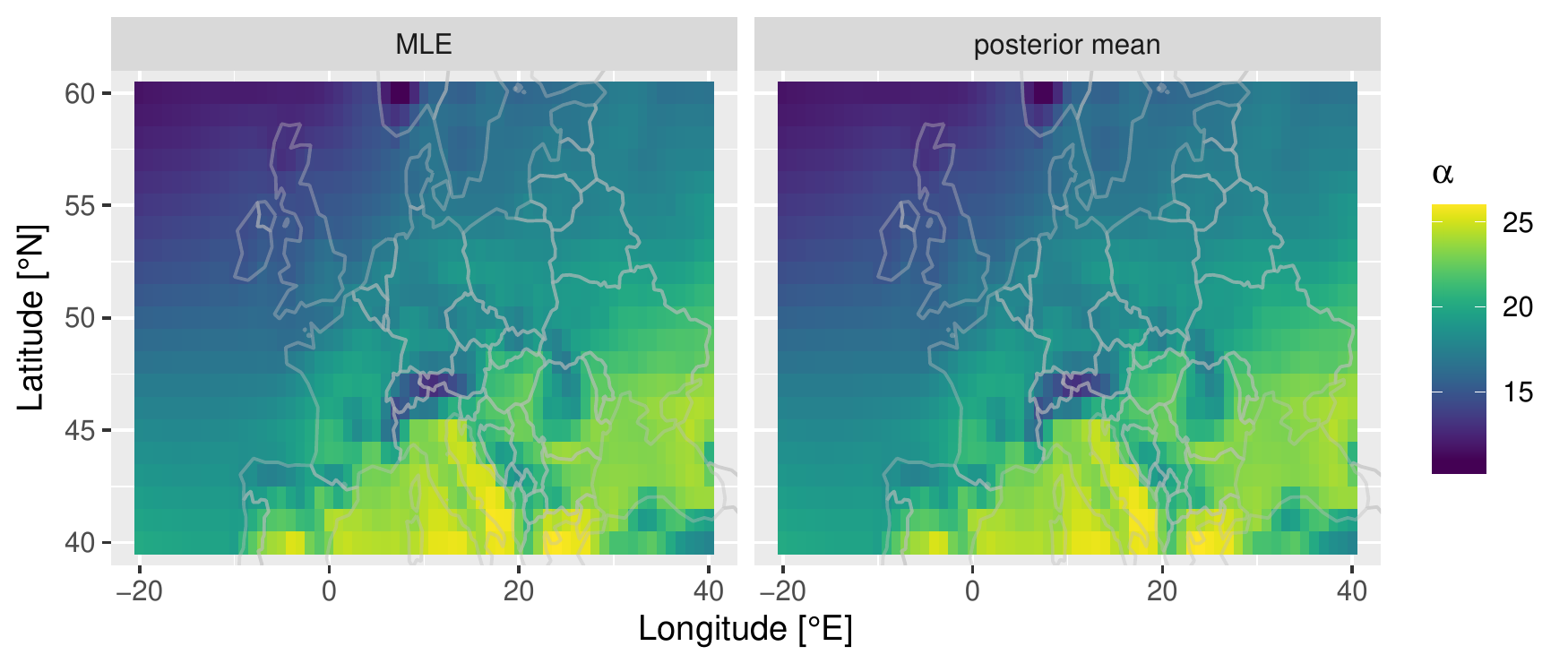}\\
\hspace{-5pt}\includegraphics[width=0.935\linewidth]{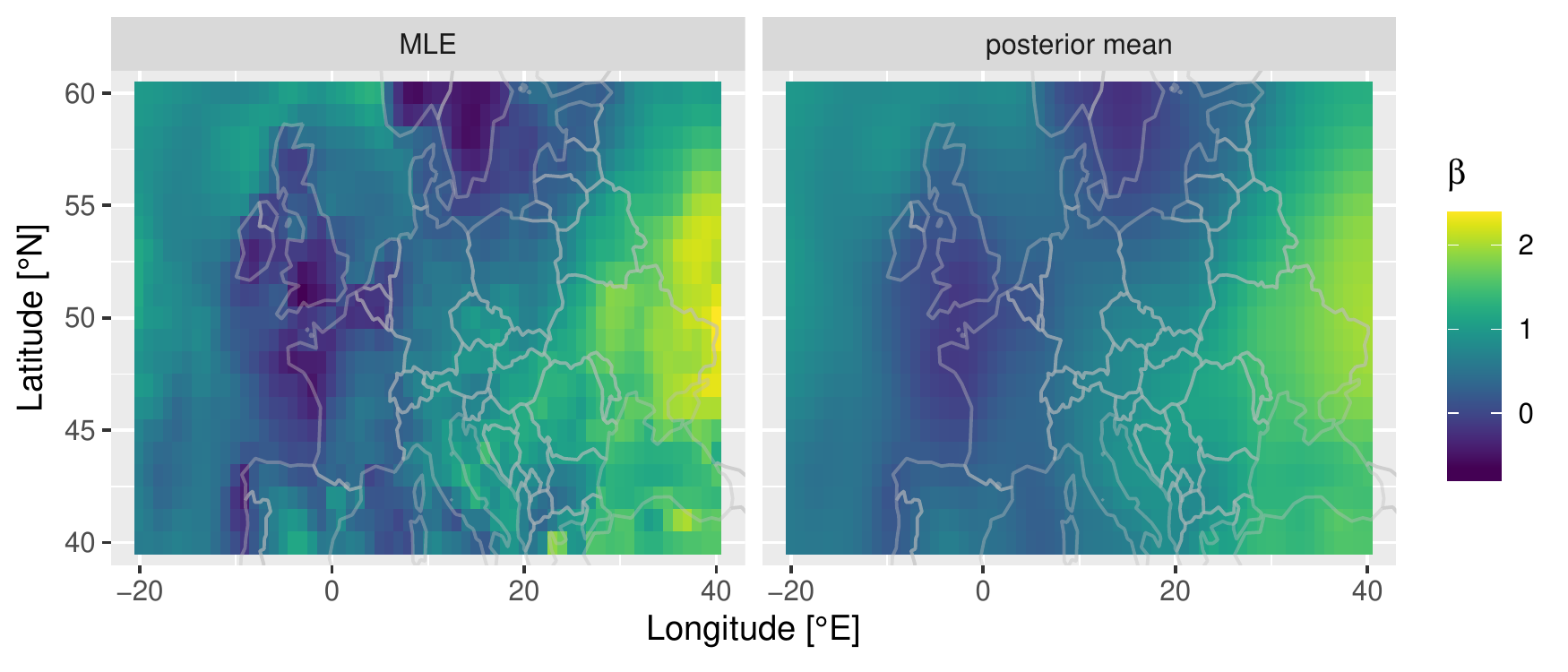}\\
\hspace{4pt}\includegraphics[width=0.963\linewidth]{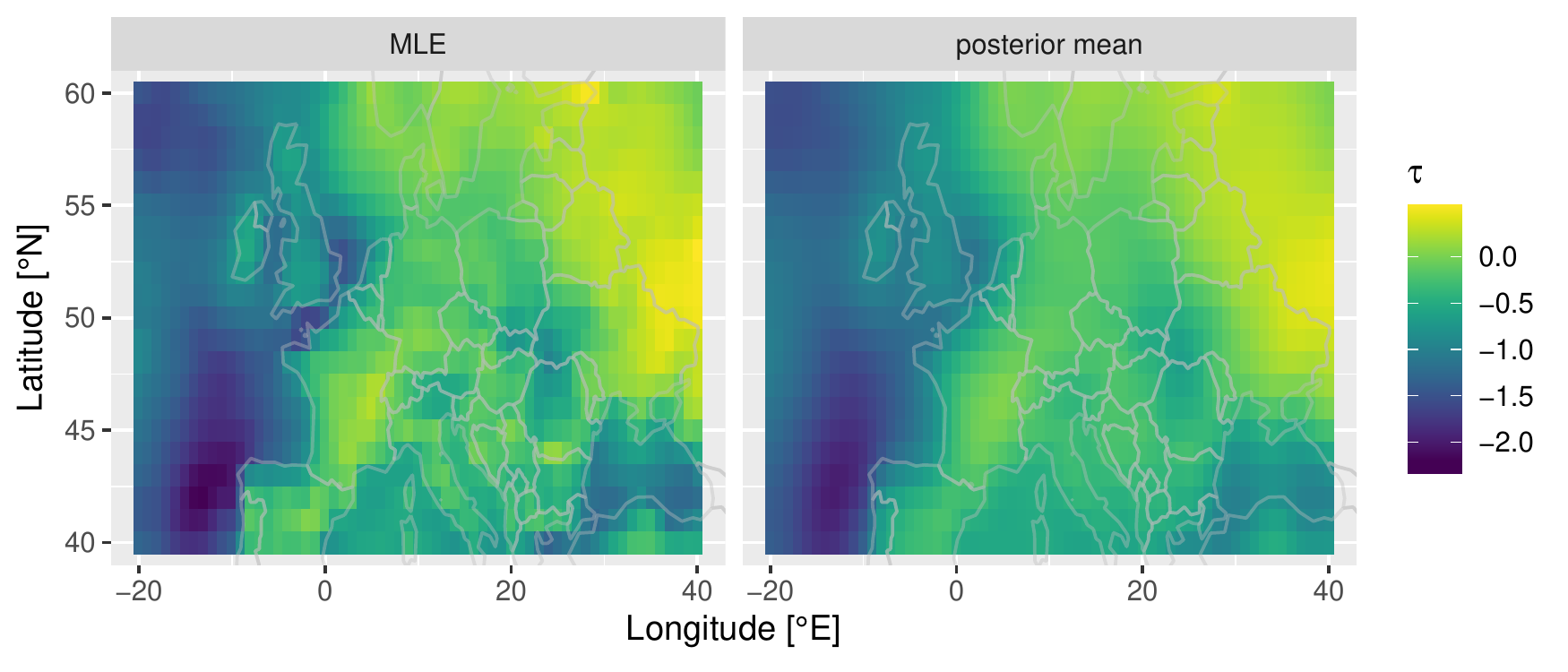}
\caption{Pointwise maximum likelihood estimates (left) of the regression parameters $\alpha$ (top), $\beta$ (middle), and $\tau$ (bottom), and posterior expectations inferred from the extended LGM with a spatially correlated prior (right). The results were obtained from Max-and-Smooth  
based on the first Gaussian approximation.}
\label{fig:etapost}
\end{figure}

Figure \ref{fig:etapost} compares the posterior means of the latent fields
$\fat{\eta}_l = \fat{u}_l + \fat{\epsilon}_l$ for $l \in \{\alpha, \beta, \tau\}$
(conditional on the posterior mode of the hyperparameters) with the corresponding maximum
likelihood estimates.
We used the first Gaussian approximation (ML estimates and observed information) for Figure \ref{fig:etapost}, noting that the corresponding plots for the second Gaussian approximation (mean and variance of normalized likelihood) are almost indistinguishable.
The intercept parameters of the regression model in the vector $\fat{\alpha}$ are not smoothed very much
compared to the corresponding ML estimates. 
This is because the maximum likelihood estimates of $\fat{\alpha}$ are generally well-constrained by the data, as indicated by
the low sampling uncertainty derived from the observed information matrix. 
In contrast, $\fat{\beta}$ and $\fat{\tau}$ exhibit considerable smoothing. 
The pointwise posterior standard deviations of the latent fields (not shown) are smaller
than the sampling standard deviations of the maximum likelihood estimates: they are on average about 5\%
smaller for $\alpha$ and around 60\% smaller for $\beta$ and $\tau$.

We then test the performance of the spatial regression model using cross-validated
predictions of surface temperatures calculated using equation \eqref{eq:mos}, integrated
over the posterior distributions of $\fat{\alpha}$, $\fat{\beta}$ and $\fat{\tau}$.
In practice, forecast and observation data are typically available at all locations in the historical dataset, but no data are available for the present year, for which bias-corrected forecasts are required. We mimic this setting by applying "leave-one-year-out" cross-validation: We leave out all forecasts and observations from one year, fitting the regression parameters using data from the remaining years, and then using the fitted parameters to forecast the temperature at all grid points for the left-out year.
For each year and grid point, we generated $N_s = 1000$ samples from the posterior
predictive distribution by repeating the following algorithm $N_s$ times:
\begin{itemize}
\item Sample hyperparameters $\fat{\theta}^{(s)}$ independently from their marginal posteriors, i.e., $\pi(\tau_{u,l} | \fat{y}_{-t})$, and $\pi(\tau_{\epsilon,l} | \fat{y}_{-t})$, where $l \in \{\alpha,\beta, \tau\}$ and $\fat{y}_{-t}$ is forecast and observation data with year $t$ being left out of the training dataset. Sampling of hyperparameters is done by grid sampling, i.e., after evaluating the marginal density of each hyperparameter at 21 equally spaced points, one of the 21 values is sampled with probability proportional to the marginal density. 
\item Draw an independent sample $\fat{\eta}^{(s)}$ of each of the latent fields $\fat{\eta}_\alpha$, $\fat{\eta}_\beta$ and $\fat{\eta}_\tau$ from their conditional posterior distribution $\pi(\fat{\eta} | \hat{\fat{\eta}},\fat{\theta}^{(s)})$.
\item Using the sampled regression parameter in $\fat{\eta}^{(s)}$, simulate a spatial field of responses $\fat{y}_t$ from the regression model conditional on $\fat{\eta}^{(s)}_{\alpha}$, $\fat{\eta}^{(s)}_{\beta}$, and $\fat{\eta}^{(s)}_{\tau}$, and using the covariates $\fat{f}_t$ from the left-out year $t$.
\end{itemize}

The procedure is repeated leaving each year out in turn, ultimately resulting in $N_s$
out-of-sample posterior predictive samples 
per grid point for each year. 
The predictive samples
are compared to the verifying observations by the mean squared error (MSE) of the
posterior predictive mean, by the continuous ranked probability score (CRPS, see \citet{winkler1969scoring,
gneiting2007scoringrules}) of
the posterior predictive distribution, by the average widths of the central 95\%
posterior prediction interval (W95), and by the average coverage frequencies of
the posterior predictive $5$, $50$ and $95$-percentiles (COV05, COV50, COV95, respectively).
The CRPS is preferred here because it can be evaluated solely from samples drawn from the posterior predictive distribution, without having to know the predictive distribution in closed form, using the following relation: 
\begin{equation}
\mbox{CRPS}(\{x_1, \dots, x_{N_s}\}, y) = \frac{1}{N_s}\sum_{1\le i \le N_s} |y - x_i| - \frac{1}{2N_s^2} \sum_{1 \le i,j \le N_s} |x_i - x_j|
\end{equation}
where $x_1, \dots, x_{N_s}$ are samples from the posterior predictive distribution, and $y$ is the corresponding observed temperature value \citep[see][Equation~(20)]{gneiting2007scoringrules}. Lower CRPS values indicate better forecasts.
The different measures of sharpness, reliability and accuracy of the posterior
predictive distribution are compared to two benchmark predictions. We draw the
same number of samples from the classical predictive distributions using the pointwise ML estimates for the linear regression models.
This forecast scheme is denoted MLE and is
used to characterize the performance of the forecast post-processing method achieved when spatial correlation in the regression parameters is ignored. 
An even simpler benchmark model is constructed by sampling $1000$ times from a Gaussian distribution
which has the mean and standard deviation (taken over time) of the observations $\fat{y}_i$, calculated separately at each grid point $i$. 
This benchmark prediction is almost constant between the years, with slight differences only due to the leave-one-out procedure.
This forecast scheme is denoted CLIM (for climatology) and quantifies the average predictability of atmospheric surface temperature if no further forecast information is available. 
The prediction scheme CLIM is thus used to quantify the merit of the forecast information available from the numerical model. 
Finally, we use the acronyms SPAT1 and SPAT2 to denote the posterior predictive distribution derived from the extended LGM with a spatial prior, inferred using Max-and-Smooth based on the first and second Gaussian approximations, respectively. Due to computational limitations we have not implemented an exact Bayesian inference using an MCMC sampler. 
\begin{table}
\centering
\caption{Comparison of leave-one-year-out predictive performance of the four forecast schemes, CLIM, MLE, SPAT1, SPAT2 (descriptions in main text), by mean squared error of the predictive mean (MSE), continuous ranked probability score (CRPS), average width of the central 95\% prediction interval (W95), and average coverage frequencies of the predictive 5, 50, and 95-percentiles (COV05, COV50, COV95, respectively).}
\label{tab:performance}
\begin{tabular}{ccccc}
\toprule 
 & CLIM & MLE & SPAT1 & SPAT2 \\
\midrule
MSE & 0.8329 & 0.7843 & 0.7747 & 0.7742 \\
CRPS & 0.4952 & 0.4843 & 0.4817 & 0.4814 \\
W95 & 3.31 & 3.35 & 3.08 & 3.08 \\
COV05 & 0.056 & 0.049 & 0.06 & 0.06 \\
COV50 & 0.50 & 0.51 & 0.51 & 0.51 \\
COV95 & 0.93 & 0.94 & 0.93 & 0.93 \\ 
\bottomrule
\end{tabular}
\end{table}
Table \ref{tab:performance} shows that spatially smoothing the regression estimates using a hierarchical prior leads to improvements in mean squared error of the predictive mean, and CRPS of the predictive distribution. Simultaneously, forecast uncertainty, as indicated by the narrower average prediction intervals, is reduced by our method. Coverage frequencies are quite good overall. 

The seemingly small improvement in MSE and CRPS of SPAT1/2 compared to MLE may suggest a rather negligible improvement of our approach. However, the magnitude of this improvement has to be compared to the improvement of MLE over CLIM. Informally speaking, the improvement of MLE over CLIM quantifies the advantage of a forecaster who has access to the numerical model forecast compared to a forecaster who only knows the climatological distribution. Due to the inherent unpredictability of the atmosphere on long time scales, this advantage tends to be small in seasonal climate prediction. The improvement of SPAT1/2 over MLE thus has to be judged relative to the improvement of MLE over CLIM. The improvement of SPAT1 and SPAT2 over MLE relative to the improvement of MLE over CLIM is about 20\% for the MSE, and about 25\% for the CRPS. Both improvements are substantial.

To further assess statistical significance of the improvement of SPAT1 over MLE, we approximated the sampling distribution of the average CRPS difference between the MLE and SPAT1 predictions using block bootstrapping. We treat all data as independent in time because the time increment between successive samples is one year, long enough to ignore the temporal correlation in small scale meteorological settings. To account for spatial correlation, we divided the spatial domain into a number of $S$ rectangular non-overlapping blocks. This yields a total of $23 \times S$ blocks. We resample these blocks $23 \times S$ times with replacement, and then create a bootstrap sample of the average CRPS difference. The bootstrap distribution was estimated from 500 replicates. When using $S=3$ spatial blocks (i.e. assuming only 3 effective spatial degrees of freedom) we obtain a bootstrap mean of $0.0025$ and boostrap standard deviation of $0.0013$ for the CRPS difference, and a bootstrap p-value of $0.032$. For $S=10$ spatial degrees of freedom, we obtain a bootstrap mean of $0.0025$, bootstrap standard deviation of $0.0012$, and a bootstrap p-value of $0.006$. For $S>10$ the bootstrap p-value is essentially zero. Based on visual inspection of maps of CRPS differences, a conservative estimate of the spatial degrees of freedom is at least 20, so the improvement of SPAT over MLE, albeit small, can thus be considered statistically significant with high confidence. 

Finally, it should be noted that evaluating the marginal posterior distributions of the 6 hyperparameters at 21 values each, drawing 1000 samples of the spatial fields of the regression parameters from their posterior distributions, and calculating 1000 posterior predictive samples of atmospheric temperature over the entire grid takes less than 1 minute on a standard laptop. 

\section{Discussion}\label{sec:discussion}

In this paper, we have introduced Max-and-Smooth, 
a new two-step approximate posterior sampling scheme for extended LGMs with independent data replicates. Extended LGMs include a diverse range of important applications, such as regression models with spatial or temporal effects in regression coefficients and error variances. Our proposed Max-and-Smooth approach 
is fast and well-suited for this class of LGMs with a complex and high-dimensional latent structure and multivariate link functions. The first step of the inference scheme involves approximating the likelihood function around its mode by its asymptotic Gaussian density, and we also explored another Gaussian approximation which uses the mean and variance of the normalized likelihood. We suggest that the adequacy of the Gaussian approximation be explored case by case. The second step involves Bayesian inference for the latent parameters and the hyperparameters of the model such that the uncertainty from the first step is correctly propagated into the uncertainty of the posterior distributions. Max-and-Smooth  
contrasts with the \texttt{INLA} software, which is designed for LGMs with a univariate link function but not for extended LGMs.

Our approach scales well with the dimensions of the multivariate link function and the latent parameter vector given that the precision matrices of the Gaussian priors at the latent level are sparse. This is important for high-dimensional applications, e.g., models with high-dimensional temporal and spatial effects. Additionally, exploiting conjugacy of the Gaussian--Gaussian pseudo model (similar to the INLA method), posterior samples from the marginal density of the hyperparameters can be generated efficiently with low sample autocorrelation. 
For our approach to be successful, we need the Gaussian likelihood approximation in the first step of our two-step inference approach to be accurate. 
It turns out that models with non-Gaussian likelihood functions can often be aggregated over groups of independently replicated observations, such that the joint likelihood function of the group is close to a multivariate Gaussian likelihood. 
In our paper, we have provided evidence that Max-and-Smooth  
indeed provides accurate and reliable inference in various settings. 
The computational cost of our proposed inference scheme is close to being insensitive to the number of independent data replicates. 
In the case of a high-dimensional latent vector modeled with sparse precision matrices and a large number of independent data replicates, the speedup is substantial in comparison to a Markov chain Monte Carlo inference scheme which samples from the exact posterior density.  
A case study comparing Max-and-Smooth with \texttt{INLA} (in a simplified scenario with a univariate link function) showed that \texttt{INLA} seems to do better at approximating marginal posteriors of latent variables, whereas Max-and-Smooth produces better approximations of marginal posteriors of hyperparameters. The observed differences between posterior approximations are minor, and the extent to which the results generalize to different applications will be subject of future research.
 
Max-and-Smooth is designed for a rich class of flexible models. 
It is straightforward to implement, and when sparse precision matrices are used at the latent level, high-dimensional latent vectors can be handled. 
This paves the way for having an additional set of complex models that are feasible for the analysis of complex and high-dimensional datasets. 
Future research involves exploring ways to drop the conditional independence assumption at the observation level, and computing the Gaussian approximation based on the normalized likelihood when analytical results for the mean and variance are not available, and when the dimension of the parameter vector within groups is high.



\bibliographystyle{aps-nameyear} 



\end{document}



\baselineskip12pt


\maketitle

\section{An approximate inference scheme for LGMs with a multivariate link function}
\label{sec:inferscheme}

In this section, we show the details of Max-and-Smooth, in particular, Step 2 which involves smoothing the estimates of the latent parameters jointly. Max-and-Smooth is used to infer the parameters of the extend LGM presented in Section 3.2 of the main paper, and it is based on the approximations to the posterior density in equations (3.1) and (3.2) in the main paper.  
We consider a model that is such that either $\hat{\fat{\eta}}$ or $\tilde{\fat{\eta}}$ is treated as the data. The proposed data density is either $\textrm{N}(\hat{\fat{\eta}}|\fat{\eta},Q_{\eta y}^{-1})$ where $Q_{\eta y}$ is known and $Q_{\eta y}=\Sigma_{\eta y}^{-1}$, or $\textrm{N}(\tilde{\fat{\eta}}|\fat{\eta},\tilde{Q}_{\eta y}^{-1})$ where $\tilde{Q}_{\eta y}$ is known and $\tilde{Q}_{\eta y}=\Omega^{-1}_{\eta y}$. The numerical values of the matrices, $\Sigma_{\eta y}$ and $\Omega_{\eta y}$ are evaluated from the already observed data and these two matrices stem from the matrices $\Sigma_{\eta y_i}$ and $\Omega_{\eta y_i}$, $i=1,...,G$, respectively, which are defined in Section 3.3 in the main paper. 
The model for $\hat{\fat{\eta}}$ can be written hierarchically as 
\begin{equation*}
\begin{aligned}
\pi(\fat{\hat{\eta}}|\fat{\eta},Q_{\eta y}, \fat{\theta}) &= \textrm{N}(\fat{\hat{\eta}}|\fat{\eta},Q_{\eta y}^{-1}), \\
\pi(\fat{\eta}|\fat{\nu},\fat{\theta}) &= \textrm{N}(\fat{\eta}|Z\fat{\nu},Q_{\epsilon}^{-1}), \\ 
\pi(\fat{\nu}|\fat{\theta}) &= \textrm{N}(\fat{\nu}|\fat{\mu}_\nu,Q_{\nu}^{-1})
\end{aligned}
\end{equation*}
and $\pi(\fat\theta)$ is the prior density for $\fat\theta$ as before.
The model for $\tilde{\fat{\eta}}$ has the same structure, however, $\fat{\hat{\eta}}$ and $Q_{\eta y}$ are replaced with
$\fat{\tilde{\eta}}$ and $\tilde{Q}_{\eta y}$, respectively.
The posterior density for the model above is given by
\begin{equation*}
\begin{aligned}
\pi(\fat{\eta},\fat{\nu},\fat{\theta}|\fat{\hat{\eta}}) & 
\propto \pi(\fat{\theta})\pi(\fat{\eta},\fat{\nu}|\fat{\theta})
\pi(\fat{\hat{\eta}}|\fat{\eta},Q_{\eta y}, \fat{\theta})\\
&\propto \pi(\fat{\theta})\pi(\fat{\eta},\fat{\nu}|\fat{\theta})\textrm{N}(\fat{\hat{\eta}}|\fat{\eta},Q^{-1}_{\eta y}) \\
&\propto \pi(\fat{\theta})\pi(\fat{\eta},\fat{\nu}|\fat{\theta})\textrm{N}(\fat{\eta}|\fat{\hat{\eta}},Q^{-1}_{\eta y}) \\
&\propto \pi(\fat{\theta})\pi(\fat{\eta},\fat{\nu}|\fat{\theta})\hat{L}(\fat{\eta}|\fat{\hat{\eta}},\Sigma_{\eta y}).
\end{aligned}
\end{equation*}
The above posterior density stems from looking at it as a function of $\fat{\eta}$ and taking $\hat{\fat{\eta}}$ as a fixed quantity, which gives 
$\textrm{N}(\fat{\hat{\eta}}|\fat{\eta},Q^{-1}_{\eta y})=\textrm{N}(\fat{\eta}|\fat{\hat{\eta}},Q^{-1}_{\eta y})$.
So, the above posterior density is exactly the same as 
the approximated posterior density in equation (3.1) in the main paper.
The model presented in this section is a Gaussian--Gaussian model and it is more convenient to approach the inference for the
unknown parameters through this model than the original model
with the approximated likelihood function
even though the two posterior densities are the same.
In particular, it will be useful to assume that $\fat{\hat{\eta}}$ is Gaussian when
deriving the marginal posterior density of $\fat\theta$ and
the conditional posterior density of $(\fat{\eta}, \fat{\nu})$ given $\fat{\theta}$.
To tackle the inference for $(\fat{\eta, \nu, \theta})$ the joint prior density of $\fat{x} = (\fat{\eta}\trp,\fat{\nu}\trp)\trp$ conditional on $\fat{\theta}$ is derived. It is given by
$$
\pi(\fat{x}|\fat{\theta})=\textrm{N}\Bigg(\fat{x}\Bigg|\begin{bmatrix}Z\fat{\mu}_\nu \\ \fat{\mu}_\nu\end{bmatrix}, \begin{bmatrix}Q_\epsilon & -Q_\epsilon Z \\ -Z\trp Q_\epsilon & Q_\nu+Z\trp Q_\epsilon Z\end{bmatrix}^{-1}\Bigg).
$$
The covariance matrix of $\fat{x}$ given $\fat{\theta}$ is 
$$
\Sigma_x = Q_x^{-1}=\begin{bmatrix}Q_{\epsilon}^{-1}+Z Q_\nu^{-1}Z\trp & ZQ_\nu^{-1} \\ Q_\nu^{-1}Z\trp & Q_\nu^{-1}\end{bmatrix}.
$$
Let $p = \dim(\fat{\eta})=MG$, $q = \dim(\fat{\nu})$ and $B=\begin{bmatrix}I_{p\times p} & 0_{p\times q}\end{bmatrix}$.
Because $\hat{\fat{\eta}}$ can be expressed as $\hat{\fat{\eta}} = \fat{\eta} + \fat{e} =  B \fat{x} + \fat{e}$,
where $\fat{e} \sim \textrm{N}(\fat{0},Q_{\eta y}^{-1})$,
then the joint density of 
$(\hat{\fat{\eta}}\trp,\fat{x}\trp)\trp=(\hat{\fat{\eta}}\trp,\fat{\eta}\trp,\fat{\nu}\trp)\trp$
is given by
$$
\pi\Bigg(\begin{bmatrix}\hat{\fat{\eta}} \\ \fat{x}\end{bmatrix}\Bigg) = \textrm{N}\Bigg(\begin{bmatrix}\hat{\fat{\eta}} \\ \fat{x}\end{bmatrix}\Bigg|\begin{bmatrix}B\fat{\mu}_x \\ \fat{\mu}_x\end{bmatrix},\begin{bmatrix}Q_{\eta y}^{-1}+BQ_x^{-1}B\trp & BQ_x^{-1} \\ Q_x^{-1}B\trp & Q_x^{-1} \end{bmatrix}\Bigg).
$$
The mean vector and the covariance matrix of $(\hat{\fat{\eta}},\fat{x})$ can, respectively, be written as
$$
\begin{bmatrix}
Z\fat{\mu}_\nu \\
Z\fat{\mu}_\nu \\
\fat{\mu}_\nu \\
\end{bmatrix},
\quad
\begin{bmatrix}
Q_{\eta y}^{-1}+
Q_\epsilon^{-1}+ZQ_\nu^{-1}Z\trp
& Q_\epsilon^{-1}+ZQ_\nu^{-1}Z\trp
& ZQ_{\nu}^{-1}
\\
Q_\epsilon^{-1}+ZQ_\nu^{-1}Z\trp
& Q_\epsilon^{-1}+ZQ_\nu^{-1}Z\trp
& ZQ_\nu^{-1}
\\
Q_{\nu}^{-1}Z\trp
& Q_\nu^{-1}Z\trp
& Q_\nu^{-1}
\end{bmatrix}.
$$
The precision matrix of $(\hat{\fat{\eta}},\fat{x})$ is
$$
Q_{\hat{\eta}x} = \Sigma_{\hat{\eta}x}^{-1}=\begin{bmatrix}Q_{\eta y} & -Q_{\eta y} B \\ -B\trp Q_{\eta y}  & Q_x+B\trp Q_{\eta y} B\end{bmatrix}
$$
and it can be written as
$$
Q_{\hat{\eta}x} =
\begin{bmatrix}
Q_{\eta y} & -Q_{\eta y} & 0 \\ 
-Q_{\eta y}  & Q_\epsilon +Q_{\eta y}  
& -Q_{\epsilon}Z
\\
0  & -Z\trp Q_{\epsilon} &
Q_{\nu} +Z\trp Q_{\epsilon}Z 
\end{bmatrix}.
$$
The marginal density of $\hat{\fat{\eta}}$ given $\fat{\theta}$ is
$$
\pi(\hat{\fat{\eta}}|\fat{\theta})=\textrm{N}(\hat{\fat{\eta}}|B\fat{\mu}_x,Q_{\eta y}^{-1}+BQ_{x}^{-1}B\trp)=\textrm{N}(\hat{\fat{\eta}}|Z\fat{\mu}_\nu, Q_{\eta y}^{-1}+Q_{\epsilon}^{-1}+ZQ_\nu^{-1}Z\trp).
$$
The marginal posterior density of $\fat{\theta}$
given $\hat{\fat{\eta}}$ is
$
\pi(\fat{\theta}|\hat{\fat{\eta}})\propto \pi(\fat{\theta})\pi(\hat{\fat{\eta}}|\fat{\theta})
$
and it can be represented as
\begin{equation}
\pi(\fat{\theta}|\hat{\fat{\eta}}) 
\propto
\pi(\fat{\theta})
\frac{\pi(\hat{\fat{\eta}}|\fat{x},\fat{\theta})\pi(\fat{x}|\fat{\theta})}
{\pi(\fat{x}|\hat{\fat{\eta}},\fat{\theta})},
\label{eq:marginaltheta1}
\end{equation}
using the fact that 
$\pi(\hat{\fat{\eta}}|\fat{\theta})\pi(\fat{x}|\hat{\fat{\eta}},\fat{\theta})
=\pi(\hat{\fat{\eta}}|\fat{x},\fat{\theta})\pi(\fat{x}|\fat{\theta}).
$
The densities $\pi(\hat{\fat{\eta}}|\fat{x},\fat{\theta})$ and $\pi(\fat{x}|\fat{\theta})$ have
precision matrices $Q_{\eta y}$ and $Q_{x}$, respectively. The precision matrix of $\pi(\fat{x}|\hat{\fat{\eta}},\fat{\theta})$ is given below.
Note that the density $\pi(\fat{\theta}|\hat{\fat{\eta}})$ can be evaluated with any value for $\fat{x}$ because it does not depend on $\fat{x}$. 
This fact can be used to simplify calculations of the marginal density in (\ref{eq:marginaltheta1}) by
setting $\fat{x}$ equal to a convenient value, e.g., $\fat{x}=\fat{0}$. 
The conditional posterior density of $\fat{x}$
given $\hat{\fat{\eta}}$ and $\fat{\theta}$ is
\begin{equation}
\pi(\fat{x}|\hat{\fat{\eta}},\fat{\theta})=\textrm{N}(\fat{x}|Q_{x|\hat{\eta}}^{-1}(Q_x\fat{\mu}_x+B\trp Q_{\eta y}\hat{\fat{\eta}}),Q_{x|\hat{\eta}}^{-1}),
\label{eq:pixeta}
\end{equation}
where
\begin{equation}
Q_{x|\hat{\eta}}=Q_x + B\trp Q_{\eta y}B
=\begin{bmatrix}Q_\epsilon + Q_{\eta y} & -Q_\epsilon Z \\ -Z^{\textrm{T}} Q_\epsilon & Q_\nu+Z\trp Q_\epsilon Z\end{bmatrix}
\label{eq:Qxeta}
\end{equation}
and
\begin{equation}
Q_x\fat{\mu}_x+B\trp Q_{\eta y}\hat{\fat{\eta}}
=
\begin{bmatrix}Q_{\eta y}\hat{\fat{\eta}}   \\
Q_\nu\fat{\mu}_{\nu}\end{bmatrix}.
\label{eq:meanxeta}
\end{equation}
To sample from the posterior density of $(\fat{x},\fat{\theta})$, first, a sample from $\pi(\fat{\theta} | \hat{\fat{\eta}})$, which is an approximation of $\pi(\fat{\theta}|\fat{y})$, is taken. Then a sample from $\pi(\fat{x}|\hat{\fat{\eta}},\fat{\theta})$, which is an approximation of $\pi(\fat{x}|\fat{y},\fat{\theta})$, is taken.

In order to gain insight into how $\hat{\fat{\eta}}$ is used to update our prior knowledge of $\fat{\eta}$ and
$\fat{\nu}$ then the conditional distribution of $\fat{\eta}$ given $\hat{\fat{\eta}}$ and $\fat{\theta}$ is explored
as well as the conditional distribution of $\fat{\nu}$ given $\hat{\fat{\eta}}$ and $\fat{\theta}$. The density of the conditional distribution of $\fat{\eta}$ can be found by using the fact that 
$$
\pi(\fat{\eta}|\hat{\fat{\eta}},\fat{\theta})\propto \pi(\fat{\eta}|\fat{\theta})\pi(\hat{\fat{\eta}}|\fat{\eta})
$$
where 
$\pi(\hat{\fat{\eta}}|\fat{\eta}) = \textrm{N}(\hat{\fat{\eta}}|\fat{\eta},Q^{-1}_{\eta y})$ and
$\pi(\fat{\eta}|\fat{\theta}) = \textrm{N}(\fat{\eta}|Z\fat{\mu}_\nu,Q^{-1}_{\epsilon}+ZQ^{-1}_\nu Z^{\textrm{T}})$. The result is a Gaussian density with mean $\fat{\mu}_{\eta|\hat{\eta}}$ and precision matrix $Q_{\eta|\hat{\eta}}$, where
$$
\fat{\mu}_{\eta|\hat{\eta}} =
Q^{-1}_{\eta|\hat{\eta}}\{(Q^{-1}_\epsilon+ZQ^{-1}_\nu Z\trp)^{-1}Z\fat{\mu}_{\nu}
+ Q_{\eta y}\hat{\fat{\eta}} \},
\quad
Q_{\eta|\hat{\eta}} =
(Q^{-1}_\epsilon+ZQ^{-1}_\nu Z\trp)^{-1}
+ Q_{\eta y}.
$$
It is clear from the above formula for the posterior mean $\fat{\mu}_{\eta|\hat{\eta}}$, 
that it is a linear combination of the prior mean $Z\fat{\mu}_{\nu}$ and the maximum 
likelihood estimates in $\hat{\fat{\eta}}$. The posterior mean stems from Bayesian 
regression where the design matrix is an identity matrix, the error precision matrix is $Q_{\eta y}$, and the prior precision matrix is $Q_\eta = (Q_\epsilon^{-1}+ZQ_\nu^{-1}Z\trp)^{-1}$.
The prior precision matrix assigns weights to the prior mean $Z\fat{\mu}_{\nu}$ while
the precision matrix $Q_{\eta y}$ assigns weights to $\hat{\fat{\eta}}$. The larger the weights in $Q_{\eta y}$ with respect to $Q_{\eta}$, the greater is the influence of $\hat{\fat{\eta}}$ on the posterior mean of $\fat{\eta}$. If the prior mean of $\fat{\mu}_{\nu}$ is equal to zero then the posterior mean
of $\fat{\eta}$ is simply
$$
\fat{\mu}_{\eta|\hat{\eta}} =
Q^{-1}_{\eta|\hat{\eta}}Q_{\eta y}\hat{\fat{\eta}}
=\{Q_{\eta y}^{-1}(Q^{-1}_\epsilon+ZQ^{-1}_\nu Z\trp)^{-1}
+ I \}^{-1}\hat{\fat{\eta}}=
(Q_{\eta y}^{-1}Q_{\eta}
+ I )^{-1}\hat{\fat{\eta}},
$$
so, if the elements of $Q_{\eta y}$ are large with respect to the elements of $Q_\eta$, $\fat{\mu}_{\eta|\hat{\eta}}$ will be closer to $\hat{\fat{\eta}}$.

The conditional density of $\fat{\nu}$ given $\hat{\fat{\eta}}$
can be found by using the fact that 
$$
\pi(\fat{\nu}|\hat{\fat{\eta}},\fat{\theta})\propto \pi(\fat{\nu}|\fat{\theta})\pi(\hat{\fat{\eta}}|\fat{\nu},\fat{\theta}),
$$
where 
$\pi(\hat{\fat{\eta}}|\fat{\nu},\fat{\theta}) = \textrm{N}(\hat{\fat{\eta}}|Z\fat{\nu},Q^{-1}_{\eta y}+Q^{-1}_{\epsilon})$ and
$\pi(\fat{\nu}|\fat{\theta}) = \textrm{N}(\fat{\nu}|\fat{\mu}_\nu,Q^{-1}_\nu)$. The result is a Gaussian density with mean $\fat{\mu}_{\nu|\hat{\eta}}$ and precision matrix $Q_{\nu|\hat{\eta}}$ where
$$
\fat{\mu}_{\nu|\hat{\eta}} =
Q^{-1}_{\nu|\hat{\eta}}
\{Q_{\nu}\fat{\mu}_\nu
+ Z\trp (Q^{-1}_\epsilon
+Q^{-1}_{\eta y})^{-1}\hat{\fat{\eta}}\},
\quad
Q_{\nu|\hat{\eta}} = Q_{\nu}+
Z\trp (Q^{-1}_\epsilon + Q^{-1}_{\eta y} )^{-1}Z.
$$
The formula for $\fat{\mu}_{\nu|\hat{\eta}}$ shows that the posterior mean for $\fat{\nu}$
is a linear combination of the prior mean $\fat{\mu}_{\nu}$ and the maximum likelihood estimates in $\hat{\fat{\eta}}$.
This posterior mean is the result of a Bayesian regression with design matrix $Z$, the error precision matrix $(Q^{-1}_\epsilon + Q^{-1}_{\eta y})^{-1}$ and the prior precision matrix $Q_\nu$. The larger the weights in $(Q^{-1}_\epsilon + Q^{-1}_{\eta y})^{-1}$ are with respect to $Q_{\nu}$, the greater is the influence of $\hat{\fat{\eta}}$ on the posterior mean $\fat{\mu}_{\nu|\hat{\eta}}$. If the prior mean of $\fat{\mu}_{\nu}$ is equal to zero then the posterior mean
of $\fat{\nu}$ is
$$
\fat{\mu}_{\nu|\hat{\eta}} =
\{Q_{\nu}+
Z\trp (Q^{-1}_\epsilon + Q^{-1}_{\eta y} )^{-1}Z\}^{-1}
Z\trp (Q^{-1}_\epsilon
+Q^{-1}_{\eta y})^{-1}\hat{\fat{\eta}},
$$
thus, in this case $\fat{\mu}_{\nu|\hat{\eta}}$ is solely a linear combination of $\hat{\fat{\eta}}$,
and the role of the prior precision matrix $Q_\nu$ in $\fat{\mu}_{\nu|\hat{\eta}}$ can be seen.

The marginal posterior density of $\fat{\theta}$ given $\hat{\fat{\eta}}$ 
is proportional to $\pi(\fat{\theta})\pi(\hat{\fat{\eta}}|\fat{\theta})$ and it
can be represented in terms of $\fat{\nu}$, that is,
\begin{equation}
\pi(\fat{\theta}|\hat{\fat{\eta}}) 
\propto
\pi(\fat{\theta})
\frac{\pi(\hat{\fat{\eta}}|\fat{\nu},\fat{\theta})\pi(\fat{\nu}|\fat{\theta})}
{\pi(\fat{\nu}|\hat{\fat{\eta}},\fat{\theta})},
\label{eq:marginaltheta2}
\end{equation}
using the fact that 
$
\pi(\hat{\fat{\eta}}|\fat{\theta})\pi(\fat{\nu}|\hat{\fat{\eta}},\fat{\theta})
=\pi(\hat{\fat{\eta}}|\fat{\nu},\fat{\theta})\pi(\fat{\nu}|\fat{\theta})
$, 
where $\pi(\hat{\fat{\eta}}|\fat{\nu},\fat{\theta})$, 
$\pi(\fat{\nu}|\fat{\theta})$ and 
$\pi(\fat{\nu}|\hat{\fat{\eta}},\fat{\theta})$
are as above. Representing $\pi(\fat{\theta}|\hat{\fat{\eta}})$ in terms of $\fat{\nu}$ as in (\ref{eq:marginaltheta2})
is useful for LGMs with a univariate link function. Furthermore, this setup is relevant to LGMs with a multivariate link function
when the approximated likelihood induces independence in such a way that one or more of the $\fat{\eta}_m$ vectors in $\fat{\eta}$ 
can be inferred independently of the other vectors in $\fat{\eta}$.

\section{Examples: simulated data}

\subsection{Log-variance on a lattice}

This example is presented in the main paper. 
In this section the details of the statistical model
and its inference are given. Results are given in 
Section 4 of the main paper. 

\subsubsection{Statistical model}
\label{subsubsec:statmodel1}

Let $y_{i,t}$ be the observed variable at lattice point $i$ and time $t$, respectively. The index $i$ corresponds to the lattice point with horizontal coordinates $i_1$ and vertical coordinates $i_2$. We assume that the observation at lattice point $i$ and time point $t$ follows the Gaussian distribution with mean zero 
and variance $\sigma^2_i$, i.e.,
$$
y_{i,t}\sim \textrm{N}(0,\sigma^2_i), \quad \sigma^2_i > 0, 
$$ 
where $i \in \{1,...,N\}$, $t\in \{1,...,T\}$,  
$N$ is the number of lattice points and 
$T$ is the number of observations at each lattice point.
Here the ``groups'' mentioned in Section 2.2 in the main paper represent the observed variables at each lattice point, thus, the number of groups is $G=N$.
The data density for $\fat{y}_i=(y_{i,1},...,y_{i,T})\trp$ is then  
\begin{equation}
\label{eq:datadens1}
\pi(\fat{y}_i|\sigma^2_i) = \prod_{t=1}^{T}\frac{1}{\sqrt{2 \pi \sigma^2_i}} \exp\bigg(-\frac{y_{i,t}^2}{2\sigma^2_i}\bigg).
\end{equation}
The parameter $\sigma^2_i$ is transformed to $x_i = \log(\sigma^2_i)$, so $x_i \in \mathbb{R}$.
In terms of $x_i$, this density can be written as
\begin{equation}
\label{eq:datadens1}
\pi(\fat{y}_i|x_i) = (2\pi)^{-T/2}\exp(-(T/2) x_i) \exp\bigg(-\frac{\exp(-x_i)}{2} \sum_{t=1}^{T} y_{i,t}^2  \bigg).
\end{equation}

The parameters are collected into the vector $\fat{x}=(x_1,...,x_N)$, and modeled at the latent level as a Gaussian Markov random field (GMRFs) on a regular lattice\citep{rue2005gaussian}. 
The precision matrix of $\fat{x}$ is $Q_{x} = \tau Q_{u}$ where 
\begin{equation}
Q_{u} = R_{n_1}\otimes I_{n_2} +  I_{n_1} \otimes R_{n_2},
\label{eq:qumatrix_b}
\end{equation}
$\tau$ is a precision parameter, $\otimes$ denotes the Kronecker product, $n_1$ and $n_2$ are the dimensions of the lattice ($N=n_1 n_2$),
$I_{m}$ is an $m$ dimensional identity matrix, and
\begin{equation}
R_{m} = \begin{bmatrix} 2 & -1 & & & & & \\ 
                           -1 & 2 & -1 & & & & \\
                           & -1 & 2 & -1 & &  & \\
                           & & \vdots & \vdots & \vdots & & \\
                           & & & -1 & 2 & -1 &  \\
                           &  & & & -1 & 2 & -1  \\                            
                           & & & & & -1 & 2 \end{bmatrix}
\label{eq:precisionr_b}
\end{equation}
is an $m \times m$ matrix. Note that $R_m$ is different from the $R$ matrix defined in \citet[Section 3.3.1, p. 95]{rue2005gaussian},
in (\ref{eq:precisionr_b}) the first and the last diagonal elements are equal to $2$ as opposed to $1$ and, as a result, the precision matrix in (\ref{eq:qumatrix_b}) is full rank as 
opposed to having rank $N-1$ as the $Q$ matrix defined in \citet[Section 3.3.2, p. 107]{rue2005gaussian}. The precision matrix in (\ref{eq:qumatrix_b}) corresponds to a random process defined on a lattice such that the variables on the points surrounding it are all equal to zero.

Here $\tau$ is set as the hyperparameter. At the hyperparameter level we assign a gamma prior density to $\tau$, such that
$$
\tau \sim \textrm{Gamma}(10,10).
$$  
In our simulation study based on this model, we set the value of the hyperparameter equal to $\tau = 1.0$,
and the models were inferred with the unstructured model term set equal to zero, i.e., $\fat{\epsilon}=\fat{0}$. 
Furthermore, the size of the lattice was set equal to $10 \times 10$, $20 \times 20$, $35 \times 35$, and $50 \times 50$, so $N=100$, $N=400$, $N=1225$ and $N=2500$, 
and the number of observations per lattice point is set equal to $T=10,20,50,100$. 

\subsubsection{Inference}
\label{sec:samplescheme1}

We use Max-and-Smooth to infer $\fat{x}$ and $\tau$. 
The likelihood function for $x_i$ was approximated with two Gaussian approximations. The first one has mean equal to the
mode of the likelihood function, namely
$$
\hat{x}_i = \log\bigg(\frac{1}{T}\sum_{t=1}^{T}y_{i,t}^2 \bigg)
$$
and variance equal to the inverse of the observed information,
$$
\mathcal{I}^{-1}_{x y_i} = \frac{2}{T}.
$$
The likelihood function was also approximated with a Gaussian density with mean and variance equal to those of the normalized likelihood function. Here the normalized likelihood function is a log-inverse-gamma density, which has the general form
$$
\pi(x_i)=\frac{\gamma^{\alpha}}{\Gamma(\alpha)} \exp\{-\alpha x_i - \gamma \exp(-x_i)   \}
$$
and its mean and variance are
$$
\textrm{E}(x_i) = -\psi(\alpha)+\log(\gamma), \quad
\textrm{var}(x_i) = \psi'(\alpha)
$$
where $\psi(\cdot)$ and $\psi'(\cdot)$ are the digamma and trigamma functions, respectively. The normalized likelihood 
function of $x_i$ is a log-inverse-gamma density with parameters $\alpha=T/2$ and $\gamma=0.5\sum_{t=1}^{T}y^2_{i,t}$,
and its mean is
$$
\textrm{E}(x_i|\fat{y}_i) = -\psi(T/2)+\log\left(0.5\sum_{t=1}^{T}y^2_{i,t}\right) = \hat{x}_i +\log(T/2) - \psi(T/2),
$$
and its variance is $\textrm{var}(x_i|\fat{y}_i) = \psi'(T/2)$.
Denote the above mean by $\tilde{x}_i$.
Thus, the likelihood function was approximated with both
$L(x_i)\approx c_i\hat{L}(x_i)= c_i\textrm{N}(x_i|\hat{x}_i,2/T)$,
and
$L(x_i)\approx \tilde{c}_i\tilde{L}(x_i)= \tilde{c}_i\textrm{N}(x_i|\tilde{x}_i,\psi'(T/2))$, where $c_i$ and $\tilde{c}_i$ are constants independent of $x_i$.

The inference for the above model involves modeling the data in 
$\hat{\fat{x}}=(\hat{x}_1,...,\hat{x}_N)$ (or in 
$\tilde{\fat{x}}=(\tilde{x}_1,...,\tilde{x}_N)$ ) with a pseudo LGM with a univariate link function.
This model is a Gaussian--Gaussian model, and it can be inferred by using a slightly modified version of the inference scheme in 
Section \ref{sec:inferscheme} and Section 3.3 in the main paper.
The data density for $\hat{x}_i$ in the pseudo model is given by  
\begin{equation}
\pi(\hat{x}_i|x_i) = \textrm{N}(\hat{x}_i|x_i,2/T)= \frac{1}{\sqrt{2 \pi (2/T) }} \exp\bigg\{-\frac{(\hat{x}_{i}-x_{i})^2}{2(2/T)}\bigg\},
\label{eq:datadens2}
\end{equation}
and the latent level and the hyperparameter level of the pseudo model are the same as the ones in
the original model. The data density for $\tilde{x}_i$ in the pseudo model is $\pi(\tilde{x}_i|x_i) = \textrm{N}(\tilde{x}_i|x_i,\psi'(T/2))$.

The inference scheme below is based on the pseudo model and is presented in terms of $\hat{\fat{x}}$ as opposed to $\tilde{\fat{x}}$. However, the pseudo model below can be modified to be applicable for $\tilde{\fat{x}}$, in particular, $\hat{\fat{x}}$ is replaced with $\tilde{\fat{x}}$ and the variance, $2/T$, is replaced with the variance $\psi'(T/2)$.   
The likelihood of $\fat{x}$, $L(\fat{x})$, 
is approximated with $\hat{L}(\fat{x})$, that is 
$$
L(\fat{x})=\prod_{i=1}^N L(x_i) \approx c\hat{L}(\fat{x}) =\prod_{i=1}^N c_i \hat{L}(x_h) = c \textrm{N}(\fat{x}|\hat{\fat{x}}, (2/T)I)
$$
where $c=\prod_{i=1}^{N}c_i$.
Here $\hat{\fat{x}}$ is taken as the data and the covariance matrix $(2/T)I$ is fixed. The approximate data density is then 
$$
\pi(\fat{\hat{x}}|\fat{x}) = \textrm{N}(\fat{\hat{x}}|\fat{x},(2/T)I).
$$ 

Here the hyperparameter is $\theta=\tau$.
To sample from the posterior density of $(\fat{x},\theta)$, we first draw a sample from
the marginal posterior density of $\theta$ and
and then we sample from the  posterior density of $\fat{x}$
conditional on $\hat{\fat{x}}$ and $\theta$.  
When equation (\ref{eq:marginaltheta2}) is applied to the model 
with $\hat{\fat{\eta}}=\hat{\fat{x}}$ and $\fat{\nu}=\fat{x}$
then 
$$
\pi(\theta|\hat{\fat{x}})\propto \pi(\theta)\pi(\hat{\fat{x}}|\theta)
=
\pi(\theta)
\frac{\pi(\hat{\fat{x}}|\fat{x},\theta)\pi(\fat{x}|\theta)}
{\pi(\fat{x}|\hat{\fat{x}},\theta)},
$$
where   
$$
\pi(\hat{\fat{x}}|\fat{x},\theta) = \textrm{N}(\hat{\fat{x}}|\fat{x},(2/T)I),
$$
$$
\pi(\fat{x}|\hat{\fat{x}},\theta)
=\textrm{N}(\fat{x}|\fat{\mu}_{x|\hat{x}},\Sigma_{x|\hat{x}}),
$$
$$
\Sigma_{x|\hat{x}} = \{Q_{x} + (T/2)I\}^{-1}=\{\theta Q_{u} + (T/2)I\}^{-1},
$$
$$
\hat{\fat{\mu}}_{x|\hat{x}} = \Sigma_{x|\hat{x}}(T/2)\hat{\fat{x}}=\{(2/T)\theta Q_{u} + I\}^{-1}\hat{\fat{x}},
$$
and the prior density for $\fat{x}$, $\pi(\fat{x}|\theta)$, is Gaussian with mean zero and precision matrix specified with equation (\ref{eq:qumatrix_b}). Samples from the posterior density of $\fat{x}$
conditional on $\hat{\fat{x}}$ and $\theta$ are obtained
from the Gaussian density
$\pi(\fat{x}|\hat{\fat{x}},\theta)
=\textrm{N}(\fat{x}|\fat{\mu}_{x|\hat{x}},\Sigma_{x|\hat{x}})$
presented above. 
 
\subsection{Linear regression on a lattice}

\subsubsection{Statistical model}
\label{subsubsec:statmodel1}

Let $y_{i,t}$ and $f_{i,t}$ be the response and the predictor at lattice point $i$ and time $t$, respectively. The data are assumed to follow a Gaussian 
distribution with mean 
$$
\alpha_i + \beta_i (f_{i,t}-\bar{f}_i),
$$ 
where $\bar{f}_i=n^{-1}\sum_{t=1}^{T}f_{i,t} $, and variance $\sigma^2_i$, i.e.,
$$
y_{i,t}\sim \textrm{N}(\alpha_i + \beta_i (f_{i,t}-\bar{f}_i),\sigma^2_i),\quad\alpha_i,\beta_i \in \mathbb{R}, \quad \sigma^2_i > 0, 
$$ 
where $i\in \{1,...,N\}$, $t\in \{1,...,T\}$, $N$ is the number of lattice points and $T$ is the number of observations at each lattice point. The lattice points form the groups, so the number of groups is $G=N$. The data density for $\fat{y}_i=(y_{i1},...,y_{i T})\trp$ is then  
\begin{equation}
\label{eq:datadens1}
\pi(\fat{y}_i|\alpha_i,\beta_i,\sigma^2_i) = \prod_{t=1}^{T}\frac{1}{\sqrt{2 \pi \sigma^2_i}} \exp\bigg[-\frac{\{y_{i,t}-\alpha_i - \beta_i (f_{i,t}-\bar{f}_i)\}^2}{2\sigma^2_i}\bigg].
\end{equation}
The parameter $\sigma^2_i$ is transformed to $\tau_i = \log(\sigma^2_i)$, so $\tau_i \in \mathbb{R}$.

The parameters are collected into the vectors $\fat{\alpha}=(\alpha_1,...,\alpha_N)$, $\fat{\beta}=(\beta_1,...,\beta_N)$ and $\fat{\tau}=(\tau_1,...,\tau_N)$.  
The latent level consists of three linear predictors for $\fat{\alpha}$, $\fat{\beta}$ and $\fat{\tau}$ of the form 
\begin{equation}
\label{eq:LatentLevel}
\begin{aligned}
\fat{\alpha} &=  \fat{u}_{\alpha} + \fat{\epsilon}_{\alpha}, \\
\fat{\beta} &=  \fat{u}_{\beta} + \fat{\epsilon}_{\beta}, \\
\fat{\tau} &=  \fat{u}_{\tau}+ \fat{\epsilon}_{\tau},
\end{aligned}
\end{equation}
where $\fat{u}_\alpha$, $\fat{u}_\beta$ 
and $\fat{u}_\tau$ are modeled as first-order intrinsic Gaussian Markov random fields (IGMRFs) on a regular lattice \citep[Section 3.3.2, pp. 104--108]{rue2005gaussian}. 
Selecting Gaussian Markov random fields as priors for $\fat{\alpha}$, $\fat{\beta}$ and $\fat{\tau}$ 
is justified as spatial dependancy between elements close in space is expected. 
This model choice is beneficial for the estimation of these parameters as it allows borrowing strength from neighboring
lattice points.   

A priori the vectors $\fat{u}_{\alpha}$, $\fat{u}_{\beta}$ and $\fat{u}_{\tau}$ are assumed to be independent. Their precisions matrices are
$Q_{u,\alpha}$, $Q_{u,\beta}$ and $Q_{u,\tau}$ where $Q_{u,l} = \sigma_{u,l}^{-2}Q_{u}$, $l \in \{\alpha,\beta,\tau \}$, and $Q_u$ is the precision matrix given by equation (\ref{eq:qumatrix}) below, and $\sigma_{u,\alpha}$, $\sigma_{u,\beta}$ and $\sigma_{u,\tau}$ are unknown hyperparameters.  
The dimensions of the lattice are $n_1$ and $n_2$ and $N=n_1 n_2$. The form of $Q_u$ is 
\begin{equation}
Q_{u} = R_{n_1}\otimes I_{n_2} +  I_{n_1} \otimes R_{n_2}
\label{eq:qumatrix}
\end{equation}
where $\otimes$ denotes the Kronecker product, $I_{m}$ is an $m$ dimensional identity matrix and  
\begin{equation}
R_{m} = \begin{bmatrix} 1 & -1 & & & & & \\ 
                           -1 & 2 & -1 & & & & \\
                           & -1 & 2 & -1 & &  & \\
                           & & \vdots & \vdots & \vdots & & \\
                           & & & -1 & 2 & -1 &  \\
                           &  & & & -1 & 2 & -1  \\                            
                           & & & & & -1 & 1 \end{bmatrix}
\label{eq:precisionr}
\end{equation}
is an $m \times m$ matrix. Let
$x_{i_1,i_2}$ be the element of $\fat{x}$ with horizontal coordinates $i_1$ and vertical coordinates $i_2$. If $(i_1,i_2)$ is an interior lattice point then the conditional density of $x_{i_1,i_2}$ conditional on $\fat{x}_{-(i_1,i_2)}$ (the elements of $\fat{x}$ other than $x_{i_1,i_2}$) is Gaussian with mean
$$
\frac{1}{4}(x_{i_1+1,i_2}+x_{i_1-1,i_2}+x_{i_1,i_2+1}+x_{i_1,i_2-1})
$$
and variance $0.25 \sigma_{u,l}^{2}$.
The precision matrix $Q_{u}$ is not full rank, its rank is $N-1$. By 
conditioning on $\sum_i u_{l,i}=\fat{1}\trp \fat{u}_l=u_l^*$ then the density for $\fat{u}_l$ can be specified with the proper density
$\pi(\fat{u}_l)=\pi(\fat{u}_l|u_l^*)\pi(u_l^*)$ where $\pi(u_l^*)$ is a Gaussian density with mean zero and precision $\gamma$, and
\begin{equation}
\log \pi(\fat{u}_l|u_l^*) = -\frac{(N-1)}{2}\log 2(\pi) - \frac{(N-1)}{2}\log(\sigma_{u,l}^{2}) + \frac{1}{2}\sum_{i=2}^{N}\log(\lambda_i) - \frac{\sigma_{u,l}^{-2}}{2}\fat{u}_l\trp Q_{u}\fat{u}_l,
\label{eq:densityx}
\end{equation}
where $\lambda_i$ is the $i$-th eigenvalue of $Q_{u}$,
$$
\lambda_{i} = 2(1 - \cos(\pi(i_1-1)/n_1)) + 2(1 - \cos(\pi(i_2-1)/n_2)), 
$$  
$i_1=1,...,n_1$, $i_2=1,...,n_2$, $i = i_1 + n_1(i_2-1)$. Note that the first eigenvalue ($i=1$) is excluded in the above density.   
By letting $\gamma \rightarrow 0$ then $\pi(\fat{u}_l) \propto \pi(\fat{u}_l|u_l^*)$, and $\pi(\fat{u}_l|u_l^*)$ can be used 
to infer $\fat{u}_l$ and $\sigma_{u,l}$.

The three vectors $\fat{\epsilon}_{\alpha}$, $\fat{\epsilon}_{\beta}$ and $\fat{\epsilon}_{\tau}$ contain unstructured model errors with variances $\sigma_{\epsilon,\alpha}^2,\sigma_{\epsilon,\beta}^2$ and $\sigma_{\epsilon,\tau}^2$, respectively. The prior distributions for $\fat{\epsilon}_{\alpha}$, $\fat{\epsilon}_{\beta}$ and $\fat{\epsilon}_{\tau}$ can be written as 
$$
\fat{\epsilon}_{\alpha}{\sim}\textrm{N}(0,\sigma_{\epsilon,\alpha}^2\text{I}), \quad \fat{\epsilon}_{\beta}{\sim}\textrm{N}(0,\sigma_{\epsilon,\beta}^2\text{I}), \quad \fat{\epsilon}_{\tau}{\sim}\textrm{N}(0,\sigma_{\epsilon,\tau}^2\text{I}).
$$
At the hyperparameter level an exponential prior density is assigned to each of the parameters 
$\sigma_{u,\alpha}$, $\sigma_{\epsilon,\alpha}$, $\sigma_{u,\beta}$, $\sigma_{\epsilon,\beta}$, $\sigma_{u,\tau}$ and $\sigma_{\epsilon,\tau}$. 
The decision to select independent exponential distributions for these parameters is motivated by the penalized complexity (PC) prior framework presented in \cite{simpson2017penalising}.
The PC prior framework can be used to design prior densities for hyperparameters like the standard deviation parameters of random effects.
It is based on the Kullback-Leibler divergence and is designed to support simpler models.
One of the principles of this framework is such that in the case of standard deviation parameters their PC prior densities support the case of them being equal to zero. 
If the data suggest that a given parameter is equal to zero then the PC prior density
pulls the posterior mass toward zero. On the other hand, evidence for a more complex model 
with a standard deviation parameter different from
zero needs to be pulled by the data. 

\subsubsection{Inference}
\label{sec:samplescheme1}

We apply Max-and-Smooth to infer the unknown parameters of
the extended LGM above, see Section \ref{sec:inferscheme} and Section 3.3 in the main paper. The extended LGM model has a trivariate link function.
Let $\fat{\psi}_i = (\alpha_i, \beta_i, \sigma^2_i)\trp$ and let
$g:\mathbb{R}^2 \times \mathbb{R}_{+} \rightarrow \mathbb{R}^3$ be a trivariate link function such that $g(x_1, x_2, x_3) = (x_1, x_2, \log(x_3))\trp$, and $g^{-1}(u_1, u_2, u_3)=(u_1, u_2, \exp(u_3))\trp$. 
Let 
$$
\fat{\eta}_i:=(\alpha_i, \beta_i,\tau_i)\trp =g(\fat{\psi_i})=(\alpha_i, \beta_i, \log(\sigma^2_i))\trp.
$$ 
Now define $\hat{\fat{\psi}}_i:=(\hat{\alpha}_i, \hat{\beta}_i, \hat{\sigma}^2_i)\trp$ as 
the maximum likelihood estimate of $\fat{\psi}_i$ based on the data density for lattice point $i$ given by (\ref{eq:datadens1})
which is computed separately from the other lattice points. 
Let $F_i$ be the design matrix with $(1,f_{i,t}-\bar{f}_i)$ as its $t$-th row. Then $(\hat{\alpha}_i, \hat{\beta}_i)\trp = (F_i\trp F_i)^{-1}F_i\trp \fat{y}_i$
and $\hat{\sigma}^2_i=T^{-1}\sum_{t=1}^{T}\{y_{i,t}-\hat\alpha_i - \hat\beta_i (f_{i,t}-\bar{f}_i)\}^2$.   

Instead of using $\fat{y}_i$ as the data, we treat  
$$
\hat{\fat{\eta}}_i=(\hat{\alpha}_i, \hat{\beta}_i, \hat{\tau}_i)\trp=g(\hat{\fat{\psi}}_i)=(\hat{\alpha}_i, \hat{\beta}_i, \log(\hat{\sigma}^2_i))\trp
$$ 
as the data for lattice point $i$. 
Let $H_{\eta_i}$ denote the Hessian matrix corresponding to the logarithm of the likelihood evaluated at the mode $\hat{\fat{\eta}}_i$ where
$$
H_{\eta_i}=\nabla^2 \log (L( \fat{\eta}_i|\fat{y}_i))|_{\eta_i=\hat{\eta}_i}.
$$
The observed information matrix, $\mathcal{I}_{\eta y_i}$, is equal to the negative Hessian matrix, $\mathcal{I}_{\eta y_i}=-H_{\eta_i}$, and it is given by
$$
\mathcal{I}_{\eta y_i} =\begin{bmatrix}\exp(-\hat{\tau}_i)(F_i\trp F_i)_{1,1} & \exp(-\hat{\tau}_i)(F_i\trp F_i)_{1,2} & 0 \\ 
                            \exp(-\hat{\tau}_i)(F_i\trp F_i)_{2,1} & \exp(-\hat{\tau}_i)(F_i\trp F_i)_{2,2} & 0 \\ 
                            0 & 0 & T/2 \end{bmatrix}.
$$
The likelihood function for $\fat{\eta}_i$ and its approximation are such that
$$
L(\fat{\eta}_i)=\pi(\fat{y}_i|\fat{\eta}_i)\approx c_i\hat{L}(\fat{\eta}_i)=c_i\pi(\fat{\eta}_i|\hat{\fat{\eta}}_i) = c_i\textrm{N}(\fat{\eta}_i|\hat{\fat{\eta}}_i,\mathcal{I}_{\eta y_i}^{-1})
$$
where $c_i$ is a constant independent of $\fat{\eta}_i$.   
Now let $\fat{\eta} = (\fat{\alpha}\trp, \fat{\beta}\trp, \fat{\tau}\trp)\trp$ with $\fat{\alpha}$, $\fat{\beta}$ and $\fat{\tau}$ as before,  
then the likelihood of $\fat{\eta}$, $L(\fat{\eta})$, is approximated with $\hat{L}(\fat{\eta})$,
that is 
$$
L(\fat{\eta})=\prod_{i=1}^N L(\fat{\eta}_i) \approx c\hat{L}(\fat{\eta}) =\prod_{i=1}^N c_i \hat{L}(\fat{\eta}_i) = c \textrm{N}(\fat{\eta}|\hat{\fat{\eta}}, Q_{\eta y}^{-1})
$$
where $\hat{\fat{\eta}}=(\hat{\fat{\alpha}}\trp,\hat{\fat{\beta}}\trp, \hat{\fat{\tau}}\trp)\trp$ and $c=\prod_{i=1}^{N}c_i$.
Define $\hat{\fat{\eta}}^*:=(\hat{\alpha}_1,\hat{\beta}_1,\hat{\tau}_1,...,\hat{\alpha}_N,\hat{\beta}_N,\hat{\tau}_N)\trp$ as a
rearrangement of $\hat{\fat{\eta}}$ then  
$$(Q_{\eta y}^*)^{-1}=\textrm{bdiag}((\mathcal{I}_{\eta y_1})^{-1},...,(\mathcal{I}_{\eta y_N})^{-1}),$$
is known, and it is possible to rearrange $(Q_{\eta y}^*)^{-1}$ accordingly to get $Q_{\eta y}^{-1}$.  

Here $\hat{\fat{\eta}}$ is taken as the data and $Q_{\eta y}^{-1}$ as fixed. The approximate data density is then 
$$
\pi(\fat{\hat{\eta}}|\fat{\eta}, \fat{\nu}, \theta) = \textrm{N}(\fat{\hat{\eta}}|\fat{\eta},Q_{\eta y}^{-1})
$$
where $\fat{\nu}$ are all unknown parameters that are assigned a Gaussian prior distribution and $\fat{\theta}$ are all unknown hyperparameters. 
The vectors $\fat{x}$, $\fat{\eta}$, $\fat{\nu}$ and $\fat{\theta}$ are such that 
$$
\fat{x}\trp=(\fat{\eta}\trp,\fat{\nu}\trp), \quad \fat{\eta}\trp = (\fat\alpha\trp,\fat\beta\trp,\fat\tau\trp), \quad \fat{\nu}\trp = (\fat{u}_{\alpha}\trp,\fat{u}_{\beta}\trp,\fat{u}_{\tau}\trp)
$$ 
and  
$$\fat{\theta}=(\theta_1,...,\theta_6)\trp = (\sigma_{u,\alpha}, \sigma_{\epsilon,\alpha}, \sigma_{u,\beta}, \sigma_{\epsilon,\beta}, \sigma_{u,\tau}, \sigma_{\epsilon,\tau})\trp.$$
To sample from the posterior density of $(\fat{x},\fat{\theta})$, we first draw a sample from  
$$
\pi(\fat{\theta}|\hat{\fat{\eta}})\propto \pi(\fat{\theta})\pi(\hat{\fat{\eta}}|\fat{\theta})
=
\pi(\fat{\theta})
\frac{\pi(\hat{\fat{\eta}}|\fat{x},\fat{\theta})\pi(\fat{x}|\fat{\theta})}
{\pi(\fat{x}|\hat{\fat{\eta}},\fat{\theta})},
$$
and then we sample from the conditional posterior density of $\fat{x}$
given $\hat{\fat{\eta}}$ and $\fat{\theta}$. This conditional density, $\pi(\fat{x}|\hat{\fat{\eta}},\fat{\theta})$,
is specified in Section \ref{sec:inferscheme}. 
Since the posterior density for $\fat{\theta}$ is such that the subvector 
$(\sigma_{u,\tau},\sigma_{\epsilon,\tau})$ is independent of the subvector 
$(\sigma_{u,\alpha},\sigma_{\epsilon,\alpha},\sigma_{u,\beta},\sigma_{\epsilon,\beta})$
then it is feasible to use grid sampling based on a two dimensional grid for 
$(\sigma_{u,\tau},\sigma_{\epsilon,\tau})$,
and a four dimensional grid for 
$(\sigma_{u,\alpha},\sigma_{\epsilon,\alpha},\sigma_{u,\beta},\sigma_{\epsilon,\beta})$.
This approach yields independent samples from the marginal posterior density of $\fat{\theta}$.
Furthermore, as the samples from the full conditional posterior density of $\fat{x}$
are drawn from a Gaussian density then the sampling scheme yields independent
samples from the approximated posterior density. As a result, there is no autocorrelation between the samples, and therefore, the proposed sampling scheme will be efficient. 

Alternatively, a Metropolis step can be used to sample from the marginal posterior of $\fat{\theta}$.
This requires the parameters to be defined on the real-line and calls for a transformation of the
form $\fat{\kappa} = (\log\theta_1,...,\log\theta_6)\trp$.
Samples from $\pi(\fat{\kappa}|\hat{\fat{\eta}})$ are generated with a Metropolis step where the proposal distribution is 
$$
q(\fat{\kappa}^*|\fat{\kappa}^{(l)},Q_\kappa^{-1}) = \textrm{N}(\fat{\kappa}^*|\fat{\kappa}^{(l)},Q_\kappa^{-1})
$$
where $\fat{\kappa}^{(l)}$ is the value of $\fat{\kappa}$ in the $l$-th iteration.
If the sampling scheme proposed by
\citet{roberts1997weak} is used then $Q_\kappa$ is set equal to the negative Hessian matrix of $\pi(\fat{\kappa}|\hat{\fat{\eta}})$ times $\dim(\fat{\kappa})/2.382^2$.
The prior density of each ${\kappa}_k$ is
$$
\pi(\kappa_k) = {\lambda_k}\exp\{-\lambda_k \exp(\kappa_k) + \kappa_k\}
$$
where $k\in \{1,...,6\}$. 
The prior distribution of $\fat{\kappa}$ can now be written as 
$$
\pi(\fat{\kappa}) = \prod_{k=1}^{6}\pi(\kappa_k)
$$
and the marginal posterior density of $\fat{\kappa}$ becomes
$$
\pi(\fat{\kappa}|\hat{\fat{\eta}})\propto
\pi(\fat{\kappa})
\frac{\pi(\hat{\fat{\eta}}|\fat{x},\fat{\kappa})\pi(\fat{x}|\fat{\kappa})}
{\pi(\fat{x}|\fat{\kappa},\hat{\fat{\eta}})}.
$$

We can also apply Max-and-Smooth to infer 
$\fat{\alpha}$, $\fat{\beta}$ and $\fat{\tau}$ by approximating the likelihood function of the parameters of the $i$-th lattice point with a Gaussian density that has the same mean and variance as the 
normalized likelihood function of $(\alpha_i,\beta_i,\tau_i)$. In the case of a normalized likelihood function stemming from a general Gaussian regression model with $p$ regression coefficients in $\fat{\beta}$ and error log-variance $\tau$, it can be shown that the covariance between $\fat{\beta}$ and $\tau$ is zero, thus, under the Gaussian approximation $\fat{\beta}$ and $\tau$ are independent.  
The marginal likelihood function of $\tau$ is a log-inverse-gamma density with parameters $\alpha=(T-p)/2$ and $\gamma=s^{2}(T-p)/2$ where
$$
s^{2}=(T-p)^{-1}
\sum_{t=1}^{T}(y_{t} - \fat{f}_{t} \hat{\fat{\beta}})^2 
$$
and $\fat{f}_{t}$ contains the covariates of the $t$-th observation and it is also the $t$-th row of the design matrix $F$. The mean and variance of $\tau$
stemming from the normalized likelihood function are
$$
\textrm{E}(\tau)=\log(s^2)+\log((T-p)/2)-\psi((T-p)/2), \quad
\textrm{var}(\tau)=\psi'((T-p)/2).
$$ 
Furthermore, the marginal likelihood function of $\fat{\beta}$ is a multivariate $t$-density with parameters 
$$
\fat{\mu}=(F\trp F)^{-1}F\trp \fat{y}, \quad 
\Sigma=s^2(F\trp F)^{-1}.
$$
The mean and variance of $\fat{\beta}$ are
$$
\textrm{E}(\fat{\beta})=(F\trp F)^{-1}F\trp \fat{y}, \quad
\textrm{var}(\fat{\beta})=\frac{(T-p)}{(T-p-2)}s^2(F\trp F)^{-1}.
$$ 
Therefore, the Gaussian approximation for $\fat{\beta}$ based on the mean and variance of the normalized likelihood function will have the same mean as the Gaussian approximation for $\fat{\beta}$ based on ML estimates and the observed information while 
its variance is larger by a factor $T(T-p-2)^{-1}$.

\subsubsection{Results}
\label{subsubsec:results1}

\begin{figure}[h]
\centering
\includegraphics[width=0.99\textwidth]{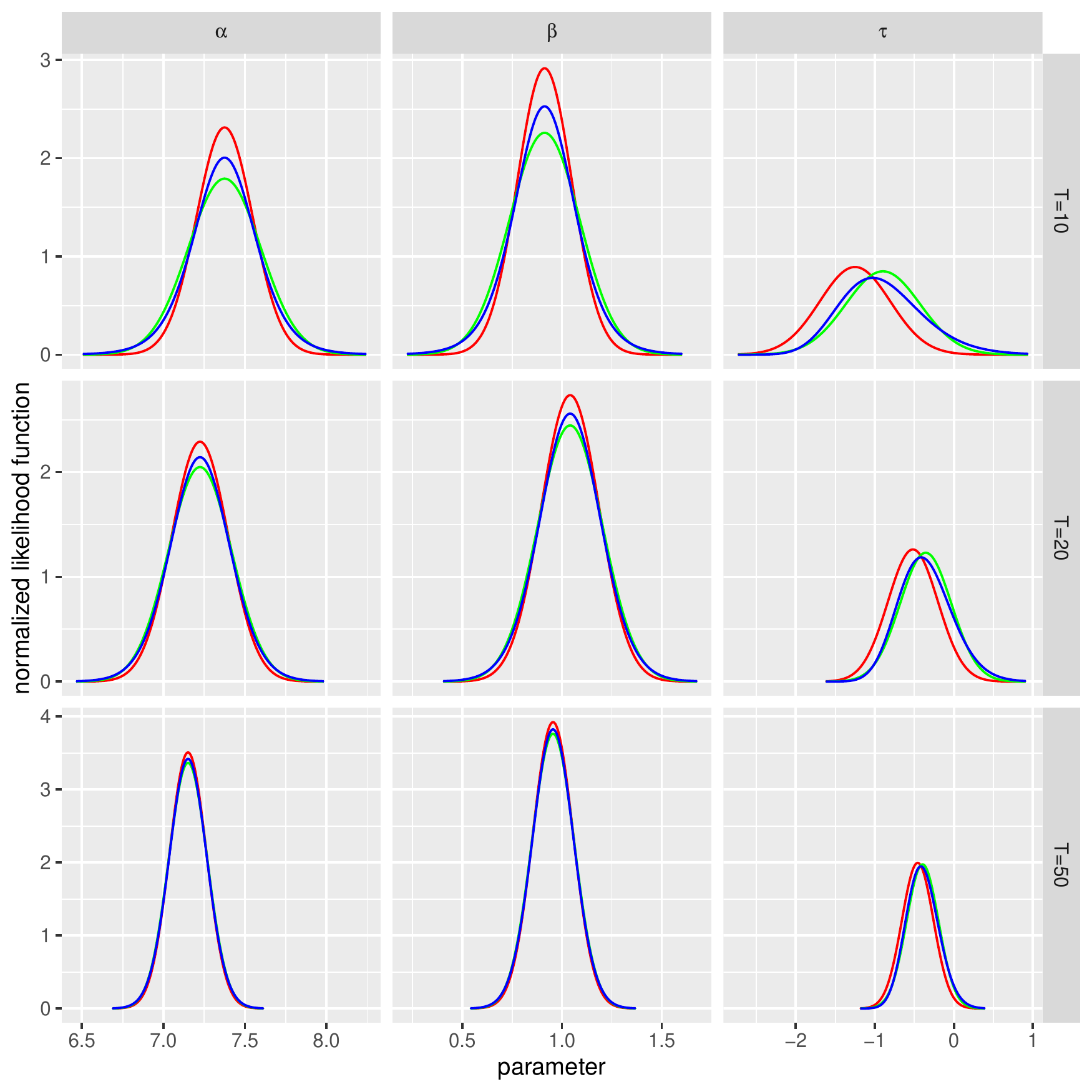}
\caption{Linear regression on a lattice. The marginal normalized likelihood functions (blue lines) of $\alpha$ (left column of panel), $\beta$ (middle column of panel) and $\tau$ (right column of panel), along with the marginal densities of the Gaussian approximations based on (a) the ML estimates and the inverse of the observed information (red lines), and (b) the mean and variance of the normalized likelihood function (green lines). The top, middle and bottom row panels show the results for $T=10,20,50$, respectively.}
\label{fig:normlikhood}
\end{figure}

In this subsection we simulate data from a latent Gaussian model representing a linear regression models on a lattice. The purpose of this simulation is to evaluate the Gaussian approximation to the likelihood function,
and to check whether the marginal posterior intervals capture the true values of the inferred parameters. 
The dimension of the lattice is $61 \times 61$. 
Figure \ref{fig:normlikhood} reflects the likelihood function of $\alpha$, $\beta$ and $\tau$ 
for a single lattice point with respect to the Gaussian approximations of the likelihood function based on the ML estimates and the observed information matrix, and based on the mean and variance of the normalized likelihood function.
The part of the likelihood function corresponding to $\alpha$, $\beta$ and 
$\tau$ at the particular lattice point is normalized by integrating over the three variables. 
The normalized likelihood function is treated as a probability density and based on it 
the marginal probability functions of $\alpha$, 
$\beta$ and $\tau$ can be derived. The marginal normalized likelihood functions of $\alpha$ and $\beta$ are 
scaled $t$-densities with $T-2$ degrees of freedom. The marginal 
normalized likelihood function of $\tau$ is a log-inverse-gamma density. This density is also referred to as a log-scaled-inverse $\chi^2$ density with $T-2$ degrees of freedom. 
Figure \ref{fig:normlikhood} shows the marginal normalized likelihood functions of $\alpha$, $\beta$ and $\tau$ along with the marginal densities of the trivariate Gaussian densities that are used to approximated the likelihood function
in the case of $T=10,20,50$.
The Gaussian approximation based on the ML estimates is 
reasonably good in the cases of $\alpha$ and $\beta$ when
$T=20$ and $T=50$ as expected since the marginal normalized likelihood 
functions are $t$-densities with $18$ and $48$ degrees of freedom.
The Gaussian approximation is not as good for $\alpha$ and $\beta$ when
$T=10$ as that case boils down to approximating a
$t$-density with $8$ degrees of freedom with a Gaussian density. 
In the case of the log-variance, $\tau$, the Gaussian approximation is 
acceptable but it misses the right skewness of the 
marginal normalized likelihood function of $\tau$.
Figure \ref{fig:normlikhood} reveals that this skewness decreases rapidly as
the sample size increases.
A comparison of the results shown in Figure \ref{fig:normlikhood} reveals that the Gaussian approximation that has the same mean and variance as the normalized likelihood function gives a slightly better approximation than the Gaussian approximation that is based on the ML estimate and the observed information matrix. This is most obvious for $\tau$ when $T=10$ and $T=20$, but it is also visible for $\alpha$ and $\beta$ when $T=10$ and $T=20$. However, the difference between the two Gaussian approximations to the normalized likelihood function is small when $T=50$.

\begin{figure}[h]
\centering
\includegraphics[width=0.99\textwidth]{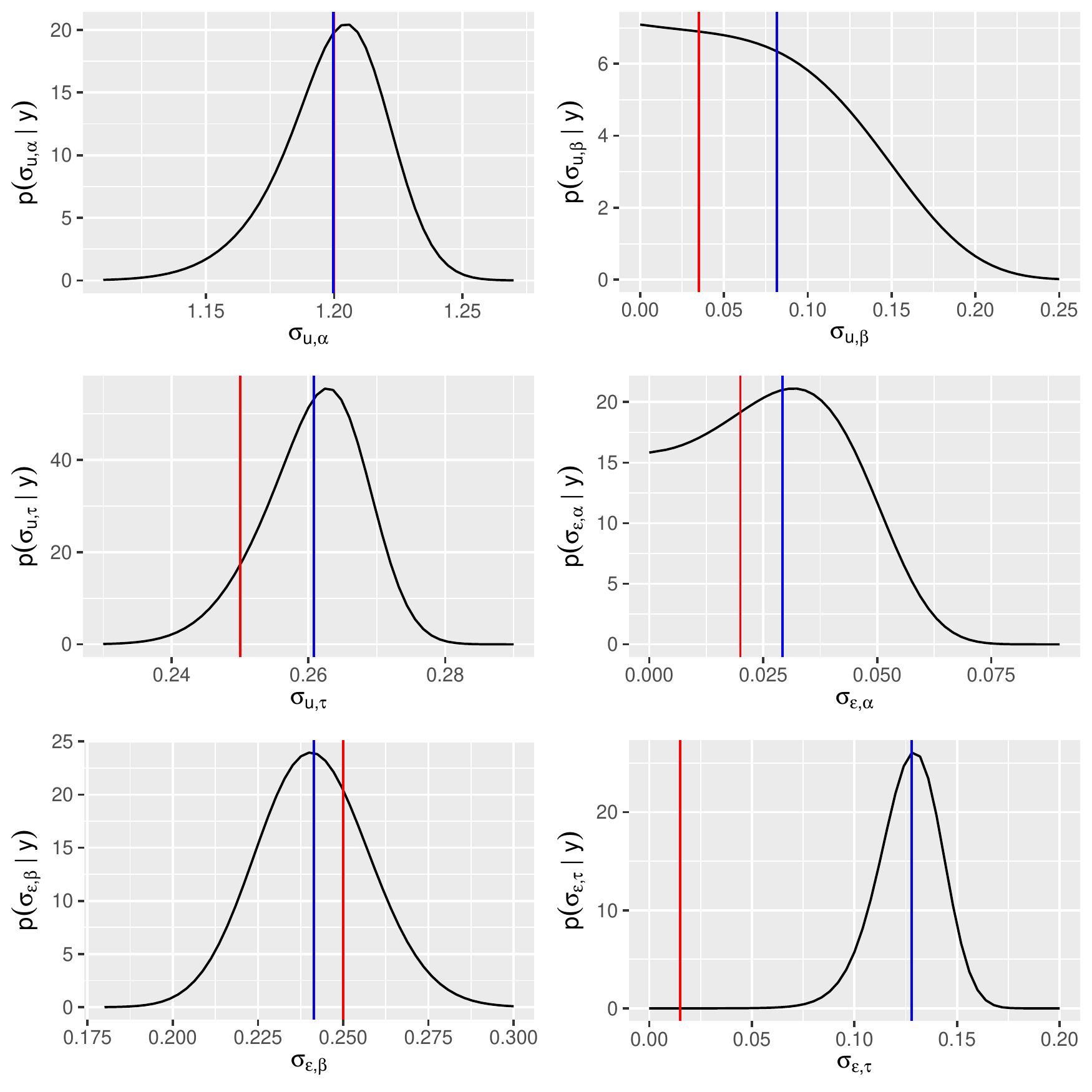} 
\caption{Linear regression on a lattice. The marginal posterior densities of the hyperparameters (black lines). Top panel: $\sigma_{u,\alpha}$ (left), $\sigma_{\epsilon,\alpha}$ (right). Middle panel: $\sigma_{u,\beta}$ (left), $\sigma_{\epsilon,\beta}$ (right). Bottom panel: $\sigma_{u,\tau }$ (left), $\sigma_{\epsilon,\tau}$ (right). The solid vertical red lines show the true values while the solid vertical blue line show the posterior means. In the case of $\sigma_{u,\alpha}$, the true value and the posterior mean are close.}
\label{fig:margin_post_dens}
\end{figure}

Figure \ref{fig:margin_post_dens} shows the marginal posterior densities of the hyperparameters and it can be seen that the
marginal posterior densities capture the true values of $\sigma_{u,\alpha}$, $\sigma_{u,\beta}$ and $\sigma_{u,\tau}$ quite well.
The marginal posterior densities of $\sigma_{\epsilon,\alpha}$, 
$\sigma_{\epsilon,\beta}$ and $\sigma_{\epsilon,\tau}$
in Figure \ref{fig:margin_post_dens} show that it is more difficult to determine 
the values of $\sigma_{\epsilon,\alpha}$, $\sigma_{\epsilon,\beta}$ 
and $\sigma_{\epsilon,\tau}$ compared to  
$\sigma_{u,\alpha}$, $\sigma_{u,\beta}$ and $\sigma_{u,\tau}$. In particular, the marginal posterior 
density of $\sigma_{\epsilon,\tau}$ gives negligible probability mass over the true value
and the posterior mean is much larger than the true value. 

\begin{figure}[h]
\centering
\includegraphics[width=0.99\textwidth]{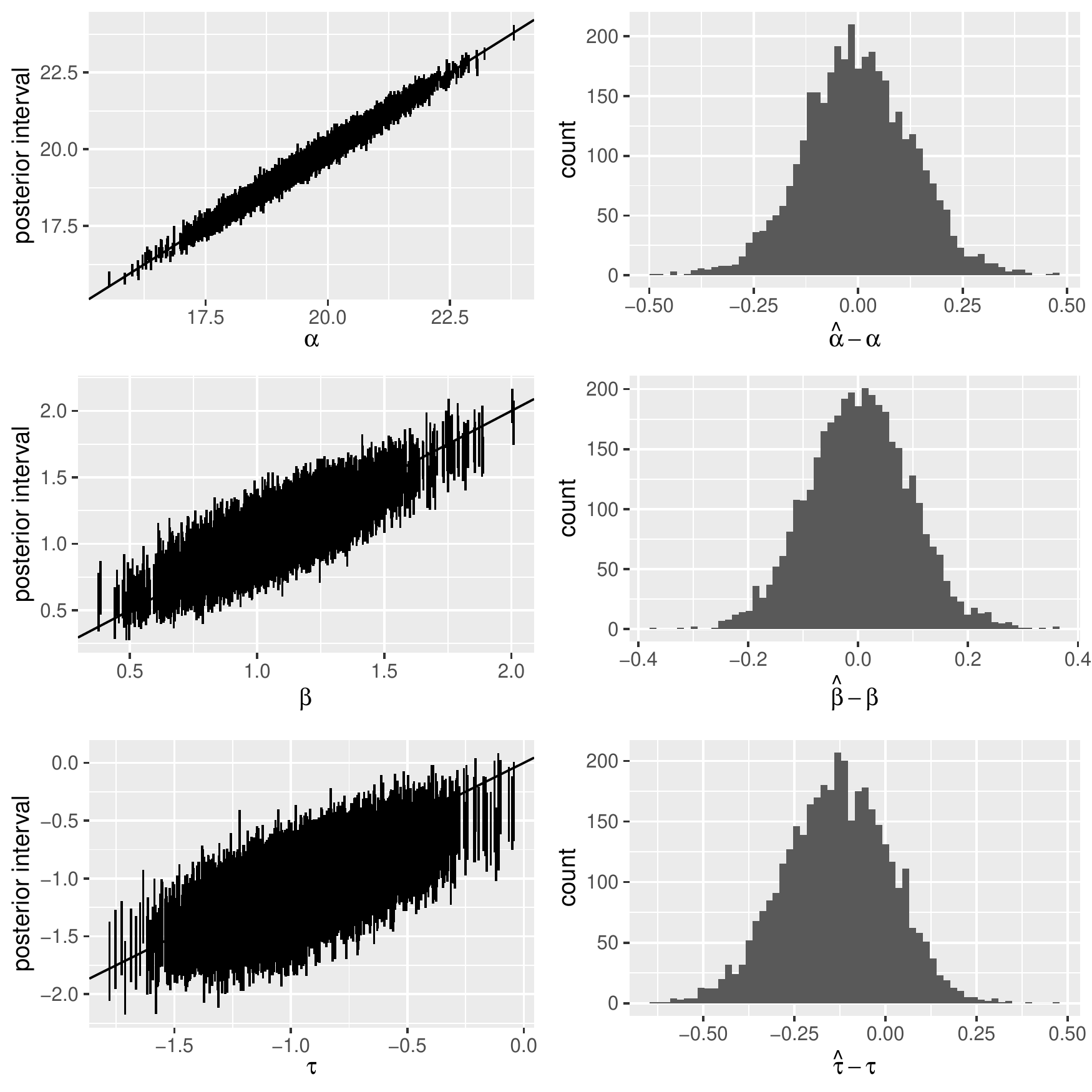} 
\caption{Linear regression on a lattice. The elements in $\fat\alpha$ (top row of panel): the 95\% posterior intervals versus the true values (left column of panel) and a histogram of the posterior means minus the true values (right column of panel). The middle row and the bottom row of the panel show the same for the elements in $\fat\beta$ and the elements in $\fat\tau$, respectively.}
\label{fig:intervals}
\end{figure}

Figure \ref{fig:intervals} shows 95\% posterior credible intervals of the elements in $\fat{\alpha}$, $\fat{\beta}$
and $\fat\tau$ as a function of their true values as well as histograms of the difference between the 
posterior means and the true values. Since the data are simulated, the coverage of the 95\% posterior
credible intervals is evaluated for each of the vectors $\fat{\alpha}$, $\fat{\beta}$ and $\fat{\tau}$
by calculating the proportion of the credible intervals within each vector that capture the 
corresponding true value. The proportions for $\fat{\alpha}$, $\fat{\beta}$ and $\fat{\tau}$
are $92.4$\%, $93.9$\% and $89.0$\%, respectively. The difference between the 
posterior means and the true values of $\fat{\alpha}$ and $\fat{\beta}$ are centred around zero
while this same difference of the elements $\fat{\tau}$ is centred around a value close to $-0.1337$. 
Thus, the posterior means are more likely to give an underestimate of the true value of the elements in $\fat{\tau}$.     
This could be due to the fact that the Gaussian approximation for the marginal normalized likelihood density for 
a given $\tau$ has a mean that is smaller than the mean of the actual marginal normalized likelihood density for $\tau$.
The Gaussian approximation based on the mean and variance of the normalized likelihood function gives mean that is greater than the mean of the Gaussian approximation based on the ML estimates by $0.1393$ for all lattice points $i$. When the Gaussian approximation based on the mean and variance of the normalized likelihood function is applied (results not shown) then the bias defined above is close to zero, which supports using this approximation when the normalized likelihood function is skewed.  
This skewness is most likely observed when the number of replicates with in a group is small.

\subsection{Linear regression model with temporally varying coefficients}

Linear regression models for time referenced data consisting of variable of interest and corresponding predictors
can sometimes be improved by allowing the regression coefficients
to vary over time.
Examples of data of this type are data on real estate prices \citep{gelfand2007multilevel} 
and spatio-temporal fillet yield data \citep{margeirsson2010decision}. A general setup
for models with temporally varying regression coefficients can be found
in \cite{west1999bayesian} and \cite{prado2010time}.

\subsubsection{Statistical model}

Here $y_{t,k}$ denotes the $k$-th observation of the variable of interest at time $t$, and $\fat{x}_{t,k}$ is a vector of $p$ predictors for the $k$-th observation. 
The data within a given time period are assumed to follow 
the $t$-distribution with location parameter $\mu_{t,k} =\fat{x}_{t,k}^{\trp}\fat{\beta}_{t}$, scale parameter $\sigma_{t}$ and degrees of freedom parameter $\vartheta_t$, i.e.,
$$
y_{t,k}\sim t_{\vartheta_t}(\fat{x}_{t,k}^{\trp}\fat{\beta}_{t},\sigma_t^2), \quad \fat{\beta}_{t} \in \mathbb{R}^p, \quad \sigma_t > 0, \quad \vartheta_t > 0,  
$$ 
where $k\in \{1,...,n_t\}$, $t \in \{1,...,T\}$ and $n_t$ is the number of observations 
at time $t$. Each time point $t$ forms a group, so, here the number of groups is $G=T$. 
The data density for the data at time $t$, $\fat{y}_t=(y_{t,1},...,y_{t,n_t})\trp$, is given by  
\begin{equation}
\label{eq:datadens3}
\pi(\fat{y}_t|\fat{\beta}_{t},\sigma_t,\vartheta_t) = \prod_{k=1}^{n_t}
\frac{\Gamma((\vartheta_t+1)/2)}{\Gamma(\vartheta_t/2)\sqrt{\vartheta_t \pi}\sigma_t} \left\{1 + \frac{1}{\vartheta_t} \left(\frac{y_{t,k}-\fat{x}_{t,k}^{\trp}\fat{\beta}_{t}}{\sigma_t}    \right)^2    \right\}^{-(\vartheta_t+1)/2}.
\end{equation}
The parameters $\sigma_t$ and $\vartheta_t$ are transformed to $\tau_t = \log(\sigma_t)$ and $\varphi_t = \log(\vartheta_t)$, so $\tau_t, \varphi_t \in \mathbb{R}$.

The parameters are collected into the vectors $\fat{\beta}_s =(\beta_{1,s},...,\beta_{T,s})$, $s=1,...,p$, $\fat{\tau}=(\tau_1,...,\tau_T)$ and $\fat{\varphi}=(\varphi_1,...,\varphi_T)$.  
The latent level consists of linear predictors for these $p+2$, vectors,
\begin{equation}
\label{eq:LatentLevel4}
\begin{aligned}
\fat{\beta}_s &= \fat{u}_{\beta,s} + \fat{\epsilon}_{\beta,s}, \quad s = 1,...,p, \\
\fat{\tau} &=    \fat{u}_{\tau} + \fat{\epsilon}_{\tau}, \\
\fat{\varphi} &=    \fat{u}_{\varphi}+ \fat{\epsilon}_{\varphi},
\end{aligned}
\end{equation}
where $\fat{u}_{\beta,s}$, $\fat{u}_\tau$ and $\fat{u}_\varphi$ are modeled with random walk processes and $\fat{\epsilon}_{\beta,s}$, $\fat{\epsilon}_{\tau}$ and $\fat{\epsilon}_{\varphi}$ are 
unstructured random effects. For example, the model for $\fat{u}_\tau$ is such that
$u_{t,\tau}=u_{t-1,\tau} + v_t$ where  $v_t$ are independent Gaussian random variables
with mean zero and variance $\sigma^2_{u,\tau}$, and the precision matrix of
$\fat{u}_{\tau}$ is $\sigma^{-2}_{u,\tau} R_T$ 
where $R_T$ is specified by equation (\ref{eq:precisionr}) above. 

\subsubsection{Inference}

To infer the unknown parameters in the model, we apply Max-and-Smooth. Since the degrees of freedom parameter is difficult to infer, we propose using a prior density that is multiplied to
the likelihood function. In particular, we propose using a log-gamma prior density with parameters $\alpha$ and
$\gamma$. Its density is given by
$$
\pi(\varphi_{t})=\frac{\gamma^{\alpha}}{\Gamma(\alpha)}\exp\{\alpha \varphi_{t}-\gamma\exp(\varphi_{t})\},
$$
and its mean and variance are
$$
\textrm{E}(\varphi_t)=\psi(\alpha)-\log(\gamma), \quad \textrm{var}(\varphi_t)=\psi'(\alpha). 
$$
Thus, the generalized likelihood function of $(\fat{\beta}_{t},\tau_t,\varphi_t)$ is given by
\begin{equation}
L(\fat{\beta}_{t},\tau_t,\varphi_t|\fat{y}_t)=\pi(\fat{y}_t|\fat{\beta}_{t},\tau_t,\varphi_t) \pi(\varphi_{t}),
\label{eq:datadens3b}
\end{equation}
where $\sigma_t$ and $\vartheta_{t}$ are replaced by $\exp(\tau_{t})$ and $\exp(\varphi_{t})$, respectively, in the data density for $\fat{y}_t$ given by (\ref{eq:datadens3}).
Here the maximum likelihood estimate of $(\beta_{t,1},...,\beta_{t,p},\tau_t,\varphi_t)$, evaluated by maximizing the above generalized likelihood function, are modeled as an LGM with $p+2$ linear predictors with
the Gaussian--Gaussian model described in Section 3.3 of the main paper and the scheme in Section \ref{sec:inferscheme} is used for inference.

Let $\fat{\psi}_t = (\beta_{t,1},...,\beta_{t,p},\sigma_t, \vartheta_t)\trp$ and let
$g:\mathbb{R}^{p} \times \mathbb{R}_{+}^2 \rightarrow \mathbb{R}^{p+2}$ be a $(p+2)$-variate link function such that $g(x_1, ..., x_{p+2}) = (x_1, ..., x_p, \log(x_{p+1}), \log(x_{p+2}))\trp$. 
Then the inverse of $g$ is $g^{-1}(u_1, ..., u_{p+2})=(u_1, ..., u_p, \exp(u_{p+1}), \exp(u_{p+2}))\trp$. Furthermore, let 
$$
\fat{\eta}_t:=(\beta_{t,1},...,\beta_{t,p}, \tau_t,\varphi_t)\trp =g(\fat{\psi_t})=(\beta_{t,1},...,\beta_{t,p}, \log(\sigma_t), \log(\vartheta_t))\trp.
$$ 
Now define $\hat{\fat{\eta}}_t:=(\hat{\beta}_{t,1}, ..., \hat{\beta}_{t,p}, \hat{\tau}_t,\hat{\varphi_t})\trp$ 
as the maximum likelihood estimate of the 
generalized likelihood function at time $t$ given by (\ref{eq:datadens3b}) 
which is computed separately from the other time points. Instead of using $\fat{y}_t$ as the data, we treat $\hat{\fat{\eta}}_t$
as the data at time $t$. The Hessian matrix corresponding to the logarithm of the generalized likelihood function evaluated at the mode $\hat{\fat{\eta}}_t$ is given by 
$$
H_{\eta_t}=\nabla^2 \log (L( \fat{\eta}_t|\fat{y}_t))|_{\eta_t=\hat{\eta}_t}.
$$
The observed information matrix, $\mathcal{I}_{\eta y_t}$, is equal to the negative Hessian matrix, i.e., $\mathcal{I}_{\eta y_t}=-H_{\eta_t}$. 
The likelihood function for $\fat{\eta}_t$ and its approximation are such that
$$
L(\fat{\eta}_t)=\pi(\fat{y}_t|\fat{\eta}_t)\approx c_t \hat{L}(\fat{\eta}_t)=c_t\pi(\fat{\eta}_t|\hat{\fat{\eta}}_t) = c_t\textrm{N}(\fat{\eta}_t|\hat{\fat{\eta}}_t,\mathcal{I}_{\eta y_t}^{-1})
$$
where $c_t$ is a constant independent of $\fat{\eta}_t$.
Now let $\fat{\eta} = (\fat{\beta}_{1}\trp, ..., \fat{\beta}_{p}\trp, \fat{\tau}\trp, \fat{\varphi}\trp)\trp$ with $\fat{\beta}_1$, ..., $\fat{\beta}_p$, $\fat{\tau}$ and $\fat{\varphi}$ as before, then the likelihood of $\fat{\eta}$, $L(\fat{\eta})$, is approximated with $\hat{L}(\fat{\eta})$,
that is 
$$
L(\fat{\eta})=\prod_{i=1}^T L(\fat{\eta}_t) \approx c\hat{L}(\fat{\eta}) =\prod_{i=1}^T c_t\hat{L}(\fat{\eta}_t) = c\textrm{N}(\fat{\eta}|\hat{\fat{\eta}}, Q_{\eta y}^{-1})
$$
where $\hat{\fat{\eta}}=(\hat{\fat{\beta}}_1\trp, ..., \hat{\fat{\beta}}_p\trp, \hat{\fat{\tau}}\trp, \hat{\fat{\varphi}}\trp)\trp$ and $c=\prod_{t=1}^{T}c_t$.
Define 
$$
\hat{\fat{\eta}}^*:=(\hat{\beta}_{1,1}, ..., \hat{\beta}_{1,p}, \hat{\tau}_1,\hat{\varphi}_1,...,\hat{\beta}_{T,1}, ..., \hat{\beta}_{T,p}, \hat{\tau}_T,\hat{\varphi}_T)\trp
$$ 
as a rearrangement of $\hat{\fat{\eta}}$ then  
$$(Q_{\eta y}^*)^{-1}=\textrm{bdiag}(\mathcal{I}_{\eta y_1}^{-1},...,\mathcal{I}_{\eta y_T}^{-1}),$$
is known, and it is possible to rearrange $(Q_{\eta y}^*)^{-1}$ accordingly to get $Q_{\eta y}^{-1}$.  

In the psuedo Gaussian--Gaussian model that is used for inference, $\hat{\fat{\eta}}$ is taken as the data and $Q_{\eta y}^{-1}$ as fixed.
The approximate data distribution of $\hat{\fat{\eta}}$ is  
$$
\pi(\fat{\hat{\eta}}|\fat{\eta}, \fat{\nu}, \fat{\theta}) = \textrm{N}(\fat{\hat{\eta}}|\fat{\eta},Q_{\eta y}^{-1})
$$
where $\fat{\nu}$ are all unknown parameters that are assigned a Gaussian prior distribution and $\fat{\theta}$ are all unknown hyperparameters. 
The vectors $\fat{x}$, $\fat{\eta}$, $\fat{\nu}$ and $\fat{\theta}$ are such that 
$$
\fat{x}\trp=(\fat{\eta}\trp,\fat{\nu}\trp), \quad \fat{\eta}\trp = (\fat\beta_1\trp,...,\fat\beta_p\trp,\fat\tau\trp,\fat\varphi\trp), \quad \fat{\nu}\trp = (\fat{u}_{\beta,1}\trp,...,
\fat{u}_{\beta,p}\trp, \fat{u}_{\tau}\trp,\fat{u}_{\varphi}\trp)
$$ 
and  
$$
\fat{\theta}=(\theta_1,...,\theta_{p+2})\trp = (\sigma_{u,\beta,1}, ..., \sigma_{u,\beta,p},\sigma_{u, \tau},\sigma_{u, \varphi})\trp.
$$
Posterior samples are generated with the sampling schemes described in Section \ref{sec:inferscheme}, however, further details will be omitted.

\subsubsection{Results}

A simulation study was conducted using the model above with $p=1$, so, at time $t$ the parameters are $\fat{\eta}_t=(\beta_{t,1},\tau_t,\varphi_t)$. The likelihood function of $(\beta_{t,1},\tau_t,\varphi_t)$ for a particular time point was explored. The true values of the parameters in this simulation were $\beta_{t,1}=10$, $\tau_t=0$ ($\sigma_t=1$) and $\varphi_t=2.0794$ ($\vartheta_t=8$). The degrees of freedom parameter can be difficult to infer. To regulate the log-degrees of freedom parameter, $\varphi_t$, we opt for a log-gamma prior density. Assume that according to prior knowledge the degrees of freedom are in the interval from $1$ to $30$. Thus, we select a log-gamma prior density with parameters $\alpha=2$ and $\gamma=0.2$, which has 95\% of its probability mass over the interval $(0.1915,3.327)$. In terms of the degrees of freedom parameter, $\vartheta_t$, this translates to the interval $(1.211,27.858)$. 

\begin{figure}[h]
\centering
\includegraphics[width=0.99\textwidth]{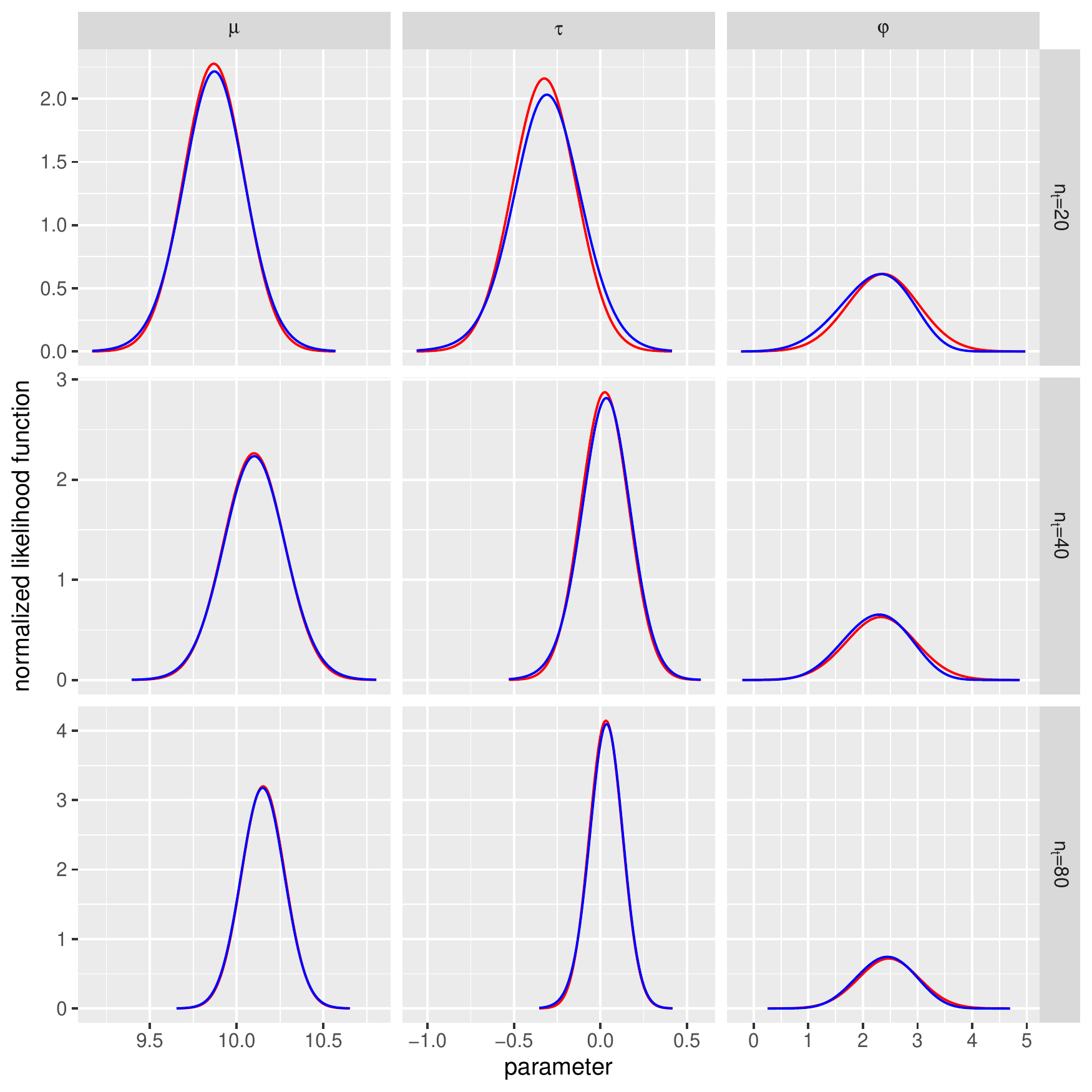} 
\caption{Linear regression model with temporally varying regression coefficients. In this model the error terms follow a $t$-distribution, and its scale and degree of freedom parameters also vary with time. The marginal normalized generalized likelihood functions (blue lines) of $\beta_{t,1}$ (left column of panel), $\tau_t=\log(\sigma_t)$ (middle column of panel) and $\varphi_t=\log(\vartheta_t)$ (right column of panel) are shown along with the marginal Gaussian approximations based on the ML estimates and the observed information matrix (red lines). The top row, middle row and bottom row of the panel show the results for $n_t=20,40,80$, respectively.}
\label{fig:normlikhood3}
\end{figure}

Figure \ref{fig:normlikhood3} shows the normalized generalized likelihood functions through the marginal normalized generalized likelihood functions of $\beta_{t,1}$, $\tau_t=\log(\sigma_t)$ and $\varphi_t=\log(\vartheta_t)$ when the sample size at time $t$, $n_t$, is $20$, $40$ and $80$. The Gaussian approximation based on the ML estimates and the observed information matrix is also shown in Figure \ref{fig:normlikhood3} in terms of the marginal Gaussian approximations of $\beta_{t,1}$, $\tau_t$ and $\varphi_t$. The Gaussian approximation for $\beta_{t,1}$ is very good for the three sample sizes and for $\tau_t$ and $\varphi_t$ when $n_t=40,80$. The Gaussian approximation is reasonably good for $\tau_t$ and $\varphi_t$ when $n_t=20$, however, a small degree of skewness is apparent in the two marginal generalized likelihood functions in this case. This skewness vanishes as the sample size increases. The generalized likelihood function is influenced by the selected log-gamma prior density for $\varphi_t$. The prior density for $\varphi_t$ can be useful for regularizing the inference for the unknown parameters when the sample size is $80$ or less since in these cases the unregularized likelihood function provides a limited amount of information about the degrees of freedom parameter which affects the information about the other parameters. 

\subsection{Spatio-temporal model for count data}

\subsubsection{Statistical model}

In this section $y_{i,t}$ denotes the observed count at lattice point $i$ and time $t$. We assume that the counts follow a Poisson distribution, and that $y_{i,t}$ has mean $\lambda_{i,t}$, i.e.,
$$
y_{i,t} \sim \text{Poisson}(\lambda_{i,t}), \quad \lambda_{i,t}>0,  
$$
where $t\in \{1,...,T\}$, $T$ is the number of time points, $i\in \{1,...,N\}$, $N$ is the number of lattice points, and the lattice has dimensions $n_1$ and $n_2$, so, $N=n_1 n_2$. The spatio-temporal cell at lattice point $i$ and time $t$ forms a group, so, the total number of spatio-temporal groups is $G=NT$. The mean of $y_{i,t}$ is transformed to $\eta_{i,t}=\log(\lambda_{i,t})$, $\eta_{i,t}\in\mathbb{R}$. The data density of $y_{i,t}$ is given by
$$
\pi(y_{i,t}|\eta_{i,t})= \frac{1}{y_{i,t}!}\exp\{\eta_{i,t}y_{i,t}-\exp(\eta_{i,t})\}.
$$  
The parameters are collected into vectors by grouping first over the lattice points, i.e., $\fat{\eta}_t=(\eta_{1,t},...,\eta_{N,t})$, $t\in \{1,...,T\}$. The latent level consists of the following model
$$
\fat{\eta}_t=\fat{u}_t+\fat{\epsilon}_t, \quad 
\fat{u}_{t+1}=\fat{u}_t+\fat{v}_t,
$$
so, $\fat{u}_{t}$ is modeled as a vector random walk where the vector $\fat{v}_t$ is independent of $\fat{u}_t$,
and $\fat{v}_t$ modeled as a Gaussian random vector with mean zero and precision matrix $\sigma^{-2}_{u,\eta}Q_u$ where $Q_u$ is given by equation (\ref{eq:qumatrix}) and $\sigma^{2}_{u,\eta}$ is an unknown variance parameter. In fact, $\fat{v}_t$ is modeled as a first-order IGMRF \citep[Section 3.3.2, pp. 104--108]{rue2005gaussian}, and $\sigma^{2}_{u,\eta}/4$ is the conditional variance of an element of $\fat{v}_t$ located at an interior point of the lattice.

\subsubsection{Inference}

Max-and-Smooth is applied to infer $\fat{\eta}$ and $\sigma_{u,\eta}$. To stabilize the likelihood function of $\eta_{i,t}$ in the case of a small $y_{i,t}$, it is multiplied by a log-gamma prior density with parameters $\alpha$ and $\gamma$, namely,
$$
\pi(\eta_{i,t})=\frac{\gamma^\alpha}{\Gamma(\alpha)}\exp\{\alpha \eta_{i,t}-\gamma\exp(\eta_{i,t})\}.
$$
The mean and variance of this density are $\psi(\alpha)-\log(\gamma)$ and $\psi'(\alpha)$, respectively.
The corresponding normalized generalized likelihood function is
$$
L(\eta_{i,t})=\frac{(\gamma+1)^{\alpha+y_{i,t}}}{\Gamma(\alpha+y_{i,t})}\exp\{(\alpha+y_{i,t})\eta_{i,t}-(\gamma+1)\exp(\eta_{i,t})\}.
$$
and it has mean and variance
$$
\tilde\eta_{i,t}=\psi(\alpha+y_{i,t}) - \log(\gamma+1), \quad
\omega^2_{\eta,i,t}=\psi'(\alpha+y_{i,t}).
$$
The mode and inverse observed information of the normalized generalized likelihood function are
$$
\hat\eta_{i,t}=\log(\alpha+y_{i,t}) - \log(\gamma+1), \quad
\sigma^2_{\eta,i,t}=(\alpha+y_{i,t})^{-1}.
$$
Thus, the two Gaussian approximations of the generalized likelihood function are,
$$
\hat{L}(\eta_{i,t})=\textrm{N}(\eta_{i,t}|\hat{\eta}_{i,t},\sigma^2_{\eta,i,t}), \quad
\tilde{L}(\eta_{i,t})=\textrm{N}(\eta_{i,t}|\tilde{\eta}_{i,t},\omega^2_{\eta,i,t}).
$$ 

In the case of $y_{i,t}>0$ the normalized likelihood function and its two Gaussian approximations are a special case of the above formulas, namely, the values of $\alpha$ and $\gamma$ are both set equal to zero. The normalized likelihood function is
$$
L(\eta_{i,t})=\frac{1}{\Gamma(y_{i,t})}\exp\{y_{i,t}\eta_{i,t}-\exp(\eta_{i,t})\}.
$$
and it has mean and variance
$$
\tilde\eta_{i,t}=\psi(y_{i,t}), \quad
\omega^2_{\eta,i,t}=\psi'(y_{i,t}).
$$
The mode and inverse observed information of the normalized likelihood function are
$$
\hat\eta_{i,t}=\log(y_{i,t}), \quad
\sigma^2_{\eta,i,t}=(y_{i,t})^{-1}.
$$
The two Gaussian approximations of the normalized likelihood function are specified with the means and variances above. 

Now let $\fat{\eta}=(\fat{\eta}_1\trp,...,\fat{\eta}_T\trp)\trp$, and let $L(\fat{\eta})$ denote either the likelihood function or the generalized likelihood function of $\fat{\eta}$.
$L(\fat{\eta})$ is approximated with either $\hat{L}(\fat{\eta})$ or $\tilde{L}(\fat{\eta})$ where
$$
L(\fat{\eta})=\prod_{t=1}^T \prod_{i=1}^N L(\eta_{i,t}) \approx c\hat{L}(\fat{\eta}) =\prod_{t=1}^T \prod_{i=1}^N c_{i,t}\hat{L}(\eta_{i,t}) = \prod_{t=1}^T \prod_{i=1}^N c_{i,t} \textrm{N}(\eta_{i,t}|\hat{\eta}_{i,t}, \sigma^2_{\eta,i,t})
$$
and 
$$
L(\fat{\eta})=\prod_{t=1}^T \prod_{i=1}^N L(\eta_{i,t}) \approx c\tilde{L}(\fat{\eta}) =\prod_{t=1}^T \prod_{i=1}^N c_{i,t}\tilde{L}(\eta_{i,t}) = \prod_{t=1}^T \prod_{i=1}^N c_{i,t} \textrm{N}(\eta_{i,t}|\tilde{\eta}_{i,t}, \omega^2_{\eta,i,t}).
$$

The Gaussian--Gaussian pseudo model treats either
$\hat{\eta}_{i,t}$ or $\tilde{\eta}_{i,t}$ as the data,
and the corresponding known error variances are $\sigma^2_{\eta,i,t}$ or $\omega^2_{\eta,i,t}$ respectively. 
The approximate data densities of $\hat{\eta}_{i,t}$ and $\tilde{\eta}_{i,t}$ are  
$$
\pi(\hat{\eta}_{i,t}|\eta_{i,t},\fat{\nu},\theta) = \textrm{N}(\hat{\eta}_{i,t}|\eta_{i,t},\sigma^2_{\eta,i,t}),
\quad \pi(\tilde{\eta}_{i,t}|\eta_{i,t}, \fat{\nu},\theta) = \textrm{N}(\tilde{\eta}_{i,t}|\eta_{i,t},\omega^2_{\eta,i,t}),
$$
for $i=1,...,N$, $t=1,...,T$,
where $\fat{\nu}$ are all the unknown parameters that are assigned a Gaussian prior distribution and $\theta$ is the unknown hyperparameter. 
The vectors $\fat{x}$ and $\fat{\nu}$ are such that 
$$
\fat{x}\trp=(\fat{\eta}\trp,\fat{\nu}\trp), \quad \fat{\nu}\trp = (\fat{u}_{1}\trp,...,\fat{u}_{T}\trp),
$$ 
where $\fat{\eta}$ is as above, and $\theta=\sigma_{u,\eta}$.
Max-and-Smooth is used to infer the unknown parameters, see Section \ref{sec:inferscheme} and in Section 3.3 in the main paper. Further details will not be specified here.

\subsubsection{Results}

Here we look at two scenarios. The first scenario is such that we assume that the number of counts is small, i.e., $y_{i,t}=0$, $y_{i,t}=1$ and $y_{i,t}=2$. For the first scenario we select a log-gamma prior density for $\eta_{i,t}$ with parameters $\alpha=2$ and $\gamma=8$. The product of the prior density and the likelihood function form the generalized likelihood function of $\eta_{i,t}$. This prior density for $\eta_{i,t}$ is such that 95\% of its probability mass is over the interval $(-3.4974,-0.3618)$ which translates to the interval $(0.03028,0.69646)$ in terms of $\lambda_{i,t}$. This is an informative prior and should only be used if there is a priori high certainty of $\lambda_{i,t}$ taking values in this interval. Under the second scenario larger counts are expected, in particular, we assume that observed counts between $5$ and $100$ are expected. Furthermore, the normalized likelihood function is used, not the normalized generalized likelihood function.

\begin{figure}[h]
\centering
\includegraphics[width=0.99\textwidth]{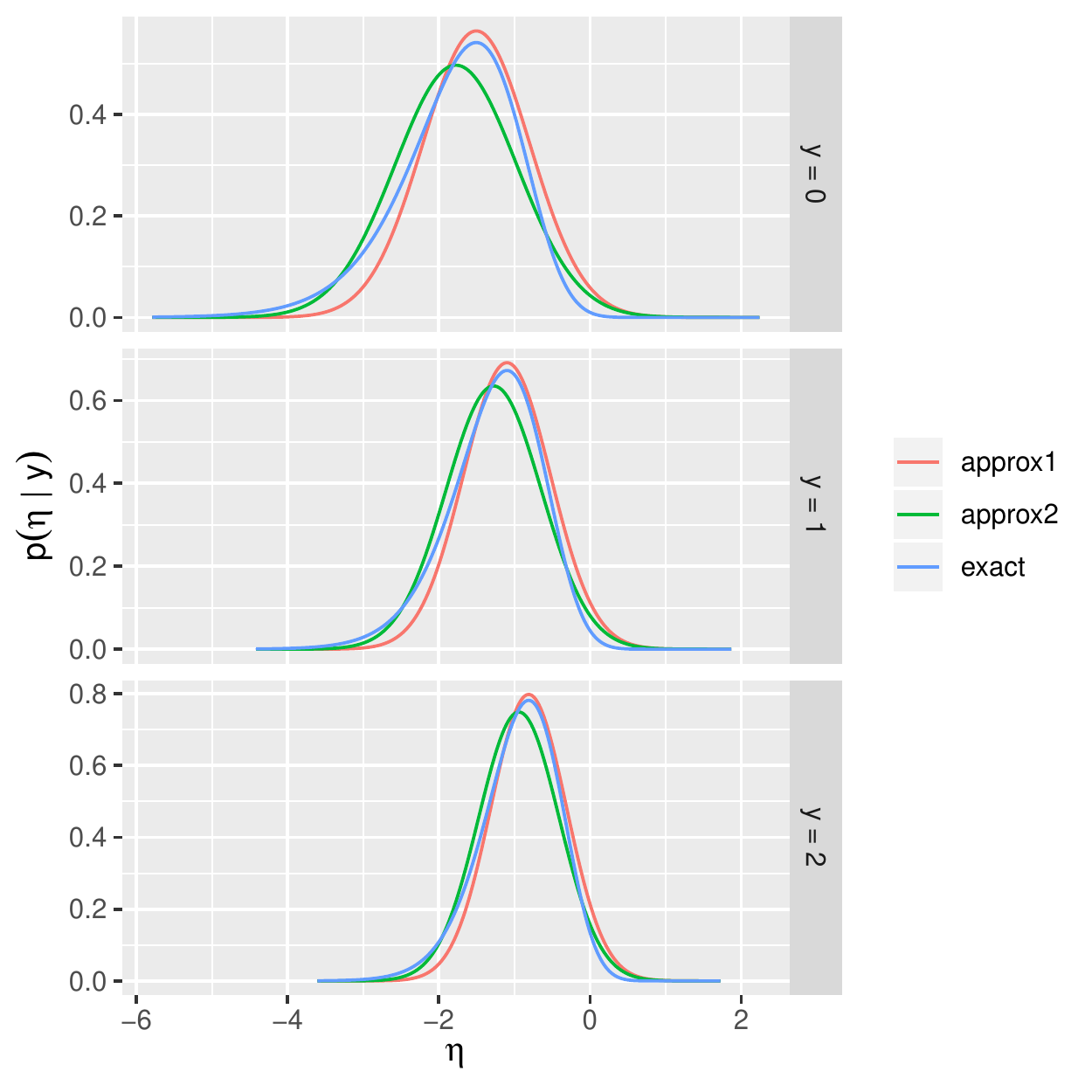}
\caption{Spatio-temporal model for count data. The normalized generalized likelihood functions of $\eta$ (blue lines) plotted along with the Gaussian approximations based on (a) the generalized ML estimates and the inverse of the observed information (red lines), and (b) the mean and variance of the normalized generalized likelihood function (green lines). The top row, the middle row and the bottom row of the panel show the results for $y=0,1,2$, respectively. The likelihood function is multiplied by a log-gamma prior density with parameters $\alpha=2$ and $\gamma=8$.}
\label{fig:normlikhood4}
\end{figure}

Figure \ref{fig:normlikhood4} shows the normalized generalized likelihood function of $\eta_{i,t}$ when the observed value is $y_{i,t}=0,1,2$, and the corresponding Gaussian approximations based on the generalized ML estimates and the inverse of the observed information, and the one based on the mean and variance of the normalized generalized likelihood function. In the case of all the three values of $y_{i,t}$, the Gaussian approximation based the mean and variance of the normalized generalized likelihood function is giving a better match since it captures better the variability of the true normalized generalized likelihood function and is closer to its tails than the Gaussian approximation based on the generalized ML estimates. 

\begin{figure}[h]
\centering
\includegraphics[width=0.99\textwidth]{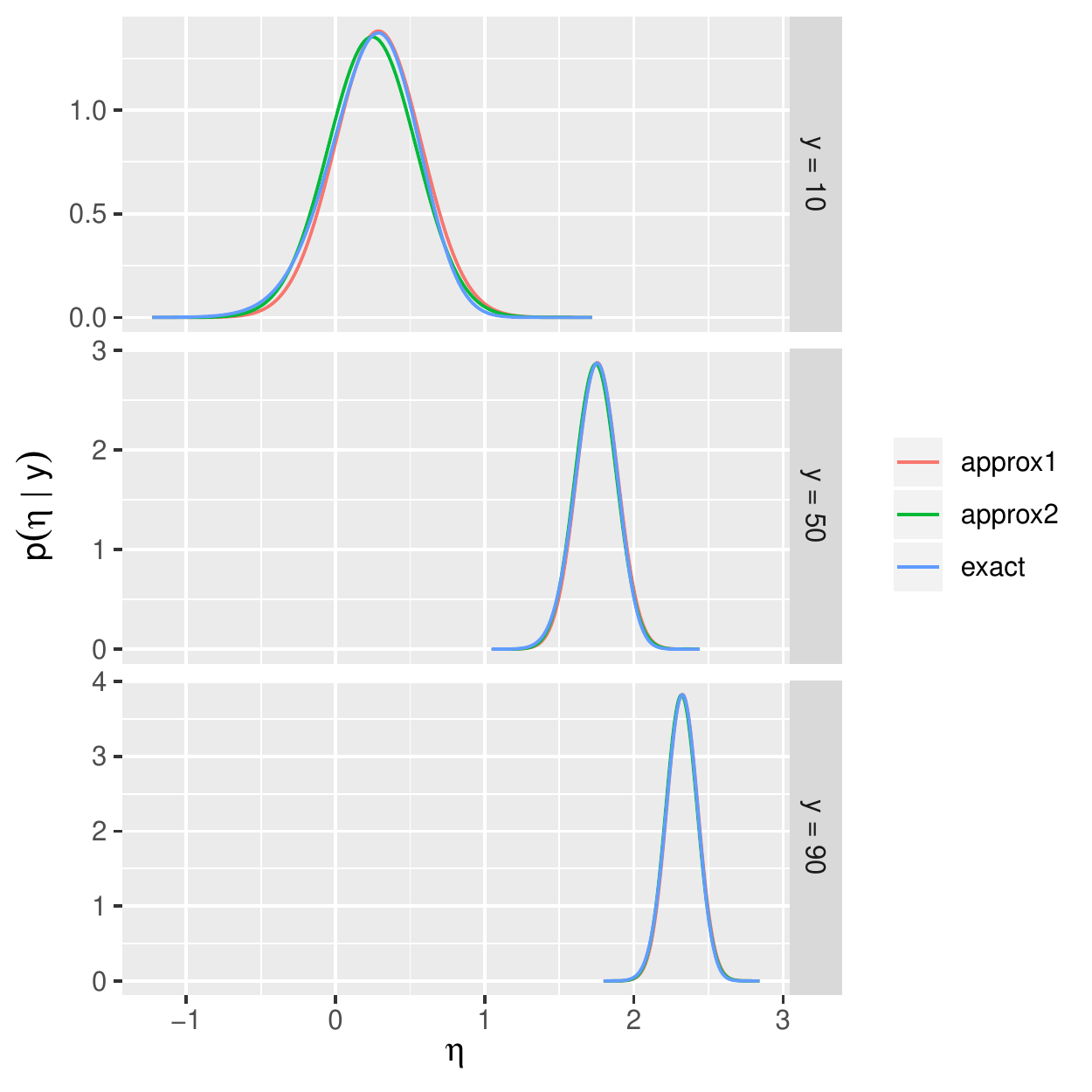}
\caption{Spatio-temporal model for count data. The normalized likelihood functions of $\eta$ (blue lines) plotted along with the Gaussian approximations based on (a) the ML estimates and the inverse of the observed information (red lines), and (b) the mean and variance of the normalized likelihood function (green lines). The top row, the middle row and the bottom row of the panel show the results for $y=10,50,90$, respectively.}
\label{fig:normlikhood5}
\end{figure}

Figure \ref{fig:normlikhood5} shows results for the second scenario using the same setup as in Figure \ref{fig:normlikhood4}. Under this scenario the observed values of $y_{i,t}$ are equal to $10$, $50$ and $90$, and the normalized likelihood function is explored as opposed to the normalized generalized likelihood function. The two Gaussian approximations match the normalized generalized likelihood function very well in the case of $y_{i,t}=50$ and $y_{i,t}=90$. In the case of $y_{i,t}=10$ the match is reasonably good for both of the Gaussian approximations. However, it is slightly better in the case of the Gaussian approximation with the mean and variance of the normalized likelihood function which has mean and standard deviation $2.252$ and $0.3243$, respectively, while the Gaussian approximation based on the ML estimates has mean and standard deviation $2.303$ and $0.3162$, respectively.

\bibliographystyle{aps-nameyear}     
